\documentclass[11pt]{article}

\usepackage[letterpaper,margin=1in]{geometry}
\usepackage{amsmath}
\usepackage{array}
\usepackage{booktabs}
\usepackage{graphicx}
\usepackage{subcaption,comment}
\usepackage{placeins}
\usepackage{pgfplots}
\usepackage{xcolor}
\usepackage[hidelinks]{hyperref}
\usepackage{microtype}

\makeatletter
\renewcommand{\@cite}[2]{\textsuperscript{#1}}
\makeatother
\pgfplotsset{compat=1.18}
\usetikzlibrary{calc}

\newcommand{\posterroot}{.}

\definecolor{mainblue}{HTML}{1F77B4}
\definecolor{mainorange}{HTML}{E08214}
\definecolor{mainred}{HTML}{B23630}
\definecolor{maingreen}{HTML}{2CA02C}
\definecolor{methodiigreen}{HTML}{6ABD45}
\definecolor{lightline}{HTML}{D7DEE8}
\definecolor{trackblue}{HTML}{3E6A99}
\definecolor{adsviolet}{HTML}{6B4FA3}
\newcommand{\rev}[1]{#1}
\newcommand{\revb}[1]{#1}
\pgfplotsset{
  paper small audio default/.style={
    only marks,mark=*,mark size=0.36pt,mark layer=like plot,
    draw=black,fill=black,opacity=0.86},
  paper small audio contrast/.style={
    only marks,mark=*,mark size=0.40pt,mark layer=like plot,
    draw=black,fill=black,opacity=0.92},
  paper small method ii default/.style={
    methodiigreen,line width=1.15pt},
  paper small method ii contrast/.style={
    methodiigreen,line width=1.15pt}
}

\expandafter\newcommand\csname sr01n86lplowridgeRank\endcsname{1}
\expandafter\newcommand\csname sr01n86lplowridgeLabel\endcsname{N86LP low ridge}
\expandafter\newcommand\csname sr01n86lplowridgeDataset\endcsname{N86LP cleaned high+low ridges}
\expandafter\newcommand\csname sr01n86lplowridgeGeometry\endcsname{near overhead}
\expandafter\newcommand\csname sr01n86lplowridgeVCat\endcsname{V constant}
\expandafter\newcommand\csname sr01n86lplowridgeHCat\endcsname{H varying}
\expandafter\newcommand\csname sr01n86lplowridgeAudio\endcsname{plateau-like + curve but scattered/ambiguous}
\expandafter\newcommand\csname sr01n86lplowridgeClosest\endcsname{2026-05-17 22:20:57.90 UTC}
\expandafter\newcommand\csname sr01n86lplowridgeXMin\endcsname{-53.8}
\expandafter\newcommand\csname sr01n86lplowridgeXMax\endcsname{58.5}
\expandafter\newcommand\csname sr01n86lplowridgeYMin\endcsname{185.6}
\expandafter\newcommand\csname sr01n86lplowridgeYMax\endcsname{340.4}
\expandafter\newcommand\csname sr01n86lplowridgeFzero\endcsname{242.5}
\expandafter\newcommand\csname sr01n86lplowridgeXStartTick\endcsname{-50}
\expandafter\newcommand\csname sr01n86lplowridgeXEndTick\endcsname{50}
\expandafter\newcommand\csname sr01n86lplowridgeYLowTick\endcsname{195}
\expandafter\newcommand\csname sr01n86lplowridgeYHighTick\endcsname{318}
\expandafter\newcommand\csname sr01n86lplowridgeAdsV\endcsname{297.5}
\expandafter\newcommand\csname sr01n86lplowridgeAdsH\endcsname{1535}
\expandafter\newcommand\csname sr01n86lplowridgeMiiV\endcsname{298.2}
\expandafter\newcommand\csname sr01n86lplowridgeMiiH\endcsname{1459}
\expandafter\newcommand\csname sr01n86lplowridgeMiV\endcsname{271.8}
\expandafter\newcommand\csname sr01n86lplowridgeMiH\endcsname{1389}
\expandafter\newcommand\csname sr01n86lplowridgeMiiVErr\endcsname{0.3\%}
\expandafter\newcommand\csname sr01n86lplowridgeMiiHErr\endcsname{-5.0\%}
\expandafter\newcommand\csname sr01n86lplowridgeMiVErr\endcsname{-8.6\%}
\expandafter\newcommand\csname sr01n86lplowridgeMiHErr\endcsname{-9.5\%}
\expandafter\newcommand\csname sr01n86lplowridgeRMSE\endcsname{2.10}
\expandafter\newcommand\csname sr01n86lplowridgeVRange\endcsname{7.5}
\expandafter\newcommand\csname sr01n86lplowridgeHRange\endcsname{234}
\expandafter\newcommand\csname sr01n86lplowridgeLoverH\endcsname{0.29}
\expandafter\newcommand\csname sr01n86lplowridgeTrackSource\endcsname{N86LP-track-EGM96\_avg.kml}
\expandafter\newcommand\csname sr01n86lplowridgeAdsAvailable\endcsname{1}
\expandafter\newcommand\csname sr01n86lplowridgeTrackAvailable\endcsname{1}
\expandafter\newcommand\csname sr01n86lplowridgeInsetXMin\endcsname{-48.2}
\expandafter\newcommand\csname sr01n86lplowridgeInsetXMax\endcsname{53.0}
\expandafter\newcommand\csname sr01n86lplowridgeInsetXTickLo\endcsname{-40}
\expandafter\newcommand\csname sr01n86lplowridgeInsetXTickHi\endcsname{40}
\expandafter\newcommand\csname sr01n86lplowridgeInsetVMin\endcsname{278}
\expandafter\newcommand\csname sr01n86lplowridgeInsetVMax\endcsname{311}
\expandafter\newcommand\csname sr01n86lplowridgeInsetHMin\endcsname{1349}
\expandafter\newcommand\csname sr01n86lplowridgeInsetHMax\endcsname{1689}
\expandafter\newcommand\csname sr01n86lplowridgeTrackXMin\endcsname{-8034}
\expandafter\newcommand\csname sr01n86lplowridgeTrackXMax\endcsname{7633}
\expandafter\newcommand\csname sr01n86lplowridgeTrackYMin\endcsname{-40}
\expandafter\newcommand\csname sr01n86lplowridgeTrackYMax\endcsname{490}
\expandafter\newcommand\csname sr01n86lplowridgeTrackXTickLo\endcsname{-5000}
\expandafter\newcommand\csname sr01n86lplowridgeTrackXTickHi\endcsname{5000}
\expandafter\newcommand\csname sr01n86lplowridgeTrackYTickLo\endcsname{0}
\expandafter\newcommand\csname sr01n86lplowridgeTrackYTickHi\endcsname{400}
\expandafter\newcommand\csname sr01n86lplowridgeTrackYTick\endcsname{400}
\expandafter\newcommand\csname sr01n86lplowridgeTrackClosestY\endcsname{444}
\expandafter\newcommand\csname sr02n9393tRank\endcsname{2}
\expandafter\newcommand\csname sr02n9393tLabel\endcsname{N9393T}
\expandafter\newcommand\csname sr02n9393tDataset\endcsname{N9393T}
\expandafter\newcommand\csname sr02n9393tGeometry\endcsname{far away/non-overhead}
\expandafter\newcommand\csname sr02n9393tVCat\endcsname{V constant}
\expandafter\newcommand\csname sr02n9393tHCat\endcsname{H varying}
\expandafter\newcommand\csname sr02n9393tAudio\endcsname{smooth plateau-like + curve}
\expandafter\newcommand\csname sr02n9393tClosest\endcsname{2026-05-17 22:04:16.52 UTC}
\expandafter\newcommand\csname sr02n9393tXMin\endcsname{-87.8}
\expandafter\newcommand\csname sr02n9393tXMax\endcsname{53.7}
\expandafter\newcommand\csname sr02n9393tYMin\endcsname{296.1}
\expandafter\newcommand\csname sr02n9393tYMax\endcsname{519.8}
\expandafter\newcommand\csname sr02n9393tFzero\endcsname{369.6}
\expandafter\newcommand\csname sr02n9393tXStartTick\endcsname{-80}
\expandafter\newcommand\csname sr02n9393tXEndTick\endcsname{50}
\expandafter\newcommand\csname sr02n9393tYLowTick\endcsname{310}
\expandafter\newcommand\csname sr02n9393tYHighTick\endcsname{488}
\expandafter\newcommand\csname sr02n9393tAdsV\endcsname{290.7}
\expandafter\newcommand\csname sr02n9393tAdsH\endcsname{2672}
\expandafter\newcommand\csname sr02n9393tMiiV\endcsname{301.0}
\expandafter\newcommand\csname sr02n9393tMiiH\endcsname{3169}
\expandafter\newcommand\csname sr02n9393tMiV\endcsname{247.6}
\expandafter\newcommand\csname sr02n9393tMiH\endcsname{1605}
\expandafter\newcommand\csname sr02n9393tMiiVErr\endcsname{3.6\%}
\expandafter\newcommand\csname sr02n9393tMiiHErr\endcsname{18.6\%}
\expandafter\newcommand\csname sr02n9393tMiVErr\endcsname{-14.8\%}
\expandafter\newcommand\csname sr02n9393tMiHErr\endcsname{-39.9\%}
\expandafter\newcommand\csname sr02n9393tRMSE\endcsname{3.77}
\expandafter\newcommand\csname sr02n9393tVRange\endcsname{6.4}
\expandafter\newcommand\csname sr02n9393tHRange\endcsname{510}
\expandafter\newcommand\csname sr02n9393tLoverH\endcsname{0.51}
\expandafter\newcommand\csname sr02n9393tTrackSource\endcsname{N9393T-track-EGM96\_avg.kml}
\expandafter\newcommand\csname sr02n9393tAdsAvailable\endcsname{1}
\expandafter\newcommand\csname sr02n9393tTrackAvailable\endcsname{1}
\expandafter\newcommand\csname sr02n9393tInsetXMin\endcsname{-80.8}
\expandafter\newcommand\csname sr02n9393tInsetXMax\endcsname{46.7}
\expandafter\newcommand\csname sr02n9393tInsetXTickLo\endcsname{-60}
\expandafter\newcommand\csname sr02n9393tInsetXTickHi\endcsname{20}
\expandafter\newcommand\csname sr02n9393tInsetVMin\endcsname{276}
\expandafter\newcommand\csname sr02n9393tInsetVMax\endcsname{305}
\expandafter\newcommand\csname sr02n9393tInsetHMin\endcsname{2245}
\expandafter\newcommand\csname sr02n9393tInsetHMax\endcsname{3114}
\expandafter\newcommand\csname sr02n9393tTrackXMin\endcsname{-11152}
\expandafter\newcommand\csname sr02n9393tTrackXMax\endcsname{7852}
\expandafter\newcommand\csname sr02n9393tTrackYMin\endcsname{-108}
\expandafter\newcommand\csname sr02n9393tTrackYMax\endcsname{1453}
\expandafter\newcommand\csname sr02n9393tTrackXTickLo\endcsname{-10000}
\expandafter\newcommand\csname sr02n9393tTrackXTickHi\endcsname{5000}
\expandafter\newcommand\csname sr02n9393tTrackYTickLo\endcsname{0}
\expandafter\newcommand\csname sr02n9393tTrackYTickHi\endcsname{1453}
\expandafter\newcommand\csname sr02n9393tTrackYTick\endcsname{1453}
\expandafter\newcommand\csname sr02n9393tTrackClosestY\endcsname{1331}
\expandafter\newcommand\csname sr03n788bmRank\endcsname{3}
\expandafter\newcommand\csname sr03n788bmLabel\endcsname{N788BM}
\expandafter\newcommand\csname sr03n788bmDataset\endcsname{N788BM pp\_pixel vs Lab Pixel}
\expandafter\newcommand\csname sr03n788bmGeometry\endcsname{far away/non-overhead}
\expandafter\newcommand\csname sr03n788bmVCat\endcsname{V constant}
\expandafter\newcommand\csname sr03n788bmHCat\endcsname{H constant}
\expandafter\newcommand\csname sr03n788bmAudio\endcsname{smooth plateau-like + curve}
\expandafter\newcommand\csname sr03n788bmClosest\endcsname{2026-05-17 19:30:19.47 UTC}
\expandafter\newcommand\csname sr03n788bmXMin\endcsname{-38.4}
\expandafter\newcommand\csname sr03n788bmXMax\endcsname{26.0}
\expandafter\newcommand\csname sr03n788bmYMin\endcsname{236.0}
\expandafter\newcommand\csname sr03n788bmYMax\endcsname{430.0}
\expandafter\newcommand\csname sr03n788bmFzero\endcsname{305.0}
\expandafter\newcommand\csname sr03n788bmXStartTick\endcsname{-40}
\expandafter\newcommand\csname sr03n788bmXEndTick\endcsname{20}
\expandafter\newcommand\csname sr03n788bmYLowTick\endcsname{248}
\expandafter\newcommand\csname sr03n788bmYHighTick\endcsname{402}
\expandafter\newcommand\csname sr03n788bmAdsV\endcsname{308.5}
\expandafter\newcommand\csname sr03n788bmAdsH\endcsname{936}
\expandafter\newcommand\csname sr03n788bmMiiV\endcsname{300.7}
\expandafter\newcommand\csname sr03n788bmMiiH\endcsname{954}
\expandafter\newcommand\csname sr03n788bmMiV\endcsname{252.9}
\expandafter\newcommand\csname sr03n788bmMiH\endcsname{461}
\expandafter\newcommand\csname sr03n788bmMiiVErr\endcsname{-2.5\%}
\expandafter\newcommand\csname sr03n788bmMiiHErr\endcsname{2.0\%}
\expandafter\newcommand\csname sr03n788bmMiVErr\endcsname{-18.0\%}
\expandafter\newcommand\csname sr03n788bmMiHErr\endcsname{-50.8\%}
\expandafter\newcommand\csname sr03n788bmRMSE\endcsname{2.86}
\expandafter\newcommand\csname sr03n788bmVRange\endcsname{4.4}
\expandafter\newcommand\csname sr03n788bmHRange\endcsname{38}
\expandafter\newcommand\csname sr03n788bmLoverH\endcsname{0.92}
\expandafter\newcommand\csname sr03n788bmTrackSource\endcsname{N788BM-track-EGM96\_avg.kml}
\expandafter\newcommand\csname sr03n788bmAdsAvailable\endcsname{1}
\expandafter\newcommand\csname sr03n788bmTrackAvailable\endcsname{1}
\expandafter\newcommand\csname sr03n788bmInsetXMin\endcsname{-35.2}
\expandafter\newcommand\csname sr03n788bmInsetXMax\endcsname{22.8}
\expandafter\newcommand\csname sr03n788bmInsetXTickLo\endcsname{-30}
\expandafter\newcommand\csname sr03n788bmInsetXTickHi\endcsname{10}
\expandafter\newcommand\csname sr03n788bmInsetVMin\endcsname{285}
\expandafter\newcommand\csname sr03n788bmInsetVMax\endcsname{316}
\expandafter\newcommand\csname sr03n788bmInsetHMin\endcsname{890}
\expandafter\newcommand\csname sr03n788bmInsetHMax\endcsname{974}
\expandafter\newcommand\csname sr03n788bmTrackXMin\endcsname{-6285}
\expandafter\newcommand\csname sr03n788bmTrackXMax\endcsname{5449}
\expandafter\newcommand\csname sr03n788bmTrackYMin\endcsname{-69}
\expandafter\newcommand\csname sr03n788bmTrackYMax\endcsname{935}
\expandafter\newcommand\csname sr03n788bmTrackXTickLo\endcsname{-4000}
\expandafter\newcommand\csname sr03n788bmTrackXTickHi\endcsname{4000}
\expandafter\newcommand\csname sr03n788bmTrackYTickLo\endcsname{0}
\expandafter\newcommand\csname sr03n788bmTrackYTickHi\endcsname{800}
\expandafter\newcommand\csname sr03n788bmTrackYTick\endcsname{800}
\expandafter\newcommand\csname sr03n788bmTrackClosestY\endcsname{854}
\expandafter\newcommand\csname sr04n784mbRank\endcsname{4}
\expandafter\newcommand\csname sr04n784mbLabel\endcsname{N784MB}
\expandafter\newcommand\csname sr04n784mbDataset\endcsname{N784MB lower ridge}
\expandafter\newcommand\csname sr04n784mbGeometry\endcsname{far away/non-overhead}
\expandafter\newcommand\csname sr04n784mbVCat\endcsname{V constant}
\expandafter\newcommand\csname sr04n784mbHCat\endcsname{H constant}
\expandafter\newcommand\csname sr04n784mbAudio\endcsname{smooth plateau-like + curve}
\expandafter\newcommand\csname sr04n784mbClosest\endcsname{2026-05-14 17:42:09 UTC}
\expandafter\newcommand\csname sr04n784mbXMin\endcsname{-42.7}
\expandafter\newcommand\csname sr04n784mbXMax\endcsname{40.8}
\expandafter\newcommand\csname sr04n784mbYMin\endcsname{179.7}
\expandafter\newcommand\csname sr04n784mbYMax\endcsname{360.9}
\expandafter\newcommand\csname sr04n784mbFzero\endcsname{245.8}
\expandafter\newcommand\csname sr04n784mbXStartTick\endcsname{-40}
\expandafter\newcommand\csname sr04n784mbXEndTick\endcsname{40}
\expandafter\newcommand\csname sr04n784mbYLowTick\endcsname{191}
\expandafter\newcommand\csname sr04n784mbYHighTick\endcsname{335}
\expandafter\newcommand\csname sr04n784mbAdsV\endcsname{328.7}
\expandafter\newcommand\csname sr04n784mbAdsH\endcsname{1947}
\expandafter\newcommand\csname sr04n784mbMiiV\endcsname{328.0}
\expandafter\newcommand\csname sr04n784mbMiiH\endcsname{1991}
\expandafter\newcommand\csname sr04n784mbMiV\endcsname{273.0}
\expandafter\newcommand\csname sr04n784mbMiH\endcsname{1196}
\expandafter\newcommand\csname sr04n784mbMiiVErr\endcsname{-0.2\%}
\expandafter\newcommand\csname sr04n784mbMiiHErr\endcsname{2.3\%}
\expandafter\newcommand\csname sr04n784mbMiVErr\endcsname{-17.0\%}
\expandafter\newcommand\csname sr04n784mbMiHErr\endcsname{-38.6\%}
\expandafter\newcommand\csname sr04n784mbRMSE\endcsname{2.70}
\expandafter\newcommand\csname sr04n784mbVRange\endcsname{2.7}
\expandafter\newcommand\csname sr04n784mbHRange\endcsname{0}
\expandafter\newcommand\csname sr04n784mbLoverH\endcsname{0.51}
\expandafter\newcommand\csname sr04n784mbTrackSource\endcsname{N784MB-track-EGM96\_avg.kml}
\expandafter\newcommand\csname sr04n784mbAdsAvailable\endcsname{1}
\expandafter\newcommand\csname sr04n784mbTrackAvailable\endcsname{1}
\expandafter\newcommand\csname sr04n784mbInsetXMin\endcsname{-38.5}
\expandafter\newcommand\csname sr04n784mbInsetXMax\endcsname{36.7}
\expandafter\newcommand\csname sr04n784mbInsetXTickLo\endcsname{-20}
\expandafter\newcommand\csname sr04n784mbInsetXTickHi\endcsname{20}
\expandafter\newcommand\csname sr04n784mbInsetVMin\endcsname{324}
\expandafter\newcommand\csname sr04n784mbInsetVMax\endcsname{335}
\expandafter\newcommand\csname sr04n784mbInsetHMin\endcsname{1927}
\expandafter\newcommand\csname sr04n784mbInsetHMax\endcsname{1967}
\expandafter\newcommand\csname sr04n784mbTrackXMin\endcsname{-8179}
\expandafter\newcommand\csname sr04n784mbTrackXMax\endcsname{8017}
\expandafter\newcommand\csname sr04n784mbTrackYMin\endcsname{-80}
\expandafter\newcommand\csname sr04n784mbTrackYMax\endcsname{1078}
\expandafter\newcommand\csname sr04n784mbTrackXTickLo\endcsname{-5000}
\expandafter\newcommand\csname sr04n784mbTrackXTickHi\endcsname{5000}
\expandafter\newcommand\csname sr04n784mbTrackYTickLo\endcsname{200}
\expandafter\newcommand\csname sr04n784mbTrackYTickHi\endcsname{800}
\expandafter\newcommand\csname sr04n784mbTrackYTick\endcsname{800}
\expandafter\newcommand\csname sr04n784mbTrackClosestY\endcsname{995}
\expandafter\newcommand\csname sr05n413akRank\endcsname{5}
\expandafter\newcommand\csname sr05n413akLabel\endcsname{N413AK}
\expandafter\newcommand\csname sr05n413akDataset\endcsname{N413AK}
\expandafter\newcommand\csname sr05n413akGeometry\endcsname{near overhead}
\expandafter\newcommand\csname sr05n413akVCat\endcsname{V constant}
\expandafter\newcommand\csname sr05n413akHCat\endcsname{H varying}
\expandafter\newcommand\csname sr05n413akAudio\endcsname{smooth plateau-like + curve}
\expandafter\newcommand\csname sr05n413akClosest\endcsname{2026-05-16 20:51:40.70 UTC}
\expandafter\newcommand\csname sr05n413akXMin\endcsname{-51.0}
\expandafter\newcommand\csname sr05n413akXMax\endcsname{97.1}
\expandafter\newcommand\csname sr05n413akYMin\endcsname{207.1}
\expandafter\newcommand\csname sr05n413akYMax\endcsname{335.2}
\expandafter\newcommand\csname sr05n413akFzero\endcsname{258.0}
\expandafter\newcommand\csname sr05n413akXStartTick\endcsname{-40}
\expandafter\newcommand\csname sr05n413akXEndTick\endcsname{90}
\expandafter\newcommand\csname sr05n413akYLowTick\endcsname{215}
\expandafter\newcommand\csname sr05n413akYHighTick\endcsname{317}
\expandafter\newcommand\csname sr05n413akAdsV\endcsname{229.4}
\expandafter\newcommand\csname sr05n413akAdsH\endcsname{1279}
\expandafter\newcommand\csname sr05n413akMiiV\endcsname{236.1}
\expandafter\newcommand\csname sr05n413akMiiH\endcsname{1357}
\expandafter\newcommand\csname sr05n413akMiV\endcsname{217.8}
\expandafter\newcommand\csname sr05n413akMiH\endcsname{1267}
\expandafter\newcommand\csname sr05n413akMiiVErr\endcsname{2.9\%}
\expandafter\newcommand\csname sr05n413akMiiHErr\endcsname{6.1\%}
\expandafter\newcommand\csname sr05n413akMiVErr\endcsname{-5.0\%}
\expandafter\newcommand\csname sr05n413akMiHErr\endcsname{-0.9\%}
\expandafter\newcommand\csname sr05n413akRMSE\endcsname{1.42}
\expandafter\newcommand\csname sr05n413akVRange\endcsname{7.2}
\expandafter\newcommand\csname sr05n413akHRange\endcsname{359}
\expandafter\newcommand\csname sr05n413akLoverH\endcsname{0.24}
\expandafter\newcommand\csname sr05n413akTrackSource\endcsname{N413AK-track-EGM96\_avg.kml}
\expandafter\newcommand\csname sr05n413akAdsAvailable\endcsname{1}
\expandafter\newcommand\csname sr05n413akTrackAvailable\endcsname{1}
\expandafter\newcommand\csname sr05n413akInsetXMin\endcsname{-43.7}
\expandafter\newcommand\csname sr05n413akInsetXMax\endcsname{89.7}
\expandafter\newcommand\csname sr05n413akInsetXTickLo\endcsname{-20}
\expandafter\newcommand\csname sr05n413akInsetXTickHi\endcsname{60}
\expandafter\newcommand\csname sr05n413akInsetVMin\endcsname{217}
\expandafter\newcommand\csname sr05n413akInsetVMax\endcsname{245}
\expandafter\newcommand\csname sr05n413akInsetHMin\endcsname{968}
\expandafter\newcommand\csname sr05n413akInsetHMax\endcsname{1490}
\expandafter\newcommand\csname sr05n413akTrackXMin\endcsname{-6285}
\expandafter\newcommand\csname sr05n413akTrackXMax\endcsname{9519}
\expandafter\newcommand\csname sr05n413akTrackYMin\endcsname{-63}
\expandafter\newcommand\csname sr05n413akTrackYMax\endcsname{849}
\expandafter\newcommand\csname sr05n413akTrackXTickLo\endcsname{-5000}
\expandafter\newcommand\csname sr05n413akTrackXTickHi\endcsname{9519}
\expandafter\newcommand\csname sr05n413akTrackYTickLo\endcsname{0}
\expandafter\newcommand\csname sr05n413akTrackYTickHi\endcsname{800}
\expandafter\newcommand\csname sr05n413akTrackYTick\endcsname{800}
\expandafter\newcommand\csname sr05n413akTrackClosestY\endcsname{305}
\expandafter\newcommand\csname sr06n373dfRank\endcsname{6}
\expandafter\newcommand\csname sr06n373dfLabel\endcsname{N373DF}
\expandafter\newcommand\csname sr06n373dfDataset\endcsname{N373DF lower ridge}
\expandafter\newcommand\csname sr06n373dfGeometry\endcsname{near overhead}
\expandafter\newcommand\csname sr06n373dfVCat\endcsname{V constant}
\expandafter\newcommand\csname sr06n373dfHCat\endcsname{H constant}
\expandafter\newcommand\csname sr06n373dfAudio\endcsname{smooth plateau-like + curve}
\expandafter\newcommand\csname sr06n373dfClosest\endcsname{2026-05-13 20:03:59 UTC}
\expandafter\newcommand\csname sr06n373dfXMin\endcsname{-18.6}
\expandafter\newcommand\csname sr06n373dfXMax\endcsname{25.3}
\expandafter\newcommand\csname sr06n373dfYMin\endcsname{173.4}
\expandafter\newcommand\csname sr06n373dfYMax\endcsname{318.8}
\expandafter\newcommand\csname sr06n373dfFzero\endcsname{227.8}
\expandafter\newcommand\csname sr06n373dfXStartTick\endcsname{-20}
\expandafter\newcommand\csname sr06n373dfXEndTick\endcsname{20}
\expandafter\newcommand\csname sr06n373dfYLowTick\endcsname{183}
\expandafter\newcommand\csname sr06n373dfYHighTick\endcsname{298}
\expandafter\newcommand\csname sr06n373dfAdsV\endcsname{328.1}
\expandafter\newcommand\csname sr06n373dfAdsH\endcsname{892}
\expandafter\newcommand\csname sr06n373dfMiiV\endcsname{327.9}
\expandafter\newcommand\csname sr06n373dfMiiH\endcsname{941}
\expandafter\newcommand\csname sr06n373dfMiV\endcsname{284.7}
\expandafter\newcommand\csname sr06n373dfMiH\endcsname{739}
\expandafter\newcommand\csname sr06n373dfMiiVErr\endcsname{-0.1\%}
\expandafter\newcommand\csname sr06n373dfMiiHErr\endcsname{5.5\%}
\expandafter\newcommand\csname sr06n373dfMiVErr\endcsname{-13.2\%}
\expandafter\newcommand\csname sr06n373dfMiHErr\endcsname{-17.2\%}
\expandafter\newcommand\csname sr06n373dfRMSE\endcsname{0.76}
\expandafter\newcommand\csname sr06n373dfVRange\endcsname{3.0}
\expandafter\newcommand\csname sr06n373dfHRange\endcsname{8}
\expandafter\newcommand\csname sr06n373dfLoverH\endcsname{0.23}
\expandafter\newcommand\csname sr06n373dfTrackSource\endcsname{N373DF-track-EGM96\_avg.kml}
\expandafter\newcommand\csname sr06n373dfAdsAvailable\endcsname{1}
\expandafter\newcommand\csname sr06n373dfTrackAvailable\endcsname{1}
\expandafter\newcommand\csname sr06n373dfInsetXMin\endcsname{-16.4}
\expandafter\newcommand\csname sr06n373dfInsetXMax\endcsname{23.2}
\expandafter\newcommand\csname sr06n373dfInsetXTickLo\endcsname{-10}
\expandafter\newcommand\csname sr06n373dfInsetXTickHi\endcsname{20}
\expandafter\newcommand\csname sr06n373dfInsetVMin\endcsname{324}
\expandafter\newcommand\csname sr06n373dfInsetVMax\endcsname{334}
\expandafter\newcommand\csname sr06n373dfInsetHMin\endcsname{868}
\expandafter\newcommand\csname sr06n373dfInsetHMax\endcsname{916}
\expandafter\newcommand\csname sr06n373dfTrackXMin\endcsname{-4360}
\expandafter\newcommand\csname sr06n373dfTrackXMax\endcsname{6029}
\expandafter\newcommand\csname sr06n373dfTrackYMin\endcsname{-40}
\expandafter\newcommand\csname sr06n373dfTrackYMax\endcsname{287}
\expandafter\newcommand\csname sr06n373dfTrackXTickLo\endcsname{-2000}
\expandafter\newcommand\csname sr06n373dfTrackXTickHi\endcsname{4000}
\expandafter\newcommand\csname sr06n373dfTrackYTickLo\endcsname{0}
\expandafter\newcommand\csname sr06n373dfTrackYTickHi\endcsname{287}
\expandafter\newcommand\csname sr06n373dfTrackYTick\endcsname{287}
\expandafter\newcommand\csname sr06n373dfTrackClosestY\endcsname{195}
\expandafter\newcommand\csname sr07n814srRank\endcsname{7}
\expandafter\newcommand\csname sr07n814srLabel\endcsname{N814SR}
\expandafter\newcommand\csname sr07n814srDataset\endcsname{N814SR}
\expandafter\newcommand\csname sr07n814srGeometry\endcsname{far away/non-overhead}
\expandafter\newcommand\csname sr07n814srVCat\endcsname{V varying}
\expandafter\newcommand\csname sr07n814srHCat\endcsname{H constant}
\expandafter\newcommand\csname sr07n814srAudio\endcsname{plateau-like + curve but scattered/ambiguous}
\expandafter\newcommand\csname sr07n814srClosest\endcsname{2026-05-12 20:22:47 UTC}
\expandafter\newcommand\csname sr07n814srXMin\endcsname{-34.6}
\expandafter\newcommand\csname sr07n814srXMax\endcsname{25.6}
\expandafter\newcommand\csname sr07n814srYMin\endcsname{310.8}
\expandafter\newcommand\csname sr07n814srYMax\endcsname{661.7}
\expandafter\newcommand\csname sr07n814srFzero\endcsname{437.5}
\expandafter\newcommand\csname sr07n814srXStartTick\endcsname{-30}
\expandafter\newcommand\csname sr07n814srXEndTick\endcsname{20}
\expandafter\newcommand\csname sr07n814srYLowTick\endcsname{333}
\expandafter\newcommand\csname sr07n814srYHighTick\endcsname{612}
\expandafter\newcommand\csname sr07n814srAdsV\endcsname{316.9}
\expandafter\newcommand\csname sr07n814srAdsH\endcsname{650}
\expandafter\newcommand\csname sr07n814srMiiV\endcsname{347.8}
\expandafter\newcommand\csname sr07n814srMiiH\endcsname{701}
\expandafter\newcommand\csname sr07n814srMiV\endcsname{315.4}
\expandafter\newcommand\csname sr07n814srMiH\endcsname{914}
\expandafter\newcommand\csname sr07n814srMiiVErr\endcsname{9.7\%}
\expandafter\newcommand\csname sr07n814srMiiHErr\endcsname{8.0\%}
\expandafter\newcommand\csname sr07n814srMiVErr\endcsname{-0.5\%}
\expandafter\newcommand\csname sr07n814srMiHErr\endcsname{40.7\%}
\expandafter\newcommand\csname sr07n814srRMSE\endcsname{10.17}
\expandafter\newcommand\csname sr07n814srVRange\endcsname{19.6}
\expandafter\newcommand\csname sr07n814srHRange\endcsname{48}
\expandafter\newcommand\csname sr07n814srLoverH\endcsname{1.48}
\expandafter\newcommand\csname sr07n814srTrackSource\endcsname{N814SR-track-EGM96\_avg.kml}
\expandafter\newcommand\csname sr07n814srAdsAvailable\endcsname{1}
\expandafter\newcommand\csname sr07n814srTrackAvailable\endcsname{1}
\expandafter\newcommand\csname sr07n814srInsetXMin\endcsname{-31.6}
\expandafter\newcommand\csname sr07n814srInsetXMax\endcsname{22.7}
\expandafter\newcommand\csname sr07n814srInsetXTickLo\endcsname{-20}
\expandafter\newcommand\csname sr07n814srInsetXTickHi\endcsname{10}
\expandafter\newcommand\csname sr07n814srInsetVMin\endcsname{282}
\expandafter\newcommand\csname sr07n814srInsetVMax\endcsname{340}
\expandafter\newcommand\csname sr07n814srInsetHMin\endcsname{609}
\expandafter\newcommand\csname sr07n814srInsetHMax\endcsname{692}
\expandafter\newcommand\csname sr07n814srTrackXMin\endcsname{-6935}
\expandafter\newcommand\csname sr07n814srTrackXMax\endcsname{5439}
\expandafter\newcommand\csname sr07n814srTrackYMin\endcsname{-124}
\expandafter\newcommand\csname sr07n814srTrackYMax\endcsname{1669}
\expandafter\newcommand\csname sr07n814srTrackXTickLo\endcsname{-4000}
\expandafter\newcommand\csname sr07n814srTrackXTickHi\endcsname{4000}
\expandafter\newcommand\csname sr07n814srTrackYTickLo\endcsname{0}
\expandafter\newcommand\csname sr07n814srTrackYTickHi\endcsname{1500}
\expandafter\newcommand\csname sr07n814srTrackYTick\endcsname{1500}
\expandafter\newcommand\csname sr07n814srTrackClosestY\endcsname{951}
\expandafter\newcommand\csname sr08n2933aRank\endcsname{8}
\expandafter\newcommand\csname sr08n2933aLabel\endcsname{N2933A}
\expandafter\newcommand\csname sr08n2933aDataset\endcsname{N2933A 15:50}
\expandafter\newcommand\csname sr08n2933aGeometry\endcsname{overhead}
\expandafter\newcommand\csname sr08n2933aVCat\endcsname{V varying}
\expandafter\newcommand\csname sr08n2933aHCat\endcsname{H varying}
\expandafter\newcommand\csname sr08n2933aAudio\endcsname{smooth plateau-like + curve}
\expandafter\newcommand\csname sr08n2933aClosest\endcsname{2026-05-09 22:50:35 UTC}
\expandafter\newcommand\csname sr08n2933aXMin\endcsname{-24.1}
\expandafter\newcommand\csname sr08n2933aXMax\endcsname{19.1}
\expandafter\newcommand\csname sr08n2933aYMin\endcsname{372.0}
\expandafter\newcommand\csname sr08n2933aYMax\endcsname{495.9}
\expandafter\newcommand\csname sr08n2933aFzero\endcsname{427.3}
\expandafter\newcommand\csname sr08n2933aXStartTick\endcsname{-20}
\expandafter\newcommand\csname sr08n2933aXEndTick\endcsname{20}
\expandafter\newcommand\csname sr08n2933aYLowTick\endcsname{380}
\expandafter\newcommand\csname sr08n2933aYHighTick\endcsname{477}
\expandafter\newcommand\csname sr08n2933aAdsV\endcsname{148.7}
\expandafter\newcommand\csname sr08n2933aAdsH\endcsname{289}
\expandafter\newcommand\csname sr08n2933aMiiV\endcsname{134.7}
\expandafter\newcommand\csname sr08n2933aMiiH\endcsname{233}
\expandafter\newcommand\csname sr08n2933aMiV\endcsname{126.2}
\expandafter\newcommand\csname sr08n2933aMiH\endcsname{227}
\expandafter\newcommand\csname sr08n2933aMiiVErr\endcsname{-9.4\%}
\expandafter\newcommand\csname sr08n2933aMiiHErr\endcsname{-19.5\%}
\expandafter\newcommand\csname sr08n2933aMiVErr\endcsname{-15.1\%}
\expandafter\newcommand\csname sr08n2933aMiHErr\endcsname{-21.4\%}
\expandafter\newcommand\csname sr08n2933aRMSE\endcsname{2.12}
\expandafter\newcommand\csname sr08n2933aVRange\endcsname{24.2}
\expandafter\newcommand\csname sr08n2933aHRange\endcsname{137}
\expandafter\newcommand\csname sr08n2933aLoverH\endcsname{-0.01}
\expandafter\newcommand\csname sr08n2933aTrackSource\endcsname{N2933A-track-EGM96\_avg.kml}
\expandafter\newcommand\csname sr08n2933aAdsAvailable\endcsname{1}
\expandafter\newcommand\csname sr08n2933aTrackAvailable\endcsname{1}
\expandafter\newcommand\csname sr08n2933aInsetXMin\endcsname{-22.0}
\expandafter\newcommand\csname sr08n2933aInsetXMax\endcsname{16.9}
\expandafter\newcommand\csname sr08n2933aInsetXTickLo\endcsname{-20}
\expandafter\newcommand\csname sr08n2933aInsetXTickHi\endcsname{10}
\expandafter\newcommand\csname sr08n2933aInsetVMin\endcsname{128}
\expandafter\newcommand\csname sr08n2933aInsetVMax\endcsname{157}
\expandafter\newcommand\csname sr08n2933aInsetHMin\endcsname{239}
\expandafter\newcommand\csname sr08n2933aInsetHMax\endcsname{402}
\expandafter\newcommand\csname sr08n2933aTrackXMin\endcsname{-2631}
\expandafter\newcommand\csname sr08n2933aTrackXMax\endcsname{2652}
\expandafter\newcommand\csname sr08n2933aTrackYMin\endcsname{-525}
\expandafter\newcommand\csname sr08n2933aTrackYMax\endcsname{148}
\expandafter\newcommand\csname sr08n2933aTrackXTickLo\endcsname{-2000}
\expandafter\newcommand\csname sr08n2933aTrackXTickHi\endcsname{2000}
\expandafter\newcommand\csname sr08n2933aTrackYTickLo\endcsname{-400}
\expandafter\newcommand\csname sr08n2933aTrackYTickHi\endcsname{0}
\expandafter\newcommand\csname sr08n2933aTrackYTick\endcsname{0}
\expandafter\newcommand\csname sr08n2933aTrackClosestY\endcsname{102}
\expandafter\newcommand\csname sr09n784mbRank\endcsname{9}
\expandafter\newcommand\csname sr09n784mbLabel\endcsname{N784MB}
\expandafter\newcommand\csname sr09n784mbDataset\endcsname{N784MB high ridge}
\expandafter\newcommand\csname sr09n784mbGeometry\endcsname{far away/non-overhead}
\expandafter\newcommand\csname sr09n784mbVCat\endcsname{V constant}
\expandafter\newcommand\csname sr09n784mbHCat\endcsname{H constant}
\expandafter\newcommand\csname sr09n784mbAudio\endcsname{smooth plateau-like + curve}
\expandafter\newcommand\csname sr09n784mbClosest\endcsname{2026-05-14 17:42:09 UTC}
\expandafter\newcommand\csname sr09n784mbXMin\endcsname{-39.9}
\expandafter\newcommand\csname sr09n784mbXMax\endcsname{42.2}
\expandafter\newcommand\csname sr09n784mbYMin\endcsname{239.8}
\expandafter\newcommand\csname sr09n784mbYMax\endcsname{435.1}
\expandafter\newcommand\csname sr09n784mbFzero\endcsname{307.0}
\expandafter\newcommand\csname sr09n784mbXStartTick\endcsname{-40}
\expandafter\newcommand\csname sr09n784mbXEndTick\endcsname{40}
\expandafter\newcommand\csname sr09n784mbYLowTick\endcsname{252}
\expandafter\newcommand\csname sr09n784mbYHighTick\endcsname{407}
\expandafter\newcommand\csname sr09n784mbAdsV\endcsname{328.7}
\expandafter\newcommand\csname sr09n784mbAdsH\endcsname{1947}
\expandafter\newcommand\csname sr09n784mbMiiV\endcsname{319.3}
\expandafter\newcommand\csname sr09n784mbMiiH\endcsname{1862}
\expandafter\newcommand\csname sr09n784mbMiV\endcsname{265.4}
\expandafter\newcommand\csname sr09n784mbMiH\endcsname{1213}
\expandafter\newcommand\csname sr09n784mbMiiVErr\endcsname{-2.9\%}
\expandafter\newcommand\csname sr09n784mbMiiHErr\endcsname{-4.4\%}
\expandafter\newcommand\csname sr09n784mbMiVErr\endcsname{-19.3\%}
\expandafter\newcommand\csname sr09n784mbMiHErr\endcsname{-37.7\%}
\expandafter\newcommand\csname sr09n784mbRMSE\endcsname{2.95}
\expandafter\newcommand\csname sr09n784mbVRange\endcsname{2.7}
\expandafter\newcommand\csname sr09n784mbHRange\endcsname{0}
\expandafter\newcommand\csname sr09n784mbLoverH\endcsname{0.51}
\expandafter\newcommand\csname sr09n784mbTrackSource\endcsname{N784MB-track-EGM96\_avg.kml}
\expandafter\newcommand\csname sr09n784mbAdsAvailable\endcsname{1}
\expandafter\newcommand\csname sr09n784mbTrackAvailable\endcsname{1}
\expandafter\newcommand\csname sr09n784mbInsetXMin\endcsname{-35.8}
\expandafter\newcommand\csname sr09n784mbInsetXMax\endcsname{38.1}
\expandafter\newcommand\csname sr09n784mbInsetXTickLo\endcsname{-20}
\expandafter\newcommand\csname sr09n784mbInsetXTickHi\endcsname{20}
\expandafter\newcommand\csname sr09n784mbInsetVMin\endcsname{324}
\expandafter\newcommand\csname sr09n784mbInsetVMax\endcsname{335}
\expandafter\newcommand\csname sr09n784mbInsetHMin\endcsname{1927}
\expandafter\newcommand\csname sr09n784mbInsetHMax\endcsname{1967}
\expandafter\newcommand\csname sr09n784mbTrackXMin\endcsname{-6285}
\expandafter\newcommand\csname sr09n784mbTrackXMax\endcsname{7876}
\expandafter\newcommand\csname sr09n784mbTrackYMin\endcsname{-80}
\expandafter\newcommand\csname sr09n784mbTrackYMax\endcsname{1078}
\expandafter\newcommand\csname sr09n784mbTrackXTickLo\endcsname{-4000}
\expandafter\newcommand\csname sr09n784mbTrackXTickHi\endcsname{6000}
\expandafter\newcommand\csname sr09n784mbTrackYTickLo\endcsname{200}
\expandafter\newcommand\csname sr09n784mbTrackYTickHi\endcsname{800}
\expandafter\newcommand\csname sr09n784mbTrackYTick\endcsname{800}
\expandafter\newcommand\csname sr09n784mbTrackClosestY\endcsname{995}
\expandafter\newcommand\csname sr11n60rsRank\endcsname{11}
\expandafter\newcommand\csname sr11n60rsLabel\endcsname{N60RS}
\expandafter\newcommand\csname sr11n60rsDataset\endcsname{N60RS manual vs auto}
\expandafter\newcommand\csname sr11n60rsGeometry\endcsname{far away/non-overhead}
\expandafter\newcommand\csname sr11n60rsVCat\endcsname{V varying}
\expandafter\newcommand\csname sr11n60rsHCat\endcsname{H constant}
\expandafter\newcommand\csname sr11n60rsAudio\endcsname{plateau-like + curve but scattered/ambiguous}
\expandafter\newcommand\csname sr11n60rsClosest\endcsname{2026-05-17 20:46:54.85 UTC}
\expandafter\newcommand\csname sr11n60rsXMin\endcsname{-50.7}
\expandafter\newcommand\csname sr11n60rsXMax\endcsname{40.4}
\expandafter\newcommand\csname sr11n60rsYMin\endcsname{309.7}
\expandafter\newcommand\csname sr11n60rsYMax\endcsname{497.9}
\expandafter\newcommand\csname sr11n60rsFzero\endcsname{378.7}
\expandafter\newcommand\csname sr11n60rsXStartTick\endcsname{-50}
\expandafter\newcommand\csname sr11n60rsXEndTick\endcsname{40}
\expandafter\newcommand\csname sr11n60rsYLowTick\endcsname{322}
\expandafter\newcommand\csname sr11n60rsYHighTick\endcsname{471}
\expandafter\newcommand\csname sr11n60rsAdsV\endcsname{281.5}
\expandafter\newcommand\csname sr11n60rsAdsH\endcsname{1368}
\expandafter\newcommand\csname sr11n60rsMiiV\endcsname{287.5}
\expandafter\newcommand\csname sr11n60rsMiiH\endcsname{1570}
\expandafter\newcommand\csname sr11n60rsMiV\endcsname{208.1}
\expandafter\newcommand\csname sr11n60rsMiH\endcsname{--}
\expandafter\newcommand\csname sr11n60rsMiiVErr\endcsname{2.1\%}
\expandafter\newcommand\csname sr11n60rsMiiHErr\endcsname{14.7\%}
\expandafter\newcommand\csname sr11n60rsMiVErr\endcsname{-26.1\%}
\expandafter\newcommand\csname sr11n60rsMiHErr\endcsname{--}
\expandafter\newcommand\csname sr11n60rsRMSE\endcsname{3.14}
\expandafter\newcommand\csname sr11n60rsVRange\endcsname{17.2}
\expandafter\newcommand\csname sr11n60rsHRange\endcsname{24}
\expandafter\newcommand\csname sr11n60rsLoverH\endcsname{1.81}
\expandafter\newcommand\csname sr11n60rsTrackSource\endcsname{N60RS-track-EGM96\_avg.kml}
\expandafter\newcommand\csname sr11n60rsAdsAvailable\endcsname{1}
\expandafter\newcommand\csname sr11n60rsTrackAvailable\endcsname{1}
\expandafter\newcommand\csname sr11n60rsInsetXMin\endcsname{-46.2}
\expandafter\newcommand\csname sr11n60rsInsetXMax\endcsname{35.9}
\expandafter\newcommand\csname sr11n60rsInsetXTickLo\endcsname{-40}
\expandafter\newcommand\csname sr11n60rsInsetXTickHi\endcsname{20}
\expandafter\newcommand\csname sr11n60rsInsetVMin\endcsname{266}
\expandafter\newcommand\csname sr11n60rsInsetVMax\endcsname{308}
\expandafter\newcommand\csname sr11n60rsInsetHMin\endcsname{1339}
\expandafter\newcommand\csname sr11n60rsInsetHMax\endcsname{1402}
\expandafter\newcommand\csname sr11n60rsTrackXMin\endcsname{-7071}
\expandafter\newcommand\csname sr11n60rsTrackXMax\endcsname{6346}
\expandafter\newcommand\csname sr11n60rsTrackYMin\endcsname{-200}
\expandafter\newcommand\csname sr11n60rsTrackYMax\endcsname{2701}
\expandafter\newcommand\csname sr11n60rsTrackXTickLo\endcsname{-4000}
\expandafter\newcommand\csname sr11n60rsTrackXTickHi\endcsname{4000}
\expandafter\newcommand\csname sr11n60rsTrackYTickLo\endcsname{500}
\expandafter\newcommand\csname sr11n60rsTrackYTickHi\endcsname{2000}
\expandafter\newcommand\csname sr11n60rsTrackYTick\endcsname{2000}
\expandafter\newcommand\csname sr11n60rsTrackClosestY\endcsname{2471}
\expandafter\newcommand\csname sr12n4040Rank\endcsname{12}
\expandafter\newcommand\csname sr12n4040Label\endcsname{N4040}
\expandafter\newcommand\csname sr12n4040Dataset\endcsname{N4040}
\expandafter\newcommand\csname sr12n4040Geometry\endcsname{far away/non-overhead}
\expandafter\newcommand\csname sr12n4040VCat\endcsname{V constant}
\expandafter\newcommand\csname sr12n4040HCat\endcsname{H constant}
\expandafter\newcommand\csname sr12n4040Audio\endcsname{smooth curve only / incomplete ends}
\expandafter\newcommand\csname sr12n4040Closest\endcsname{2026-05-17 23:28:14.93 UTC}
\expandafter\newcommand\csname sr12n4040XMin\endcsname{-35.9}
\expandafter\newcommand\csname sr12n4040XMax\endcsname{55.0}
\expandafter\newcommand\csname sr12n4040YMin\endcsname{343.9}
\expandafter\newcommand\csname sr12n4040YMax\endcsname{480.7}
\expandafter\newcommand\csname sr12n4040Fzero\endcsname{403.5}
\expandafter\newcommand\csname sr12n4040XStartTick\endcsname{-30}
\expandafter\newcommand\csname sr12n4040XEndTick\endcsname{50}
\expandafter\newcommand\csname sr12n4040YLowTick\endcsname{353}
\expandafter\newcommand\csname sr12n4040YHighTick\endcsname{461}
\expandafter\newcommand\csname sr12n4040AdsV\endcsname{165.0}
\expandafter\newcommand\csname sr12n4040AdsH\endcsname{850}
\expandafter\newcommand\csname sr12n4040MiiV\endcsname{162.6}
\expandafter\newcommand\csname sr12n4040MiiH\endcsname{1031}
\expandafter\newcommand\csname sr12n4040MiV\endcsname{132.0}
\expandafter\newcommand\csname sr12n4040MiH\endcsname{731}
\expandafter\newcommand\csname sr12n4040MiiVErr\endcsname{-1.4\%}
\expandafter\newcommand\csname sr12n4040MiiHErr\endcsname{21.2\%}
\expandafter\newcommand\csname sr12n4040MiVErr\endcsname{-20.0\%}
\expandafter\newcommand\csname sr12n4040MiHErr\endcsname{-14.1\%}
\expandafter\newcommand\csname sr12n4040RMSE\endcsname{3.23}
\expandafter\newcommand\csname sr12n4040VRange\endcsname{3.7}
\expandafter\newcommand\csname sr12n4040HRange\endcsname{30}
\expandafter\newcommand\csname sr12n4040LoverH\endcsname{0.69}
\expandafter\newcommand\csname sr12n4040TrackSource\endcsname{N4040-track-EGM96\_avg.kml}
\expandafter\newcommand\csname sr12n4040AdsAvailable\endcsname{1}
\expandafter\newcommand\csname sr12n4040TrackAvailable\endcsname{1}
\expandafter\newcommand\csname sr12n4040InsetXMin\endcsname{-31.4}
\expandafter\newcommand\csname sr12n4040InsetXMax\endcsname{50.5}
\expandafter\newcommand\csname sr12n4040InsetXTickLo\endcsname{-20}
\expandafter\newcommand\csname sr12n4040InsetXTickHi\endcsname{40}
\expandafter\newcommand\csname sr12n4040InsetVMin\endcsname{157}
\expandafter\newcommand\csname sr12n4040InsetVMax\endcsname{172}
\expandafter\newcommand\csname sr12n4040InsetHMin\endcsname{799}
\expandafter\newcommand\csname sr12n4040InsetHMax\endcsname{900}
\expandafter\newcommand\csname sr12n4040TrackXMin\endcsname{-3622}
\expandafter\newcommand\csname sr12n4040TrackXMax\endcsname{4209}
\expandafter\newcommand\csname sr12n4040TrackYMin\endcsname{-47}
\expandafter\newcommand\csname sr12n4040TrackYMax\endcsname{636}
\expandafter\newcommand\csname sr12n4040TrackXTickLo\endcsname{-2000}
\expandafter\newcommand\csname sr12n4040TrackXTickHi\endcsname{4000}
\expandafter\newcommand\csname sr12n4040TrackYTickLo\endcsname{0}
\expandafter\newcommand\csname sr12n4040TrackYTickHi\endcsname{600}
\expandafter\newcommand\csname sr12n4040TrackYTick\endcsname{600}
\expandafter\newcommand\csname sr12n4040TrackClosestY\endcsname{587}
\expandafter\newcommand\csname sr13n795mmRank\endcsname{13}
\expandafter\newcommand\csname sr13n795mmLabel\endcsname{N795MM}
\expandafter\newcommand\csname sr13n795mmDataset\endcsname{N795MM}
\expandafter\newcommand\csname sr13n795mmGeometry\endcsname{far away/non-overhead}
\expandafter\newcommand\csname sr13n795mmVCat\endcsname{V constant}
\expandafter\newcommand\csname sr13n795mmHCat\endcsname{H varying}
\expandafter\newcommand\csname sr13n795mmAudio\endcsname{plateau-like + curve but scattered/ambiguous}
\expandafter\newcommand\csname sr13n795mmClosest\endcsname{2026-05-17 21:45:11.05 UTC}
\expandafter\newcommand\csname sr13n795mmXMin\endcsname{-54.1}
\expandafter\newcommand\csname sr13n795mmXMax\endcsname{21.3}
\expandafter\newcommand\csname sr13n795mmYMin\endcsname{161.8}
\expandafter\newcommand\csname sr13n795mmYMax\endcsname{320.6}
\expandafter\newcommand\csname sr13n795mmFzero\endcsname{211.4}
\expandafter\newcommand\csname sr13n795mmXStartTick\endcsname{-50}
\expandafter\newcommand\csname sr13n795mmXEndTick\endcsname{20}
\expandafter\newcommand\csname sr13n795mmYLowTick\endcsname{172}
\expandafter\newcommand\csname sr13n795mmYHighTick\endcsname{291}
\expandafter\newcommand\csname sr13n795mmAdsV\endcsname{405.1}
\expandafter\newcommand\csname sr13n795mmAdsH\endcsname{4848}
\expandafter\newcommand\csname sr13n795mmMiiV\endcsname{484.3}
\expandafter\newcommand\csname sr13n795mmMiiH\endcsname{7554}
\expandafter\newcommand\csname sr13n795mmMiV\endcsname{212.9}
\expandafter\newcommand\csname sr13n795mmMiH\endcsname{1203}
\expandafter\newcommand\csname sr13n795mmMiiVErr\endcsname{19.5\%}
\expandafter\newcommand\csname sr13n795mmMiiHErr\endcsname{55.8\%}
\expandafter\newcommand\csname sr13n795mmMiVErr\endcsname{-47.4\%}
\expandafter\newcommand\csname sr13n795mmMiHErr\endcsname{-75.2\%}
\expandafter\newcommand\csname sr13n795mmRMSE\endcsname{1.99}
\expandafter\newcommand\csname sr13n795mmVRange\endcsname{2.0}
\expandafter\newcommand\csname sr13n795mmHRange\endcsname{201}
\expandafter\newcommand\csname sr13n795mmLoverH\endcsname{0.12}
\expandafter\newcommand\csname sr13n795mmTrackSource\endcsname{N795MM-track-EGM96\_avg.kml}
\expandafter\newcommand\csname sr13n795mmAdsAvailable\endcsname{1}
\expandafter\newcommand\csname sr13n795mmTrackAvailable\endcsname{1}
\expandafter\newcommand\csname sr13n795mmInsetXMin\endcsname{-50.3}
\expandafter\newcommand\csname sr13n795mmInsetXMax\endcsname{17.6}
\expandafter\newcommand\csname sr13n795mmInsetXTickLo\endcsname{-40}
\expandafter\newcommand\csname sr13n795mmInsetXTickHi\endcsname{18}
\expandafter\newcommand\csname sr13n795mmInsetVMin\endcsname{394}
\expandafter\newcommand\csname sr13n795mmInsetVMax\endcsname{411}
\expandafter\newcommand\csname sr13n795mmInsetHMin\endcsname{4623}
\expandafter\newcommand\csname sr13n795mmInsetHMax\endcsname{5057}
\expandafter\newcommand\csname sr13n795mmTrackXMin\endcsname{-10884}
\expandafter\newcommand\csname sr13n795mmTrackXMax\endcsname{6745}
\expandafter\newcommand\csname sr13n795mmTrackYMin\endcsname{-45}
\expandafter\newcommand\csname sr13n795mmTrackYMax\endcsname{605}
\expandafter\newcommand\csname sr13n795mmTrackXTickLo\endcsname{-10000}
\expandafter\newcommand\csname sr13n795mmTrackXTickHi\endcsname{5000}
\expandafter\newcommand\csname sr13n795mmTrackYTickLo\endcsname{0}
\expandafter\newcommand\csname sr13n795mmTrackYTickHi\endcsname{600}
\expandafter\newcommand\csname sr13n795mmTrackYTick\endcsname{600}
\expandafter\newcommand\csname sr13n795mmTrackClosestY\endcsname{559}
\expandafter\newcommand\csname sr14n248lfRank\endcsname{14}
\expandafter\newcommand\csname sr14n248lfLabel\endcsname{N248LF}
\expandafter\newcommand\csname sr14n248lfDataset\endcsname{N248LF}
\expandafter\newcommand\csname sr14n248lfGeometry\endcsname{far away/non-overhead}
\expandafter\newcommand\csname sr14n248lfVCat\endcsname{V constant}
\expandafter\newcommand\csname sr14n248lfHCat\endcsname{H varying}
\expandafter\newcommand\csname sr14n248lfAudio\endcsname{smooth plateau-like + curve}
\expandafter\newcommand\csname sr14n248lfClosest\endcsname{2026-05-17 16:52:02.50 UTC}
\expandafter\newcommand\csname sr14n248lfXMin\endcsname{-85.8}
\expandafter\newcommand\csname sr14n248lfXMax\endcsname{71.8}
\expandafter\newcommand\csname sr14n248lfYMin\endcsname{193.5}
\expandafter\newcommand\csname sr14n248lfYMax\endcsname{328.9}
\expandafter\newcommand\csname sr14n248lfFzero\endcsname{244.3}
\expandafter\newcommand\csname sr14n248lfXStartTick\endcsname{-80}
\expandafter\newcommand\csname sr14n248lfXEndTick\endcsname{60}
\expandafter\newcommand\csname sr14n248lfYLowTick\endcsname{202}
\expandafter\newcommand\csname sr14n248lfYHighTick\endcsname{310}
\expandafter\newcommand\csname sr14n248lfAdsV\endcsname{293.7}
\expandafter\newcommand\csname sr14n248lfAdsH\endcsname{1665}
\expandafter\newcommand\csname sr14n248lfMiiV\endcsname{291.0}
\expandafter\newcommand\csname sr14n248lfMiiH\endcsname{2100}
\expandafter\newcommand\csname sr14n248lfMiV\endcsname{233.6}
\expandafter\newcommand\csname sr14n248lfMiH\endcsname{--}
\expandafter\newcommand\csname sr14n248lfMiiVErr\endcsname{-0.9\%}
\expandafter\newcommand\csname sr14n248lfMiiHErr\endcsname{26.1\%}
\expandafter\newcommand\csname sr14n248lfMiVErr\endcsname{-20.5\%}
\expandafter\newcommand\csname sr14n248lfMiHErr\endcsname{--}
\expandafter\newcommand\csname sr14n248lfRMSE\endcsname{1.69}
\expandafter\newcommand\csname sr14n248lfVRange\endcsname{6.5}
\expandafter\newcommand\csname sr14n248lfHRange\endcsname{181}
\expandafter\newcommand\csname sr14n248lfLoverH\endcsname{1.93}
\expandafter\newcommand\csname sr14n248lfTrackSource\endcsname{N248LF-track-EGM96\_avg.kml}
\expandafter\newcommand\csname sr14n248lfAdsAvailable\endcsname{1}
\expandafter\newcommand\csname sr14n248lfTrackAvailable\endcsname{1}
\expandafter\newcommand\csname sr14n248lfInsetXMin\endcsname{-78.0}
\expandafter\newcommand\csname sr14n248lfInsetXMax\endcsname{64.0}
\expandafter\newcommand\csname sr14n248lfInsetXTickLo\endcsname{-60}
\expandafter\newcommand\csname sr14n248lfInsetXTickHi\endcsname{40}
\expandafter\newcommand\csname sr14n248lfInsetVMin\endcsname{279}
\expandafter\newcommand\csname sr14n248lfInsetVMax\endcsname{310}
\expandafter\newcommand\csname sr14n248lfInsetHMin\endcsname{1517}
\expandafter\newcommand\csname sr14n248lfInsetHMax\endcsname{1744}
\expandafter\newcommand\csname sr14n248lfTrackXMin\endcsname{-9658}
\expandafter\newcommand\csname sr14n248lfTrackXMax\endcsname{9944}
\expandafter\newcommand\csname sr14n248lfTrackYMin\endcsname{-257}
\expandafter\newcommand\csname sr14n248lfTrackYMax\endcsname{3463}
\expandafter\newcommand\csname sr14n248lfTrackXTickLo\endcsname{-5000}
\expandafter\newcommand\csname sr14n248lfTrackXTickHi\endcsname{9944}
\expandafter\newcommand\csname sr14n248lfTrackYTickLo\endcsname{0}
\expandafter\newcommand\csname sr14n248lfTrackYTickHi\endcsname{3000}
\expandafter\newcommand\csname sr14n248lfTrackYTick\endcsname{3000}
\expandafter\newcommand\csname sr14n248lfTrackClosestY\endcsname{3205}
\expandafter\newcommand\csname sr15n913tcRank\endcsname{15}
\expandafter\newcommand\csname sr15n913tcLabel\endcsname{N913TC}
\expandafter\newcommand\csname sr15n913tcDataset\endcsname{N913TC}
\expandafter\newcommand\csname sr15n913tcGeometry\endcsname{far away/non-overhead}
\expandafter\newcommand\csname sr15n913tcVCat\endcsname{V constant}
\expandafter\newcommand\csname sr15n913tcHCat\endcsname{H varying}
\expandafter\newcommand\csname sr15n913tcAudio\endcsname{smooth plateau-like + curve}
\expandafter\newcommand\csname sr15n913tcClosest\endcsname{2026-05-17 18:32:38 UTC}
\expandafter\newcommand\csname sr15n913tcXMin\endcsname{-66.4}
\expandafter\newcommand\csname sr15n913tcXMax\endcsname{31.8}
\expandafter\newcommand\csname sr15n913tcYMin\endcsname{232.3}
\expandafter\newcommand\csname sr15n913tcYMax\endcsname{412.0}
\expandafter\newcommand\csname sr15n913tcFzero\endcsname{286.0}
\expandafter\newcommand\csname sr15n913tcXStartTick\endcsname{-60}
\expandafter\newcommand\csname sr15n913tcXEndTick\endcsname{30}
\expandafter\newcommand\csname sr15n913tcYLowTick\endcsname{246}
\expandafter\newcommand\csname sr15n913tcYHighTick\endcsname{386}
\expandafter\newcommand\csname sr15n913tcAdsV\endcsname{307.3}
\expandafter\newcommand\csname sr15n913tcAdsH\endcsname{1019}
\expandafter\newcommand\csname sr15n913tcMiiV\endcsname{317.9}
\expandafter\newcommand\csname sr15n913tcMiiH\endcsname{1706}
\expandafter\newcommand\csname sr15n913tcMiV\endcsname{245.6}
\expandafter\newcommand\csname sr15n913tcMiH\endcsname{--}
\expandafter\newcommand\csname sr15n913tcMiiVErr\endcsname{3.5\%}
\expandafter\newcommand\csname sr15n913tcMiiHErr\endcsname{67.4\%}
\expandafter\newcommand\csname sr15n913tcMiVErr\endcsname{-20.1\%}
\expandafter\newcommand\csname sr15n913tcMiHErr\endcsname{--}
\expandafter\newcommand\csname sr15n913tcRMSE\endcsname{3.67}
\expandafter\newcommand\csname sr15n913tcVRange\endcsname{7.5}
\expandafter\newcommand\csname sr15n913tcHRange\endcsname{126}
\expandafter\newcommand\csname sr15n913tcLoverH\endcsname{1.67}
\expandafter\newcommand\csname sr15n913tcTrackSource\endcsname{N913TC-track-EGM96\_avg.kml}
\expandafter\newcommand\csname sr15n913tcAdsAvailable\endcsname{1}
\expandafter\newcommand\csname sr15n913tcTrackAvailable\endcsname{1}
\expandafter\newcommand\csname sr15n913tcInsetXMin\endcsname{-61.5}
\expandafter\newcommand\csname sr15n913tcInsetXMax\endcsname{26.9}
\expandafter\newcommand\csname sr15n913tcInsetXTickLo\endcsname{-40}
\expandafter\newcommand\csname sr15n913tcInsetXTickHi\endcsname{20}
\expandafter\newcommand\csname sr15n913tcInsetVMin\endcsname{291}
\expandafter\newcommand\csname sr15n913tcInsetVMax\endcsname{333}
\expandafter\newcommand\csname sr15n913tcInsetHMin\endcsname{932}
\expandafter\newcommand\csname sr15n913tcInsetHMax\endcsname{1123}
\expandafter\newcommand\csname sr15n913tcTrackXMin\endcsname{-9784}
\expandafter\newcommand\csname sr15n913tcTrackXMax\endcsname{5703}
\expandafter\newcommand\csname sr15n913tcTrackYMin\endcsname{-137}
\expandafter\newcommand\csname sr15n913tcTrackYMax\endcsname{1852}
\expandafter\newcommand\csname sr15n913tcTrackXTickLo\endcsname{-9784}
\expandafter\newcommand\csname sr15n913tcTrackXTickHi\endcsname{5000}
\expandafter\newcommand\csname sr15n913tcTrackYTickLo\endcsname{0}
\expandafter\newcommand\csname sr15n913tcTrackYTickHi\endcsname{1500}
\expandafter\newcommand\csname sr15n913tcTrackYTick\endcsname{1500}
\expandafter\newcommand\csname sr15n913tcTrackClosestY\endcsname{1704}
\expandafter\newcommand\csname sr16n413akRank\endcsname{16}
\expandafter\newcommand\csname sr16n413akLabel\endcsname{N413AK}
\expandafter\newcommand\csname sr16n413akDataset\endcsname{N413AK 17-May high+low curves}
\expandafter\newcommand\csname sr16n413akGeometry\endcsname{far away/non-overhead}
\expandafter\newcommand\csname sr16n413akVCat\endcsname{V constant}
\expandafter\newcommand\csname sr16n413akHCat\endcsname{H varying}
\expandafter\newcommand\csname sr16n413akAudio\endcsname{smooth plateau-like + curve}
\expandafter\newcommand\csname sr16n413akClosest\endcsname{2026-05-17 21:47:32 UTC}
\expandafter\newcommand\csname sr16n413akXMin\endcsname{-65.5}
\expandafter\newcommand\csname sr16n413akXMax\endcsname{86.7}
\expandafter\newcommand\csname sr16n413akYMin\endcsname{38.6}
\expandafter\newcommand\csname sr16n413akYMax\endcsname{391.8}
\expandafter\newcommand\csname sr16n413akFzero\endcsname{234.0}
\expandafter\newcommand\csname sr16n413akXStartTick\endcsname{-60}
\expandafter\newcommand\csname sr16n413akXEndTick\endcsname{80}
\expandafter\newcommand\csname sr16n413akYLowTick\endcsname{61}
\expandafter\newcommand\csname sr16n413akYHighTick\endcsname{341}
\expandafter\newcommand\csname sr16n413akAdsV\endcsname{329.1}
\expandafter\newcommand\csname sr16n413akAdsH\endcsname{1437}
\expandafter\newcommand\csname sr16n413akMiiV\endcsname{367.5}
\expandafter\newcommand\csname sr16n413akMiiH\endcsname{2182}
\expandafter\newcommand\csname sr16n413akMiV\endcsname{303.4}
\expandafter\newcommand\csname sr16n413akMiH\endcsname{1218}
\expandafter\newcommand\csname sr16n413akMiiVErr\endcsname{11.7\%}
\expandafter\newcommand\csname sr16n413akMiiHErr\endcsname{51.8\%}
\expandafter\newcommand\csname sr16n413akMiVErr\endcsname{-7.8\%}
\expandafter\newcommand\csname sr16n413akMiHErr\endcsname{-15.3\%}
\expandafter\newcommand\csname sr16n413akRMSE\endcsname{9.41}
\expandafter\newcommand\csname sr16n413akVRange\endcsname{10.0}
\expandafter\newcommand\csname sr16n413akHRange\endcsname{241}
\expandafter\newcommand\csname sr16n413akLoverH\endcsname{-0.00}
\expandafter\newcommand\csname sr16n413akTrackSource\endcsname{N413AK-track-EGM96\_avg\_2.kml}
\expandafter\newcommand\csname sr16n413akAdsAvailable\endcsname{1}
\expandafter\newcommand\csname sr16n413akTrackAvailable\endcsname{1}
\expandafter\newcommand\csname sr16n413akInsetXMin\endcsname{-57.9}
\expandafter\newcommand\csname sr16n413akInsetXMax\endcsname{79.1}
\expandafter\newcommand\csname sr16n413akInsetXTickLo\endcsname{-40}
\expandafter\newcommand\csname sr16n413akInsetXTickHi\endcsname{60}
\expandafter\newcommand\csname sr16n413akInsetVMin\endcsname{310}
\expandafter\newcommand\csname sr16n413akInsetVMax\endcsname{371}
\expandafter\newcommand\csname sr16n413akInsetHMin\endcsname{1186}
\expandafter\newcommand\csname sr16n413akInsetHMax\endcsname{1687}
\expandafter\newcommand\csname sr16n413akTrackXMin\endcsname{-10093}
\expandafter\newcommand\csname sr16n413akTrackXMax\endcsname{12142}
\expandafter\newcommand\csname sr16n413akTrackYMin\endcsname{-66}
\expandafter\newcommand\csname sr16n413akTrackYMax\endcsname{891}
\expandafter\newcommand\csname sr16n413akTrackXTickLo\endcsname{-5000}
\expandafter\newcommand\csname sr16n413akTrackXTickHi\endcsname{10000}
\expandafter\newcommand\csname sr16n413akTrackYTickLo\endcsname{0}
\expandafter\newcommand\csname sr16n413akTrackYTickHi\endcsname{800}
\expandafter\newcommand\csname sr16n413akTrackYTick\endcsname{800}
\expandafter\newcommand\csname sr16n413akTrackClosestY\endcsname{768}
\expandafter\newcommand\csname sr17n500mgRank\endcsname{17}
\expandafter\newcommand\csname sr17n500mgLabel\endcsname{N500MG}
\expandafter\newcommand\csname sr17n500mgDataset\endcsname{N500MG}
\expandafter\newcommand\csname sr17n500mgGeometry\endcsname{near overhead}
\expandafter\newcommand\csname sr17n500mgVCat\endcsname{V constant}
\expandafter\newcommand\csname sr17n500mgHCat\endcsname{H varying}
\expandafter\newcommand\csname sr17n500mgAudio\endcsname{smooth plateau-like + curve}
\expandafter\newcommand\csname sr17n500mgClosest\endcsname{2026-05-17 21:02:00.15 UTC}
\expandafter\newcommand\csname sr17n500mgXMin\endcsname{-52.6}
\expandafter\newcommand\csname sr17n500mgXMax\endcsname{18.4}
\expandafter\newcommand\csname sr17n500mgYMin\endcsname{183.2}
\expandafter\newcommand\csname sr17n500mgYMax\endcsname{322.9}
\expandafter\newcommand\csname sr17n500mgFzero\endcsname{214.5}
\expandafter\newcommand\csname sr17n500mgXStartTick\endcsname{-50}
\expandafter\newcommand\csname sr17n500mgXEndTick\endcsname{10}
\expandafter\newcommand\csname sr17n500mgYLowTick\endcsname{192}
\expandafter\newcommand\csname sr17n500mgYHighTick\endcsname{303}
\expandafter\newcommand\csname sr17n500mgAdsV\endcsname{470.7}
\expandafter\newcommand\csname sr17n500mgAdsH\endcsname{5609}
\expandafter\newcommand\csname sr17n500mgMiiV\endcsname{420.6}
\expandafter\newcommand\csname sr17n500mgMiiH\endcsname{4739}
\expandafter\newcommand\csname sr17n500mgMiV\endcsname{251.2}
\expandafter\newcommand\csname sr17n500mgMiH\endcsname{1625}
\expandafter\newcommand\csname sr17n500mgMiiVErr\endcsname{-10.6\%}
\expandafter\newcommand\csname sr17n500mgMiiHErr\endcsname{-15.5\%}
\expandafter\newcommand\csname sr17n500mgMiVErr\endcsname{-46.6\%}
\expandafter\newcommand\csname sr17n500mgMiHErr\endcsname{-71.0\%}
\expandafter\newcommand\csname sr17n500mgRMSE\endcsname{1.98}
\expandafter\newcommand\csname sr17n500mgVRange\endcsname{5.5}
\expandafter\newcommand\csname sr17n500mgHRange\endcsname{310}
\expandafter\newcommand\csname sr17n500mgLoverH\endcsname{0.03}
\expandafter\newcommand\csname sr17n500mgTrackSource\endcsname{N500MG-track-EGM96\_avg.kml}
\expandafter\newcommand\csname sr17n500mgAdsAvailable\endcsname{1}
\expandafter\newcommand\csname sr17n500mgTrackAvailable\endcsname{1}
\expandafter\newcommand\csname sr17n500mgInsetXMin\endcsname{-49.1}
\expandafter\newcommand\csname sr17n500mgInsetXMax\endcsname{14.9}
\expandafter\newcommand\csname sr17n500mgInsetXTickLo\endcsname{-40}
\expandafter\newcommand\csname sr17n500mgInsetXTickHi\endcsname{0}
\expandafter\newcommand\csname sr17n500mgInsetVMin\endcsname{434}
\expandafter\newcommand\csname sr17n500mgInsetVMax\endcsname{476}
\expandafter\newcommand\csname sr17n500mgInsetHMin\endcsname{5374}
\expandafter\newcommand\csname sr17n500mgInsetHMax\endcsname{5846}
\expandafter\newcommand\csname sr17n500mgTrackXMin\endcsname{-11692}
\expandafter\newcommand\csname sr17n500mgTrackXMax\endcsname{8220}
\expandafter\newcommand\csname sr17n500mgTrackYMin\endcsname{-40}
\expandafter\newcommand\csname sr17n500mgTrackYMax\endcsname{77}
\expandafter\newcommand\csname sr17n500mgTrackXTickLo\endcsname{-10000}
\expandafter\newcommand\csname sr17n500mgTrackXTickHi\endcsname{5000}
\expandafter\newcommand\csname sr17n500mgTrackYTickLo\endcsname{-20}
\expandafter\newcommand\csname sr17n500mgTrackYTickHi\endcsname{60}
\expandafter\newcommand\csname sr17n500mgTrackYTick\endcsname{60}
\expandafter\newcommand\csname sr17n500mgTrackClosestY\endcsname{37}
\expandafter\newcommand\csname sr18n621ksRank\endcsname{18}
\expandafter\newcommand\csname sr18n621ksLabel\endcsname{N621KS}
\expandafter\newcommand\csname sr18n621ksDataset\endcsname{N621KS}
\expandafter\newcommand\csname sr18n621ksGeometry\endcsname{far away/non-overhead}
\expandafter\newcommand\csname sr18n621ksVCat\endcsname{V constant}
\expandafter\newcommand\csname sr18n621ksHCat\endcsname{H varying}
\expandafter\newcommand\csname sr18n621ksAudio\endcsname{smooth plateau-like + curve}
\expandafter\newcommand\csname sr18n621ksClosest\endcsname{2026-05-17 23:22:15.11 UTC}
\expandafter\newcommand\csname sr18n621ksXMin\endcsname{-35.0}
\expandafter\newcommand\csname sr18n621ksXMax\endcsname{25.6}
\expandafter\newcommand\csname sr18n621ksYMin\endcsname{272.9}
\expandafter\newcommand\csname sr18n621ksYMax\endcsname{432.3}
\expandafter\newcommand\csname sr18n621ksFzero\endcsname{327.5}
\expandafter\newcommand\csname sr18n621ksXStartTick\endcsname{-30}
\expandafter\newcommand\csname sr18n621ksXEndTick\endcsname{20}
\expandafter\newcommand\csname sr18n621ksYLowTick\endcsname{283}
\expandafter\newcommand\csname sr18n621ksYHighTick\endcsname{410}
\expandafter\newcommand\csname sr18n621ksAdsV\endcsname{331.5}
\expandafter\newcommand\csname sr18n621ksAdsH\endcsname{2779}
\expandafter\newcommand\csname sr18n621ksMiiV\endcsname{283.7}
\expandafter\newcommand\csname sr18n621ksMiiH\endcsname{1760}
\expandafter\newcommand\csname sr18n621ksMiV\endcsname{189.4}
\expandafter\newcommand\csname sr18n621ksMiH\endcsname{--}
\expandafter\newcommand\csname sr18n621ksMiiVErr\endcsname{-14.4\%}
\expandafter\newcommand\csname sr18n621ksMiiHErr\endcsname{-36.7\%}
\expandafter\newcommand\csname sr18n621ksMiVErr\endcsname{-42.9\%}
\expandafter\newcommand\csname sr18n621ksMiHErr\endcsname{--}
\expandafter\newcommand\csname sr18n621ksRMSE\endcsname{3.25}
\expandafter\newcommand\csname sr18n621ksVRange\endcsname{2.5}
\expandafter\newcommand\csname sr18n621ksHRange\endcsname{279}
\expandafter\newcommand\csname sr18n621ksLoverH\endcsname{0.44}
\expandafter\newcommand\csname sr18n621ksTrackSource\endcsname{N621KS-track-EGM96\_avg.kml}
\expandafter\newcommand\csname sr18n621ksAdsAvailable\endcsname{1}
\expandafter\newcommand\csname sr18n621ksTrackAvailable\endcsname{1}
\expandafter\newcommand\csname sr18n621ksInsetXMin\endcsname{-32.0}
\expandafter\newcommand\csname sr18n621ksInsetXMax\endcsname{22.6}
\expandafter\newcommand\csname sr18n621ksInsetXTickLo\endcsname{-20}
\expandafter\newcommand\csname sr18n621ksInsetXTickHi\endcsname{10}
\expandafter\newcommand\csname sr18n621ksInsetVMin\endcsname{313}
\expandafter\newcommand\csname sr18n621ksInsetVMax\endcsname{338}
\expandafter\newcommand\csname sr18n621ksInsetHMin\endcsname{2538}
\expandafter\newcommand\csname sr18n621ksInsetHMax\endcsname{3029}
\expandafter\newcommand\csname sr18n621ksTrackXMin\endcsname{-6845}
\expandafter\newcommand\csname sr18n621ksTrackXMax\endcsname{6193}
\expandafter\newcommand\csname sr18n621ksTrackYMin\endcsname{-110}
\expandafter\newcommand\csname sr18n621ksTrackYMax\endcsname{1482}
\expandafter\newcommand\csname sr18n621ksTrackXTickLo\endcsname{-4000}
\expandafter\newcommand\csname sr18n621ksTrackXTickHi\endcsname{4000}
\expandafter\newcommand\csname sr18n621ksTrackYTickLo\endcsname{0}
\expandafter\newcommand\csname sr18n621ksTrackYTickHi\endcsname{1482}
\expandafter\newcommand\csname sr18n621ksTrackYTick\endcsname{1482}
\expandafter\newcommand\csname sr18n621ksTrackClosestY\endcsname{1212}
\expandafter\newcommand\csname sr19n4004rRank\endcsname{19}
\expandafter\newcommand\csname sr19n4004rLabel\endcsname{N4004R}
\expandafter\newcommand\csname sr19n4004rDataset\endcsname{N4004R}
\expandafter\newcommand\csname sr19n4004rGeometry\endcsname{far away/non-overhead}
\expandafter\newcommand\csname sr19n4004rVCat\endcsname{V constant}
\expandafter\newcommand\csname sr19n4004rHCat\endcsname{H varying}
\expandafter\newcommand\csname sr19n4004rAudio\endcsname{smooth curve only / incomplete ends}
\expandafter\newcommand\csname sr19n4004rClosest\endcsname{2026-05-16 23:56:40.50 UTC}
\expandafter\newcommand\csname sr19n4004rXMin\endcsname{-45.5}
\expandafter\newcommand\csname sr19n4004rXMax\endcsname{56.2}
\expandafter\newcommand\csname sr19n4004rYMin\endcsname{186.6}
\expandafter\newcommand\csname sr19n4004rYMax\endcsname{342.5}
\expandafter\newcommand\csname sr19n4004rFzero\endcsname{242.4}
\expandafter\newcommand\csname sr19n4004rXStartTick\endcsname{-40}
\expandafter\newcommand\csname sr19n4004rXEndTick\endcsname{50}
\expandafter\newcommand\csname sr19n4004rYLowTick\endcsname{196}
\expandafter\newcommand\csname sr19n4004rYHighTick\endcsname{320}
\expandafter\newcommand\csname sr19n4004rAdsV\endcsname{300.8}
\expandafter\newcommand\csname sr19n4004rAdsH\endcsname{1225}
\expandafter\newcommand\csname sr19n4004rMiiV\endcsname{290.6}
\expandafter\newcommand\csname sr19n4004rMiiH\endcsname{1069}
\expandafter\newcommand\csname sr19n4004rMiV\endcsname{270.0}
\expandafter\newcommand\csname sr19n4004rMiH\endcsname{931}
\expandafter\newcommand\csname sr19n4004rMiiVErr\endcsname{-3.4\%}
\expandafter\newcommand\csname sr19n4004rMiiHErr\endcsname{-12.7\%}
\expandafter\newcommand\csname sr19n4004rMiVErr\endcsname{-10.2\%}
\expandafter\newcommand\csname sr19n4004rMiHErr\endcsname{-24.0\%}
\expandafter\newcommand\csname sr19n4004rRMSE\endcsname{1.35}
\expandafter\newcommand\csname sr19n4004rVRange\endcsname{8.4}
\expandafter\newcommand\csname sr19n4004rHRange\endcsname{238}
\expandafter\newcommand\csname sr19n4004rLoverH\endcsname{0.62}
\expandafter\newcommand\csname sr19n4004rTrackSource\endcsname{N4004R-track-EGM96\_avg.kml}
\expandafter\newcommand\csname sr19n4004rAdsAvailable\endcsname{1}
\expandafter\newcommand\csname sr19n4004rTrackAvailable\endcsname{1}
\expandafter\newcommand\csname sr19n4004rInsetXMin\endcsname{-40.4}
\expandafter\newcommand\csname sr19n4004rInsetXMax\endcsname{51.1}
\expandafter\newcommand\csname sr19n4004rInsetXTickLo\endcsname{-20}
\expandafter\newcommand\csname sr19n4004rInsetXTickHi\endcsname{40}
\expandafter\newcommand\csname sr19n4004rInsetVMin\endcsname{278}
\expandafter\newcommand\csname sr19n4004rInsetVMax\endcsname{328}
\expandafter\newcommand\csname sr19n4004rInsetHMin\endcsname{1033}
\expandafter\newcommand\csname sr19n4004rInsetHMax\endcsname{1431}
\expandafter\newcommand\csname sr19n4004rTrackXMin\endcsname{-8046}
\expandafter\newcommand\csname sr19n4004rTrackXMax\endcsname{7888}
\expandafter\newcommand\csname sr19n4004rTrackYMin\endcsname{-93}
\expandafter\newcommand\csname sr19n4004rTrackYMax\endcsname{833}
\expandafter\newcommand\csname sr19n4004rTrackXTickLo\endcsname{-5000}
\expandafter\newcommand\csname sr19n4004rTrackXTickHi\endcsname{5000}
\expandafter\newcommand\csname sr19n4004rTrackYTickLo\endcsname{0}
\expandafter\newcommand\csname sr19n4004rTrackYTickHi\endcsname{800}
\expandafter\newcommand\csname sr19n4004rTrackYTick\endcsname{800}
\expandafter\newcommand\csname sr19n4004rTrackClosestY\endcsname{769}
\expandafter\newcommand\csname sr20n394fxRank\endcsname{20}
\expandafter\newcommand\csname sr20n394fxLabel\endcsname{N394FX}
\expandafter\newcommand\csname sr20n394fxDataset\endcsname{N394FX}
\expandafter\newcommand\csname sr20n394fxGeometry\endcsname{near overhead}
\expandafter\newcommand\csname sr20n394fxVCat\endcsname{V constant}
\expandafter\newcommand\csname sr20n394fxHCat\endcsname{H constant}
\expandafter\newcommand\csname sr20n394fxAudio\endcsname{smooth curve only / incomplete ends}
\expandafter\newcommand\csname sr20n394fxClosest\endcsname{2026-05-16 20:12:36.90 UTC}
\expandafter\newcommand\csname sr20n394fxXMin\endcsname{-15.5}
\expandafter\newcommand\csname sr20n394fxXMax\endcsname{40.6}
\expandafter\newcommand\csname sr20n394fxYMin\endcsname{241.2}
\expandafter\newcommand\csname sr20n394fxYMax\endcsname{551.8}
\expandafter\newcommand\csname sr20n394fxFzero\endcsname{355.0}
\expandafter\newcommand\csname sr20n394fxXStartTick\endcsname{-10}
\expandafter\newcommand\csname sr20n394fxXEndTick\endcsname{40}
\expandafter\newcommand\csname sr20n394fxYLowTick\endcsname{261}
\expandafter\newcommand\csname sr20n394fxYHighTick\endcsname{507}
\expandafter\newcommand\csname sr20n394fxAdsV\endcsname{463.2}
\expandafter\newcommand\csname sr20n394fxAdsH\endcsname{1688}
\expandafter\newcommand\csname sr20n394fxMiiV\endcsname{456.2}
\expandafter\newcommand\csname sr20n394fxMiiH\endcsname{1513}
\expandafter\newcommand\csname sr20n394fxMiV\endcsname{369.0}
\expandafter\newcommand\csname sr20n394fxMiH\endcsname{1173}
\expandafter\newcommand\csname sr20n394fxMiiVErr\endcsname{-1.5\%}
\expandafter\newcommand\csname sr20n394fxMiiHErr\endcsname{-10.4\%}
\expandafter\newcommand\csname sr20n394fxMiVErr\endcsname{-20.3\%}
\expandafter\newcommand\csname sr20n394fxMiHErr\endcsname{-30.5\%}
\expandafter\newcommand\csname sr20n394fxRMSE\endcsname{2.57}
\expandafter\newcommand\csname sr20n394fxVRange\endcsname{6.8}
\expandafter\newcommand\csname sr20n394fxHRange\endcsname{0}
\expandafter\newcommand\csname sr20n394fxLoverH\endcsname{0.23}
\expandafter\newcommand\csname sr20n394fxTrackSource\endcsname{N394FX-track-EGM96\_avg.kml}
\expandafter\newcommand\csname sr20n394fxAdsAvailable\endcsname{1}
\expandafter\newcommand\csname sr20n394fxTrackAvailable\endcsname{1}
\expandafter\newcommand\csname sr20n394fxInsetXMin\endcsname{-12.7}
\expandafter\newcommand\csname sr20n394fxInsetXMax\endcsname{37.8}
\expandafter\newcommand\csname sr20n394fxInsetXTickLo\endcsname{-10}
\expandafter\newcommand\csname sr20n394fxInsetXTickHi\endcsname{30}
\expandafter\newcommand\csname sr20n394fxInsetVMin\endcsname{446}
\expandafter\newcommand\csname sr20n394fxInsetVMax\endcsname{491}
\expandafter\newcommand\csname sr20n394fxInsetHMin\endcsname{1668}
\expandafter\newcommand\csname sr20n394fxInsetHMax\endcsname{1708}
\expandafter\newcommand\csname sr20n394fxTrackXMin\endcsname{-9172}
\expandafter\newcommand\csname sr20n394fxTrackXMax\endcsname{8687}
\expandafter\newcommand\csname sr20n394fxTrackYMin\endcsname{-40}
\expandafter\newcommand\csname sr20n394fxTrackYMax\endcsname{451}
\expandafter\newcommand\csname sr20n394fxTrackXTickLo\endcsname{-5000}
\expandafter\newcommand\csname sr20n394fxTrackXTickHi\endcsname{5000}
\expandafter\newcommand\csname sr20n394fxTrackYTickLo\endcsname{0}
\expandafter\newcommand\csname sr20n394fxTrackYTickHi\endcsname{400}
\expandafter\newcommand\csname sr20n394fxTrackYTick\endcsname{400}
\expandafter\newcommand\csname sr20n394fxTrackClosestY\endcsname{388}
\expandafter\newcommand\csname sr21n563fxRank\endcsname{21}
\expandafter\newcommand\csname sr21n563fxLabel\endcsname{N563FX}
\expandafter\newcommand\csname sr21n563fxDataset\endcsname{N563FX cleaned}
\expandafter\newcommand\csname sr21n563fxGeometry\endcsname{far away/non-overhead}
\expandafter\newcommand\csname sr21n563fxVCat\endcsname{V constant}
\expandafter\newcommand\csname sr21n563fxHCat\endcsname{H varying}
\expandafter\newcommand\csname sr21n563fxAudio\endcsname{plateau-like + curve but scattered/ambiguous}
\expandafter\newcommand\csname sr21n563fxClosest\endcsname{2026-05-16 18:25:00 UTC}
\expandafter\newcommand\csname sr21n563fxXMin\endcsname{-70.4}
\expandafter\newcommand\csname sr21n563fxXMax\endcsname{62.4}
\expandafter\newcommand\csname sr21n563fxYMin\endcsname{439.1}
\expandafter\newcommand\csname sr21n563fxYMax\endcsname{1433.7}
\expandafter\newcommand\csname sr21n563fxFzero\endcsname{698.3}
\expandafter\newcommand\csname sr21n563fxXStartTick\endcsname{-60}
\expandafter\newcommand\csname sr21n563fxXEndTick\endcsname{60}
\expandafter\newcommand\csname sr21n563fxYLowTick\endcsname{506}
\expandafter\newcommand\csname sr21n563fxYHighTick\endcsname{1292}
\expandafter\newcommand\csname sr21n563fxAdsV\endcsname{466.9}
\expandafter\newcommand\csname sr21n563fxAdsH\endcsname{1985}
\expandafter\newcommand\csname sr21n563fxMiiV\endcsname{532.3}
\expandafter\newcommand\csname sr21n563fxMiiH\endcsname{3135}
\expandafter\newcommand\csname sr21n563fxMiV\endcsname{512.0}
\expandafter\newcommand\csname sr21n563fxMiH\endcsname{2480}
\expandafter\newcommand\csname sr21n563fxMiiVErr\endcsname{14.0\%}
\expandafter\newcommand\csname sr21n563fxMiiHErr\endcsname{58.0\%}
\expandafter\newcommand\csname sr21n563fxMiVErr\endcsname{9.7\%}
\expandafter\newcommand\csname sr21n563fxMiHErr\endcsname{25.0\%}
\expandafter\newcommand\csname sr21n563fxRMSE\endcsname{16.55}
\expandafter\newcommand\csname sr21n563fxVRange\endcsname{10.6}
\expandafter\newcommand\csname sr21n563fxHRange\endcsname{557}
\expandafter\newcommand\csname sr21n563fxLoverH\endcsname{0.31}
\expandafter\newcommand\csname sr21n563fxTrackSource\endcsname{N563FX-track-EGM96\_avg.kml}
\expandafter\newcommand\csname sr21n563fxAdsAvailable\endcsname{1}
\expandafter\newcommand\csname sr21n563fxTrackAvailable\endcsname{1}
\expandafter\newcommand\csname sr21n563fxInsetXMin\endcsname{-63.8}
\expandafter\newcommand\csname sr21n563fxInsetXMax\endcsname{55.8}
\expandafter\newcommand\csname sr21n563fxInsetXTickLo\endcsname{-40}
\expandafter\newcommand\csname sr21n563fxInsetXTickHi\endcsname{40}
\expandafter\newcommand\csname sr21n563fxInsetVMin\endcsname{436}
\expandafter\newcommand\csname sr21n563fxInsetVMax\endcsname{505}
\expandafter\newcommand\csname sr21n563fxInsetHMin\endcsname{1569}
\expandafter\newcommand\csname sr21n563fxInsetHMax\endcsname{2495}
\expandafter\newcommand\csname sr21n563fxTrackXMin\endcsname{-15133}
\expandafter\newcommand\csname sr21n563fxTrackXMax\endcsname{13970}
\expandafter\newcommand\csname sr21n563fxTrackYMin\endcsname{-50}
\expandafter\newcommand\csname sr21n563fxTrackYMax\endcsname{681}
\expandafter\newcommand\csname sr21n563fxTrackXTickLo\endcsname{-10000}
\expandafter\newcommand\csname sr21n563fxTrackXTickHi\endcsname{10000}
\expandafter\newcommand\csname sr21n563fxTrackYTickLo\endcsname{0}
\expandafter\newcommand\csname sr21n563fxTrackYTickHi\endcsname{600}
\expandafter\newcommand\csname sr21n563fxTrackYTick\endcsname{600}
\expandafter\newcommand\csname sr21n563fxTrackClosestY\endcsname{619}
\expandafter\newcommand\csname sr22n916ktRank\endcsname{22}
\expandafter\newcommand\csname sr22n916ktLabel\endcsname{N916KT}
\expandafter\newcommand\csname sr22n916ktDataset\endcsname{N916KT}
\expandafter\newcommand\csname sr22n916ktGeometry\endcsname{far away/non-overhead}
\expandafter\newcommand\csname sr22n916ktVCat\endcsname{V constant}
\expandafter\newcommand\csname sr22n916ktHCat\endcsname{H constant}
\expandafter\newcommand\csname sr22n916ktAudio\endcsname{smooth curve only / incomplete ends}
\expandafter\newcommand\csname sr22n916ktClosest\endcsname{2026-05-16 17:13:45.20 UTC}
\expandafter\newcommand\csname sr22n916ktXMin\endcsname{-42.7}
\expandafter\newcommand\csname sr22n916ktXMax\endcsname{32.2}
\expandafter\newcommand\csname sr22n916ktYMin\endcsname{692.8}
\expandafter\newcommand\csname sr22n916ktYMax\endcsname{946.1}
\expandafter\newcommand\csname sr22n916ktFzero\endcsname{793.2}
\expandafter\newcommand\csname sr22n916ktXStartTick\endcsname{-40}
\expandafter\newcommand\csname sr22n916ktXEndTick\endcsname{30}
\expandafter\newcommand\csname sr22n916ktYLowTick\endcsname{709}
\expandafter\newcommand\csname sr22n916ktYHighTick\endcsname{910}
\expandafter\newcommand\csname sr22n916ktAdsV\endcsname{163.6}
\expandafter\newcommand\csname sr22n916ktAdsH\endcsname{889}
\expandafter\newcommand\csname sr22n916ktMiiV\endcsname{172.7}
\expandafter\newcommand\csname sr22n916ktMiiH\endcsname{1004}
\expandafter\newcommand\csname sr22n916ktMiV\endcsname{136.7}
\expandafter\newcommand\csname sr22n916ktMiH\endcsname{494}
\expandafter\newcommand\csname sr22n916ktMiiVErr\endcsname{5.6\%}
\expandafter\newcommand\csname sr22n916ktMiiHErr\endcsname{12.9\%}
\expandafter\newcommand\csname sr22n916ktMiVErr\endcsname{-16.4\%}
\expandafter\newcommand\csname sr22n916ktMiHErr\endcsname{-44.4\%}
\expandafter\newcommand\csname sr22n916ktRMSE\endcsname{4.12}
\expandafter\newcommand\csname sr22n916ktVRange\endcsname{5.1}
\expandafter\newcommand\csname sr22n916ktHRange\endcsname{12}
\expandafter\newcommand\csname sr22n916ktLoverH\endcsname{0.60}
\expandafter\newcommand\csname sr22n916ktTrackSource\endcsname{N916KT-track-EGM96\_avg.kml}
\expandafter\newcommand\csname sr22n916ktAdsAvailable\endcsname{1}
\expandafter\newcommand\csname sr22n916ktTrackAvailable\endcsname{1}
\expandafter\newcommand\csname sr22n916ktInsetXMin\endcsname{-39.0}
\expandafter\newcommand\csname sr22n916ktInsetXMax\endcsname{28.5}
\expandafter\newcommand\csname sr22n916ktInsetXTickLo\endcsname{-39}
\expandafter\newcommand\csname sr22n916ktInsetXTickHi\endcsname{20}
\expandafter\newcommand\csname sr22n916ktInsetVMin\endcsname{156}
\expandafter\newcommand\csname sr22n916ktInsetVMax\endcsname{171}
\expandafter\newcommand\csname sr22n916ktInsetHMin\endcsname{853}
\expandafter\newcommand\csname sr22n916ktInsetHMax\endcsname{916}
\expandafter\newcommand\csname sr22n916ktTrackXMin\endcsname{-3676}
\expandafter\newcommand\csname sr22n916ktTrackXMax\endcsname{3013}
\expandafter\newcommand\csname sr22n916ktTrackYMin\endcsname{-51}
\expandafter\newcommand\csname sr22n916ktTrackYMax\endcsname{690}
\expandafter\newcommand\csname sr22n916ktTrackXTickLo\endcsname{-3676}
\expandafter\newcommand\csname sr22n916ktTrackXTickHi\endcsname{2000}
\expandafter\newcommand\csname sr22n916ktTrackYTickLo\endcsname{0}
\expandafter\newcommand\csname sr22n916ktTrackYTickHi\endcsname{600}
\expandafter\newcommand\csname sr22n916ktTrackYTick\endcsname{600}
\expandafter\newcommand\csname sr22n916ktTrackClosestY\endcsname{535}
\expandafter\newcommand\csname sr23n310mgRank\endcsname{23}
\expandafter\newcommand\csname sr23n310mgLabel\endcsname{N310MG}
\expandafter\newcommand\csname sr23n310mgDataset\endcsname{N310MG}
\expandafter\newcommand\csname sr23n310mgGeometry\endcsname{far away/non-overhead}
\expandafter\newcommand\csname sr23n310mgVCat\endcsname{V constant}
\expandafter\newcommand\csname sr23n310mgHCat\endcsname{H varying}
\expandafter\newcommand\csname sr23n310mgAudio\endcsname{smooth plateau-like + curve}
\expandafter\newcommand\csname sr23n310mgClosest\endcsname{2026-05-16 17:04:44.40 UTC}
\expandafter\newcommand\csname sr23n310mgXMin\endcsname{-34.5}
\expandafter\newcommand\csname sr23n310mgXMax\endcsname{22.0}
\expandafter\newcommand\csname sr23n310mgYMin\endcsname{316.8}
\expandafter\newcommand\csname sr23n310mgYMax\endcsname{463.3}
\expandafter\newcommand\csname sr23n310mgFzero\endcsname{369.4}
\expandafter\newcommand\csname sr23n310mgXStartTick\endcsname{-30}
\expandafter\newcommand\csname sr23n310mgXEndTick\endcsname{20}
\expandafter\newcommand\csname sr23n310mgYLowTick\endcsname{326}
\expandafter\newcommand\csname sr23n310mgYHighTick\endcsname{442}
\expandafter\newcommand\csname sr23n310mgAdsV\endcsname{246.0}
\expandafter\newcommand\csname sr23n310mgAdsH\endcsname{798}
\expandafter\newcommand\csname sr23n310mgMiiV\endcsname{250.1}
\expandafter\newcommand\csname sr23n310mgMiiH\endcsname{1141}
\expandafter\newcommand\csname sr23n310mgMiV\endcsname{174.4}
\expandafter\newcommand\csname sr23n310mgMiH\endcsname{--}
\expandafter\newcommand\csname sr23n310mgMiiVErr\endcsname{1.7\%}
\expandafter\newcommand\csname sr23n310mgMiiHErr\endcsname{43.0\%}
\expandafter\newcommand\csname sr23n310mgMiVErr\endcsname{-29.1\%}
\expandafter\newcommand\csname sr23n310mgMiHErr\endcsname{--}
\expandafter\newcommand\csname sr23n310mgRMSE\endcsname{2.70}
\expandafter\newcommand\csname sr23n310mgVRange\endcsname{5.5}
\expandafter\newcommand\csname sr23n310mgHRange\endcsname{197}
\expandafter\newcommand\csname sr23n310mgLoverH\endcsname{1.85}
\expandafter\newcommand\csname sr23n310mgTrackSource\endcsname{N310MG-track-EGM96\_avg.kml}
\expandafter\newcommand\csname sr23n310mgAdsAvailable\endcsname{1}
\expandafter\newcommand\csname sr23n310mgTrackAvailable\endcsname{1}
\expandafter\newcommand\csname sr23n310mgInsetXMin\endcsname{-31.7}
\expandafter\newcommand\csname sr23n310mgInsetXMax\endcsname{19.2}
\expandafter\newcommand\csname sr23n310mgInsetXTickLo\endcsname{-20}
\expandafter\newcommand\csname sr23n310mgInsetXTickHi\endcsname{10}
\expandafter\newcommand\csname sr23n310mgInsetVMin\endcsname{229}
\expandafter\newcommand\csname sr23n310mgInsetVMax\endcsname{252}
\expandafter\newcommand\csname sr23n310mgInsetHMin\endcsname{659}
\expandafter\newcommand\csname sr23n310mgInsetHMax\endcsname{932}
\expandafter\newcommand\csname sr23n310mgTrackXMin\endcsname{-5588}
\expandafter\newcommand\csname sr23n310mgTrackXMax\endcsname{4691}
\expandafter\newcommand\csname sr23n310mgTrackYMin\endcsname{-120}
\expandafter\newcommand\csname sr23n310mgTrackYMax\endcsname{1620}
\expandafter\newcommand\csname sr23n310mgTrackXTickLo\endcsname{-4000}
\expandafter\newcommand\csname sr23n310mgTrackXTickHi\endcsname{4000}
\expandafter\newcommand\csname sr23n310mgTrackYTickLo\endcsname{0}
\expandafter\newcommand\csname sr23n310mgTrackYTickHi\endcsname{1500}
\expandafter\newcommand\csname sr23n310mgTrackYTick\endcsname{1500}
\expandafter\newcommand\csname sr23n310mgTrackClosestY\endcsname{1480}
\expandafter\newcommand\csname sr24n9658kRank\endcsname{24}
\expandafter\newcommand\csname sr24n9658kLabel\endcsname{N9658K}
\expandafter\newcommand\csname sr24n9658kDataset\endcsname{N9658K lower ridge}
\expandafter\newcommand\csname sr24n9658kGeometry\endcsname{near overhead}
\expandafter\newcommand\csname sr24n9658kVCat\endcsname{V varying}
\expandafter\newcommand\csname sr24n9658kHCat\endcsname{H varying}
\expandafter\newcommand\csname sr24n9658kAudio\endcsname{plateau-like + curve but scattered/ambiguous}
\expandafter\newcommand\csname sr24n9658kClosest\endcsname{2026-05-14 17:50:22 UTC}
\expandafter\newcommand\csname sr24n9658kXMin\endcsname{-53.0}
\expandafter\newcommand\csname sr24n9658kXMax\endcsname{51.8}
\expandafter\newcommand\csname sr24n9658kYMin\endcsname{509.6}
\expandafter\newcommand\csname sr24n9658kYMax\endcsname{728.3}
\expandafter\newcommand\csname sr24n9658kFzero\endcsname{597.0}
\expandafter\newcommand\csname sr24n9658kXStartTick\endcsname{-50}
\expandafter\newcommand\csname sr24n9658kXEndTick\endcsname{50}
\expandafter\newcommand\csname sr24n9658kYLowTick\endcsname{523}
\expandafter\newcommand\csname sr24n9658kYHighTick\endcsname{697}
\expandafter\newcommand\csname sr24n9658kAdsV\endcsname{181.5}
\expandafter\newcommand\csname sr24n9658kAdsH\endcsname{610}
\expandafter\newcommand\csname sr24n9658kMiiV\endcsname{175.3}
\expandafter\newcommand\csname sr24n9658kMiiH\endcsname{522}
\expandafter\newcommand\csname sr24n9658kMiV\endcsname{167.1}
\expandafter\newcommand\csname sr24n9658kMiH\endcsname{487}
\expandafter\newcommand\csname sr24n9658kMiiVErr\endcsname{-3.4\%}
\expandafter\newcommand\csname sr24n9658kMiiHErr\endcsname{-14.4\%}
\expandafter\newcommand\csname sr24n9658kMiVErr\endcsname{-7.9\%}
\expandafter\newcommand\csname sr24n9658kMiHErr\endcsname{-20.1\%}
\expandafter\newcommand\csname sr24n9658kRMSE\endcsname{2.29}
\expandafter\newcommand\csname sr24n9658kVRange\endcsname{44.5}
\expandafter\newcommand\csname sr24n9658kHRange\endcsname{662}
\expandafter\newcommand\csname sr24n9658kLoverH\endcsname{0.82}
\expandafter\newcommand\csname sr24n9658kTrackSource\endcsname{N9658K-track-EGM96\_avg.kml}
\expandafter\newcommand\csname sr24n9658kAdsAvailable\endcsname{1}
\expandafter\newcommand\csname sr24n9658kTrackAvailable\endcsname{1}
\expandafter\newcommand\csname sr24n9658kInsetXMin\endcsname{-47.8}
\expandafter\newcommand\csname sr24n9658kInsetXMax\endcsname{46.6}
\expandafter\newcommand\csname sr24n9658kInsetXTickLo\endcsname{-40}
\expandafter\newcommand\csname sr24n9658kInsetXTickHi\endcsname{40}
\expandafter\newcommand\csname sr24n9658kInsetVMin\endcsname{172}
\expandafter\newcommand\csname sr24n9658kInsetVMax\endcsname{200}
\expandafter\newcommand\csname sr24n9658kInsetHMin\endcsname{443}
\expandafter\newcommand\csname sr24n9658kInsetHMax\endcsname{717}
\expandafter\newcommand\csname sr24n9658kTrackXMin\endcsname{-4992}
\expandafter\newcommand\csname sr24n9658kTrackXMax\endcsname{4526}
\expandafter\newcommand\csname sr24n9658kTrackYMin\endcsname{-731}
\expandafter\newcommand\csname sr24n9658kTrackYMax\endcsname{1507}
\expandafter\newcommand\csname sr24n9658kTrackXTickLo\endcsname{-4000}
\expandafter\newcommand\csname sr24n9658kTrackXTickHi\endcsname{4000}
\expandafter\newcommand\csname sr24n9658kTrackYTickLo\endcsname{-500}
\expandafter\newcommand\csname sr24n9658kTrackYTickHi\endcsname{1000}
\expandafter\newcommand\csname sr24n9658kTrackYTick\endcsname{1000}
\expandafter\newcommand\csname sr24n9658kTrackClosestY\endcsname{503}
\expandafter\newcommand\csname sr25n9658kRank\endcsname{25}
\expandafter\newcommand\csname sr25n9658kLabel\endcsname{N9658K}
\expandafter\newcommand\csname sr25n9658kDataset\endcsname{N9658K high ridge}
\expandafter\newcommand\csname sr25n9658kGeometry\endcsname{near overhead}
\expandafter\newcommand\csname sr25n9658kVCat\endcsname{V varying}
\expandafter\newcommand\csname sr25n9658kHCat\endcsname{H varying}
\expandafter\newcommand\csname sr25n9658kAudio\endcsname{smooth plateau-like + curve}
\expandafter\newcommand\csname sr25n9658kClosest\endcsname{2026-05-14 17:50:22 UTC}
\expandafter\newcommand\csname sr25n9658kXMin\endcsname{-44.1}
\expandafter\newcommand\csname sr25n9658kXMax\endcsname{54.5}
\expandafter\newcommand\csname sr25n9658kYMin\endcsname{784.0}
\expandafter\newcommand\csname sr25n9658kYMax\endcsname{1158.1}
\expandafter\newcommand\csname sr25n9658kFzero\endcsname{935.6}
\expandafter\newcommand\csname sr25n9658kXStartTick\endcsname{-40}
\expandafter\newcommand\csname sr25n9658kXEndTick\endcsname{50}
\expandafter\newcommand\csname sr25n9658kYLowTick\endcsname{808}
\expandafter\newcommand\csname sr25n9658kYHighTick\endcsname{1105}
\expandafter\newcommand\csname sr25n9658kAdsV\endcsname{181.5}
\expandafter\newcommand\csname sr25n9658kAdsH\endcsname{610}
\expandafter\newcommand\csname sr25n9658kMiiV\endcsname{174.5}
\expandafter\newcommand\csname sr25n9658kMiiH\endcsname{507}
\expandafter\newcommand\csname sr25n9658kMiV\endcsname{157.0}
\expandafter\newcommand\csname sr25n9658kMiH\endcsname{359}
\expandafter\newcommand\csname sr25n9658kMiiVErr\endcsname{-3.8\%}
\expandafter\newcommand\csname sr25n9658kMiiHErr\endcsname{-16.8\%}
\expandafter\newcommand\csname sr25n9658kMiVErr\endcsname{-13.5\%}
\expandafter\newcommand\csname sr25n9658kMiHErr\endcsname{-41.1\%}
\expandafter\newcommand\csname sr25n9658kRMSE\endcsname{7.18}
\expandafter\newcommand\csname sr25n9658kVRange\endcsname{44.5}
\expandafter\newcommand\csname sr25n9658kHRange\endcsname{662}
\expandafter\newcommand\csname sr25n9658kLoverH\endcsname{0.82}
\expandafter\newcommand\csname sr25n9658kTrackSource\endcsname{N9658K-track-EGM96\_avg.kml}
\expandafter\newcommand\csname sr25n9658kAdsAvailable\endcsname{1}
\expandafter\newcommand\csname sr25n9658kTrackAvailable\endcsname{1}
\expandafter\newcommand\csname sr25n9658kInsetXMin\endcsname{-39.2}
\expandafter\newcommand\csname sr25n9658kInsetXMax\endcsname{49.6}
\expandafter\newcommand\csname sr25n9658kInsetXTickLo\endcsname{-20}
\expandafter\newcommand\csname sr25n9658kInsetXTickHi\endcsname{40}
\expandafter\newcommand\csname sr25n9658kInsetVMin\endcsname{172}
\expandafter\newcommand\csname sr25n9658kInsetVMax\endcsname{200}
\expandafter\newcommand\csname sr25n9658kInsetHMin\endcsname{433}
\expandafter\newcommand\csname sr25n9658kInsetHMax\endcsname{697}
\expandafter\newcommand\csname sr25n9658kTrackXMin\endcsname{-4642}
\expandafter\newcommand\csname sr25n9658kTrackXMax\endcsname{4500}
\expandafter\newcommand\csname sr25n9658kTrackYMin\endcsname{-718}
\expandafter\newcommand\csname sr25n9658kTrackYMax\endcsname{1340}
\expandafter\newcommand\csname sr25n9658kTrackXTickLo\endcsname{-4000}
\expandafter\newcommand\csname sr25n9658kTrackXTickHi\endcsname{4000}
\expandafter\newcommand\csname sr25n9658kTrackYTickLo\endcsname{-500}
\expandafter\newcommand\csname sr25n9658kTrackYTickHi\endcsname{1000}
\expandafter\newcommand\csname sr25n9658kTrackYTick\endcsname{1000}
\expandafter\newcommand\csname sr25n9658kTrackClosestY\endcsname{503}
\expandafter\newcommand\csname sr26n120rdRank\endcsname{26}
\expandafter\newcommand\csname sr26n120rdLabel\endcsname{N120RD}
\expandafter\newcommand\csname sr26n120rdDataset\endcsname{N120RD}
\expandafter\newcommand\csname sr26n120rdGeometry\endcsname{far away/non-overhead}
\expandafter\newcommand\csname sr26n120rdVCat\endcsname{V varying}
\expandafter\newcommand\csname sr26n120rdHCat\endcsname{H constant}
\expandafter\newcommand\csname sr26n120rdAudio\endcsname{smooth plateau-like + curve}
\expandafter\newcommand\csname sr26n120rdClosest\endcsname{2026-05-16 19:01:11 UTC}
\expandafter\newcommand\csname sr26n120rdXMin\endcsname{-29.6}
\expandafter\newcommand\csname sr26n120rdXMax\endcsname{45.7}
\expandafter\newcommand\csname sr26n120rdYMin\endcsname{429.1}
\expandafter\newcommand\csname sr26n120rdYMax\endcsname{687.7}
\expandafter\newcommand\csname sr26n120rdFzero\endcsname{527.1}
\expandafter\newcommand\csname sr26n120rdXStartTick\endcsname{-30}
\expandafter\newcommand\csname sr26n120rdXEndTick\endcsname{40}
\expandafter\newcommand\csname sr26n120rdYLowTick\endcsname{446}
\expandafter\newcommand\csname sr26n120rdYHighTick\endcsname{651}
\expandafter\newcommand\csname sr26n120rdAdsV\endcsname{219.7}
\expandafter\newcommand\csname sr26n120rdAdsH\endcsname{369}
\expandafter\newcommand\csname sr26n120rdMiiV\endcsname{230.0}
\expandafter\newcommand\csname sr26n120rdMiiH\endcsname{--}
\expandafter\newcommand\csname sr26n120rdMiV\endcsname{199.0}
\expandafter\newcommand\csname sr26n120rdMiH\endcsname{--}
\expandafter\newcommand\csname sr26n120rdMiiVErr\endcsname{4.7\%}
\expandafter\newcommand\csname sr26n120rdMiiHErr\endcsname{--}
\expandafter\newcommand\csname sr26n120rdMiVErr\endcsname{-9.4\%}
\expandafter\newcommand\csname sr26n120rdMiHErr\endcsname{--}
\expandafter\newcommand\csname sr26n120rdRMSE\endcsname{6.02}
\expandafter\newcommand\csname sr26n120rdVRange\endcsname{64.3}
\expandafter\newcommand\csname sr26n120rdHRange\endcsname{42}
\expandafter\newcommand\csname sr26n120rdLoverH\endcsname{2.68}
\expandafter\newcommand\csname sr26n120rdTrackSource\endcsname{N120RD-track-press\_alt\_uncorrected.kml}
\expandafter\newcommand\csname sr26n120rdAdsAvailable\endcsname{1}
\expandafter\newcommand\csname sr26n120rdTrackAvailable\endcsname{1}
\expandafter\newcommand\csname sr26n120rdInsetXMin\endcsname{-25.9}
\expandafter\newcommand\csname sr26n120rdInsetXMax\endcsname{41.9}
\expandafter\newcommand\csname sr26n120rdInsetXTickLo\endcsname{-20}
\expandafter\newcommand\csname sr26n120rdInsetXTickHi\endcsname{40}
\expandafter\newcommand\csname sr26n120rdInsetVMin\endcsname{195}
\expandafter\newcommand\csname sr26n120rdInsetVMax\endcsname{284}
\expandafter\newcommand\csname sr26n120rdInsetHMin\endcsname{288}
\expandafter\newcommand\csname sr26n120rdInsetHMax\endcsname{450}
\expandafter\newcommand\csname sr26n120rdTrackXMin\endcsname{-4441}
\expandafter\newcommand\csname sr26n120rdTrackXMax\endcsname{3811}
\expandafter\newcommand\csname sr26n120rdTrackYMin\endcsname{-1196}
\expandafter\newcommand\csname sr26n120rdTrackYMax\endcsname{1201}
\expandafter\newcommand\csname sr26n120rdTrackXTickLo\endcsname{-4000}
\expandafter\newcommand\csname sr26n120rdTrackXTickHi\endcsname{2000}
\expandafter\newcommand\csname sr26n120rdTrackYTickLo\endcsname{-1000}
\expandafter\newcommand\csname sr26n120rdTrackYTickHi\endcsname{1000}
\expandafter\newcommand\csname sr26n120rdTrackYTick\endcsname{1000}
\expandafter\newcommand\csname sr26n120rdTrackClosestY\endcsname{1036}
\expandafter\newcommand\csname sr27n8085wRank\endcsname{27}
\expandafter\newcommand\csname sr27n8085wLabel\endcsname{N8085W}
\expandafter\newcommand\csname sr27n8085wDataset\endcsname{N8085W}
\expandafter\newcommand\csname sr27n8085wGeometry\endcsname{overhead}
\expandafter\newcommand\csname sr27n8085wVCat\endcsname{V constant}
\expandafter\newcommand\csname sr27n8085wHCat\endcsname{H varying}
\expandafter\newcommand\csname sr27n8085wAudio\endcsname{smooth curve only / incomplete ends}
\expandafter\newcommand\csname sr27n8085wClosest\endcsname{2026-05-16 18:50:24 UTC}
\expandafter\newcommand\csname sr27n8085wXMin\endcsname{-48.3}
\expandafter\newcommand\csname sr27n8085wXMax\endcsname{22.8}
\expandafter\newcommand\csname sr27n8085wYMin\endcsname{331.6}
\expandafter\newcommand\csname sr27n8085wYMax\endcsname{520.4}
\expandafter\newcommand\csname sr27n8085wFzero\endcsname{388.7}
\expandafter\newcommand\csname sr27n8085wXStartTick\endcsname{-40}
\expandafter\newcommand\csname sr27n8085wXEndTick\endcsname{20}
\expandafter\newcommand\csname sr27n8085wYLowTick\endcsname{344}
\expandafter\newcommand\csname sr27n8085wYHighTick\endcsname{493}
\expandafter\newcommand\csname sr27n8085wAdsV\endcsname{237.0}
\expandafter\newcommand\csname sr27n8085wAdsH\endcsname{1471}
\expandafter\newcommand\csname sr27n8085wMiiV\endcsname{274.4}
\expandafter\newcommand\csname sr27n8085wMiiH\endcsname{1990}
\expandafter\newcommand\csname sr27n8085wMiV\endcsname{203.9}
\expandafter\newcommand\csname sr27n8085wMiH\endcsname{1011}
\expandafter\newcommand\csname sr27n8085wMiiVErr\endcsname{15.8\%}
\expandafter\newcommand\csname sr27n8085wMiiHErr\endcsname{35.3\%}
\expandafter\newcommand\csname sr27n8085wMiVErr\endcsname{-13.9\%}
\expandafter\newcommand\csname sr27n8085wMiHErr\endcsname{-31.3\%}
\expandafter\newcommand\csname sr27n8085wRMSE\endcsname{2.52}
\expandafter\newcommand\csname sr27n8085wVRange\endcsname{3.6}
\expandafter\newcommand\csname sr27n8085wHRange\endcsname{174}
\expandafter\newcommand\csname sr27n8085wLoverH\endcsname{0.01}
\expandafter\newcommand\csname sr27n8085wTrackSource\endcsname{N8085W-track-EGM96\_avg.kml}
\expandafter\newcommand\csname sr27n8085wAdsAvailable\endcsname{1}
\expandafter\newcommand\csname sr27n8085wTrackAvailable\endcsname{1}
\expandafter\newcommand\csname sr27n8085wInsetXMin\endcsname{-44.8}
\expandafter\newcommand\csname sr27n8085wInsetXMax\endcsname{19.3}
\expandafter\newcommand\csname sr27n8085wInsetXTickLo\endcsname{-40}
\expandafter\newcommand\csname sr27n8085wInsetXTickHi\endcsname{19}
\expandafter\newcommand\csname sr27n8085wInsetVMin\endcsname{229}
\expandafter\newcommand\csname sr27n8085wInsetVMax\endcsname{247}
\expandafter\newcommand\csname sr27n8085wInsetHMin\endcsname{1334}
\expandafter\newcommand\csname sr27n8085wInsetHMax\endcsname{1608}
\expandafter\newcommand\csname sr27n8085wTrackXMin\endcsname{-6166}
\expandafter\newcommand\csname sr27n8085wTrackXMax\endcsname{4543}
\expandafter\newcommand\csname sr27n8085wTrackYMin\endcsname{-40}
\expandafter\newcommand\csname sr27n8085wTrackYMax\endcsname{99}
\expandafter\newcommand\csname sr27n8085wTrackXTickLo\endcsname{-4000}
\expandafter\newcommand\csname sr27n8085wTrackXTickHi\endcsname{2000}
\expandafter\newcommand\csname sr27n8085wTrackYTickLo\endcsname{-20}
\expandafter\newcommand\csname sr27n8085wTrackYTickHi\endcsname{80}
\expandafter\newcommand\csname sr27n8085wTrackYTick\endcsname{80}
\expandafter\newcommand\csname sr27n8085wTrackClosestY\endcsname{19}
\expandafter\newcommand\csname sr30n7rrRank\endcsname{30}
\expandafter\newcommand\csname sr30n7rrLabel\endcsname{N7RR}
\expandafter\newcommand\csname sr30n7rrDataset\endcsname{N7RR}
\expandafter\newcommand\csname sr30n7rrGeometry\endcsname{far away/non-overhead}
\expandafter\newcommand\csname sr30n7rrVCat\endcsname{V constant}
\expandafter\newcommand\csname sr30n7rrHCat\endcsname{H varying}
\expandafter\newcommand\csname sr30n7rrAudio\endcsname{smooth plateau-like + curve}
\expandafter\newcommand\csname sr30n7rrClosest\endcsname{2026-05-16 17:20:42.30 UTC}
\expandafter\newcommand\csname sr30n7rrXMin\endcsname{-68.2}
\expandafter\newcommand\csname sr30n7rrXMax\endcsname{41.7}
\expandafter\newcommand\csname sr30n7rrYMin\endcsname{320.7}
\expandafter\newcommand\csname sr30n7rrYMax\endcsname{488.2}
\expandafter\newcommand\csname sr30n7rrFzero\endcsname{369.7}
\expandafter\newcommand\csname sr30n7rrXStartTick\endcsname{-60}
\expandafter\newcommand\csname sr30n7rrXEndTick\endcsname{40}
\expandafter\newcommand\csname sr30n7rrYLowTick\endcsname{332}
\expandafter\newcommand\csname sr30n7rrYHighTick\endcsname{464}
\expandafter\newcommand\csname sr30n7rrAdsV\endcsname{242.0}
\expandafter\newcommand\csname sr30n7rrAdsH\endcsname{2128}
\expandafter\newcommand\csname sr30n7rrMiiV\endcsname{314.2}
\expandafter\newcommand\csname sr30n7rrMiiH\endcsname{5515}
\expandafter\newcommand\csname sr30n7rrMiV\endcsname{180.8}
\expandafter\newcommand\csname sr30n7rrMiH\endcsname{--}
\expandafter\newcommand\csname sr30n7rrMiiVErr\endcsname{29.9\%}
\expandafter\newcommand\csname sr30n7rrMiiHErr\endcsname{159.2\%}
\expandafter\newcommand\csname sr30n7rrMiVErr\endcsname{-25.3\%}
\expandafter\newcommand\csname sr30n7rrMiHErr\endcsname{--}
\expandafter\newcommand\csname sr30n7rrRMSE\endcsname{2.01}
\expandafter\newcommand\csname sr30n7rrVRange\endcsname{6.7}
\expandafter\newcommand\csname sr30n7rrHRange\endcsname{182}
\expandafter\newcommand\csname sr30n7rrLoverH\endcsname{1.01}
\expandafter\newcommand\csname sr30n7rrTrackSource\endcsname{N7RR-track-EGM96\_avg.kml}
\expandafter\newcommand\csname sr30n7rrAdsAvailable\endcsname{1}
\expandafter\newcommand\csname sr30n7rrTrackAvailable\endcsname{1}
\expandafter\newcommand\csname sr30n7rrInsetXMin\endcsname{-62.7}
\expandafter\newcommand\csname sr30n7rrInsetXMax\endcsname{36.3}
\expandafter\newcommand\csname sr30n7rrInsetXTickLo\endcsname{-40}
\expandafter\newcommand\csname sr30n7rrInsetXTickHi\endcsname{20}
\expandafter\newcommand\csname sr30n7rrInsetVMin\endcsname{229}
\expandafter\newcommand\csname sr30n7rrInsetVMax\endcsname{259}
\expandafter\newcommand\csname sr30n7rrInsetHMin\endcsname{1936}
\expandafter\newcommand\csname sr30n7rrInsetHMax\endcsname{2266}
\expandafter\newcommand\csname sr30n7rrTrackXMin\endcsname{-8136}
\expandafter\newcommand\csname sr30n7rrTrackXMax\endcsname{5215}
\expandafter\newcommand\csname sr30n7rrTrackYMin\endcsname{-190}
\expandafter\newcommand\csname sr30n7rrTrackYMax\endcsname{2569}
\expandafter\newcommand\csname sr30n7rrTrackXTickLo\endcsname{-6000}
\expandafter\newcommand\csname sr30n7rrTrackXTickHi\endcsname{2000}
\expandafter\newcommand\csname sr30n7rrTrackYTickLo\endcsname{0}
\expandafter\newcommand\csname sr30n7rrTrackYTickHi\endcsname{2000}
\expandafter\newcommand\csname sr30n7rrTrackYTick\endcsname{2000}
\expandafter\newcommand\csname sr30n7rrTrackClosestY\endcsname{2152}
\expandafter\newcommand\csname sr31n52272Rank\endcsname{31}
\expandafter\newcommand\csname sr31n52272Label\endcsname{N52272}
\expandafter\newcommand\csname sr31n52272Dataset\endcsname{N52272}
\expandafter\newcommand\csname sr31n52272Geometry\endcsname{far away/non-overhead}
\expandafter\newcommand\csname sr31n52272VCat\endcsname{V constant}
\expandafter\newcommand\csname sr31n52272HCat\endcsname{H constant}
\expandafter\newcommand\csname sr31n52272Audio\endcsname{plateau-like + curve but scattered/ambiguous}
\expandafter\newcommand\csname sr31n52272Closest\endcsname{2026-05-13 18:18:58.60 UTC}
\expandafter\newcommand\csname sr31n52272XMin\endcsname{-48.0}
\expandafter\newcommand\csname sr31n52272XMax\endcsname{58.8}
\expandafter\newcommand\csname sr31n52272YMin\endcsname{290.4}
\expandafter\newcommand\csname sr31n52272YMax\endcsname{481.6}
\expandafter\newcommand\csname sr31n52272Fzero\endcsname{364.4}
\expandafter\newcommand\csname sr31n52272XStartTick\endcsname{-40}
\expandafter\newcommand\csname sr31n52272XEndTick\endcsname{50}
\expandafter\newcommand\csname sr31n52272YLowTick\endcsname{303}
\expandafter\newcommand\csname sr31n52272YHighTick\endcsname{454}
\expandafter\newcommand\csname sr31n52272AdsV\endcsname{175.8}
\expandafter\newcommand\csname sr31n52272AdsH\endcsname{994}
\expandafter\newcommand\csname sr31n52272MiiV\endcsname{259.4}
\expandafter\newcommand\csname sr31n52272MiiH\endcsname{1160}
\expandafter\newcommand\csname sr31n52272MiV\endcsname{220.9}
\expandafter\newcommand\csname sr31n52272MiH\endcsname{--}
\expandafter\newcommand\csname sr31n52272MiiVErr\endcsname{47.5\%}
\expandafter\newcommand\csname sr31n52272MiiHErr\endcsname{16.7\%}
\expandafter\newcommand\csname sr31n52272MiVErr\endcsname{25.7\%}
\expandafter\newcommand\csname sr31n52272MiHErr\endcsname{--}
\expandafter\newcommand\csname sr31n52272RMSE\endcsname{3.80}
\expandafter\newcommand\csname sr31n52272VRange\endcsname{13.6}
\expandafter\newcommand\csname sr31n52272HRange\endcsname{28}
\expandafter\newcommand\csname sr31n52272LoverH\endcsname{1.54}
\expandafter\newcommand\csname sr31n52272TrackSource\endcsname{N52272-track-EGM96\_avg.kml}
\expandafter\newcommand\csname sr31n52272AdsAvailable\endcsname{1}
\expandafter\newcommand\csname sr31n52272TrackAvailable\endcsname{1}
\expandafter\newcommand\csname sr31n52272InsetXMin\endcsname{-42.7}
\expandafter\newcommand\csname sr31n52272InsetXMax\endcsname{53.5}
\expandafter\newcommand\csname sr31n52272InsetXTickLo\endcsname{-20}
\expandafter\newcommand\csname sr31n52272InsetXTickHi\endcsname{40}
\expandafter\newcommand\csname sr31n52272InsetVMin\endcsname{168}
\expandafter\newcommand\csname sr31n52272InsetVMax\endcsname{190}
\expandafter\newcommand\csname sr31n52272InsetHMin\endcsname{950}
\expandafter\newcommand\csname sr31n52272InsetHMax\endcsname{1045}
\expandafter\newcommand\csname sr31n52272TrackXMin\endcsname{-4715}
\expandafter\newcommand\csname sr31n52272TrackXMax\endcsname{5034}
\expandafter\newcommand\csname sr31n52272TrackYMin\endcsname{-187}
\expandafter\newcommand\csname sr31n52272TrackYMax\endcsname{2519}
\expandafter\newcommand\csname sr31n52272TrackXTickLo\endcsname{-4000}
\expandafter\newcommand\csname sr31n52272TrackXTickHi\endcsname{4000}
\expandafter\newcommand\csname sr31n52272TrackYTickLo\endcsname{0}
\expandafter\newcommand\csname sr31n52272TrackYTickHi\endcsname{2000}
\expandafter\newcommand\csname sr31n52272TrackYTick\endcsname{2000}
\expandafter\newcommand\csname sr31n52272TrackClosestY\endcsname{1519}
\expandafter\newcommand\csname sr32n6848uRank\endcsname{32}
\expandafter\newcommand\csname sr32n6848uLabel\endcsname{N6848U}
\expandafter\newcommand\csname sr32n6848uDataset\endcsname{N6848U early segment}
\expandafter\newcommand\csname sr32n6848uGeometry\endcsname{far away/non-overhead}
\expandafter\newcommand\csname sr32n6848uVCat\endcsname{V constant}
\expandafter\newcommand\csname sr32n6848uHCat\endcsname{H constant}
\expandafter\newcommand\csname sr32n6848uAudio\endcsname{curve only + scattered/ambiguous}
\expandafter\newcommand\csname sr32n6848uClosest\endcsname{2026-05-13 18:24:13 UTC}
\expandafter\newcommand\csname sr32n6848uXMin\endcsname{-48.1}
\expandafter\newcommand\csname sr32n6848uXMax\endcsname{57.3}
\expandafter\newcommand\csname sr32n6848uYMin\endcsname{288.1}
\expandafter\newcommand\csname sr32n6848uYMax\endcsname{481.3}
\expandafter\newcommand\csname sr32n6848uFzero\endcsname{363.5}
\expandafter\newcommand\csname sr32n6848uXStartTick\endcsname{-40}
\expandafter\newcommand\csname sr32n6848uXEndTick\endcsname{50}
\expandafter\newcommand\csname sr32n6848uYLowTick\endcsname{300}
\expandafter\newcommand\csname sr32n6848uYHighTick\endcsname{454}
\expandafter\newcommand\csname sr32n6848uAdsV\endcsname{244.1}
\expandafter\newcommand\csname sr32n6848uAdsH\endcsname{1860}
\expandafter\newcommand\csname sr32n6848uMiiV\endcsname{258.4}
\expandafter\newcommand\csname sr32n6848uMiiH\endcsname{--}
\expandafter\newcommand\csname sr32n6848uMiV\endcsname{218.6}
\expandafter\newcommand\csname sr32n6848uMiH\endcsname{--}
\expandafter\newcommand\csname sr32n6848uMiiVErr\endcsname{5.9\%}
\expandafter\newcommand\csname sr32n6848uMiiHErr\endcsname{--}
\expandafter\newcommand\csname sr32n6848uMiVErr\endcsname{-10.4\%}
\expandafter\newcommand\csname sr32n6848uMiHErr\endcsname{--}
\expandafter\newcommand\csname sr32n6848uRMSE\endcsname{5.37}
\expandafter\newcommand\csname sr32n6848uVRange\endcsname{7.7}
\expandafter\newcommand\csname sr32n6848uHRange\endcsname{47}
\expandafter\newcommand\csname sr32n6848uLoverH\endcsname{13.17}
\expandafter\newcommand\csname sr32n6848uTrackSource\endcsname{N6848U-track-EGM96\_avg.kml}
\expandafter\newcommand\csname sr32n6848uAdsAvailable\endcsname{1}
\expandafter\newcommand\csname sr32n6848uTrackAvailable\endcsname{1}
\expandafter\newcommand\csname sr32n6848uInsetXMin\endcsname{-42.9}
\expandafter\newcommand\csname sr32n6848uInsetXMax\endcsname{52.1}
\expandafter\newcommand\csname sr32n6848uInsetXTickLo\endcsname{-20}
\expandafter\newcommand\csname sr32n6848uInsetXTickHi\endcsname{40}
\expandafter\newcommand\csname sr32n6848uInsetVMin\endcsname{237}
\expandafter\newcommand\csname sr32n6848uInsetVMax\endcsname{257}
\expandafter\newcommand\csname sr32n6848uInsetHMin\endcsname{1840}
\expandafter\newcommand\csname sr32n6848uInsetHMax\endcsname{1887}
\expandafter\newcommand\csname sr32n6848uTrackXMin\endcsname{-6266}
\expandafter\newcommand\csname sr32n6848uTrackXMax\endcsname{5853}
\expandafter\newcommand\csname sr32n6848uTrackYMin\endcsname{-239}
\expandafter\newcommand\csname sr32n6848uTrackYMax\endcsname{3227}
\expandafter\newcommand\csname sr32n6848uTrackXTickLo\endcsname{-4000}
\expandafter\newcommand\csname sr32n6848uTrackXTickHi\endcsname{4000}
\expandafter\newcommand\csname sr32n6848uTrackYTickLo\endcsname{0}
\expandafter\newcommand\csname sr32n6848uTrackYTickHi\endcsname{3000}
\expandafter\newcommand\csname sr32n6848uTrackYTick\endcsname{3000}
\expandafter\newcommand\csname sr32n6848uTrackClosestY\endcsname{2855}
\expandafter\newcommand\csname sr34n119rpRank\endcsname{34}
\expandafter\newcommand\csname sr34n119rpLabel\endcsname{N119RP}
\expandafter\newcommand\csname sr34n119rpDataset\endcsname{N119RP}
\expandafter\newcommand\csname sr34n119rpGeometry\endcsname{near overhead}
\expandafter\newcommand\csname sr34n119rpVCat\endcsname{V constant}
\expandafter\newcommand\csname sr34n119rpHCat\endcsname{H constant}
\expandafter\newcommand\csname sr34n119rpAudio\endcsname{smooth plateau-like + curve}
\expandafter\newcommand\csname sr34n119rpClosest\endcsname{2026-05-09 02:43:33.50 UTC}
\expandafter\newcommand\csname sr34n119rpXMin\endcsname{-53.5}
\expandafter\newcommand\csname sr34n119rpXMax\endcsname{23.8}
\expandafter\newcommand\csname sr34n119rpYMin\endcsname{109.8}
\expandafter\newcommand\csname sr34n119rpYMax\endcsname{180.5}
\expandafter\newcommand\csname sr34n119rpFzero\endcsname{129.6}
\expandafter\newcommand\csname sr34n119rpXStartTick\endcsname{-50}
\expandafter\newcommand\csname sr34n119rpXEndTick\endcsname{20}
\expandafter\newcommand\csname sr34n119rpYLowTick\endcsname{117}
\expandafter\newcommand\csname sr34n119rpYHighTick\endcsname{164}
\expandafter\newcommand\csname sr34n119rpAdsV\endcsname{267.7}
\expandafter\newcommand\csname sr34n119rpAdsH\endcsname{2023}
\expandafter\newcommand\csname sr34n119rpMiiV\endcsname{300.5}
\expandafter\newcommand\csname sr34n119rpMiiH\endcsname{2778}
\expandafter\newcommand\csname sr34n119rpMiV\endcsname{202.2}
\expandafter\newcommand\csname sr34n119rpMiH\endcsname{1180}
\expandafter\newcommand\csname sr34n119rpMiiVErr\endcsname{12.3\%}
\expandafter\newcommand\csname sr34n119rpMiiHErr\endcsname{37.3\%}
\expandafter\newcommand\csname sr34n119rpMiVErr\endcsname{-24.5\%}
\expandafter\newcommand\csname sr34n119rpMiHErr\endcsname{-41.7\%}
\expandafter\newcommand\csname sr34n119rpRMSE\endcsname{0.76}
\expandafter\newcommand\csname sr34n119rpVRange\endcsname{13.1}
\expandafter\newcommand\csname sr34n119rpHRange\endcsname{23}
\expandafter\newcommand\csname sr34n119rpLoverH\endcsname{0.23}
\expandafter\newcommand\csname sr34n119rpTrackSource\endcsname{FlightAware\_N119RP\_KHWD\_KLGB\_20260509.kml}
\expandafter\newcommand\csname sr34n119rpAdsAvailable\endcsname{1}
\expandafter\newcommand\csname sr34n119rpTrackAvailable\endcsname{1}
\expandafter\newcommand\csname sr34n119rpInsetXMin\endcsname{-49.7}
\expandafter\newcommand\csname sr34n119rpInsetXMax\endcsname{19.9}
\expandafter\newcommand\csname sr34n119rpInsetXTickLo\endcsname{-40}
\expandafter\newcommand\csname sr34n119rpInsetXTickHi\endcsname{20}
\expandafter\newcommand\csname sr34n119rpInsetVMin\endcsname{248}
\expandafter\newcommand\csname sr34n119rpInsetVMax\endcsname{273}
\expandafter\newcommand\csname sr34n119rpInsetHMin\endcsname{1980}
\expandafter\newcommand\csname sr34n119rpInsetHMax\endcsname{2043}
\expandafter\newcommand\csname sr34n119rpTrackXMin\endcsname{-4799}
\expandafter\newcommand\csname sr34n119rpTrackXMax\endcsname{5165}
\expandafter\newcommand\csname sr34n119rpTrackYMin\endcsname{-277}
\expandafter\newcommand\csname sr34n119rpTrackYMax\endcsname{3743}
\expandafter\newcommand\csname sr34n119rpTrackXTickLo\endcsname{-4000}
\expandafter\newcommand\csname sr34n119rpTrackXTickHi\endcsname{4000}
\expandafter\newcommand\csname sr34n119rpTrackYTickLo\endcsname{0}
\expandafter\newcommand\csname sr34n119rpTrackYTickHi\endcsname{3000}
\expandafter\newcommand\csname sr34n119rpTrackYTick\endcsname{3000}
\expandafter\newcommand\csname sr34n119rpTrackClosestY\endcsname{2500}
\expandafter\newcommand\csname sr35unknownc172Rank\endcsname{35}
\expandafter\newcommand\csname sr35unknownc172Label\endcsname{unknown C172}
\expandafter\newcommand\csname sr35unknownc172Dataset\endcsname{N/A C172 unknown flight number}
\expandafter\newcommand\csname sr35unknownc172Geometry\endcsname{far away/non-overhead}
\expandafter\newcommand\csname sr35unknownc172VCat\endcsname{V constant}
\expandafter\newcommand\csname sr35unknownc172HCat\endcsname{H varying}
\expandafter\newcommand\csname sr35unknownc172Audio\endcsname{smooth plateau-like + curve}
\expandafter\newcommand\csname sr35unknownc172Closest\endcsname{2026-05-17 18:29:12 UTC}
\expandafter\newcommand\csname sr35unknownc172XMin\endcsname{-83.6}
\expandafter\newcommand\csname sr35unknownc172XMax\endcsname{52.3}
\expandafter\newcommand\csname sr35unknownc172YMin\endcsname{443.2}
\expandafter\newcommand\csname sr35unknownc172YMax\endcsname{706.4}
\expandafter\newcommand\csname sr35unknownc172Fzero\endcsname{539.8}
\expandafter\newcommand\csname sr35unknownc172XStartTick\endcsname{-80}
\expandafter\newcommand\csname sr35unknownc172XEndTick\endcsname{50}
\expandafter\newcommand\csname sr35unknownc172YLowTick\endcsname{460}
\expandafter\newcommand\csname sr35unknownc172YHighTick\endcsname{669}
\expandafter\newcommand\csname sr35unknownc172AdsV\endcsname{233.0}
\expandafter\newcommand\csname sr35unknownc172AdsH\endcsname{1207}
\expandafter\newcommand\csname sr35unknownc172MiiV\endcsname{228.5}
\expandafter\newcommand\csname sr35unknownc172MiiH\endcsname{1095}
\expandafter\newcommand\csname sr35unknownc172MiV\endcsname{214.4}
\expandafter\newcommand\csname sr35unknownc172MiH\endcsname{858}
\expandafter\newcommand\csname sr35unknownc172MiiVErr\endcsname{-1.9\%}
\expandafter\newcommand\csname sr35unknownc172MiiHErr\endcsname{-9.3\%}
\expandafter\newcommand\csname sr35unknownc172MiVErr\endcsname{-8.0\%}
\expandafter\newcommand\csname sr35unknownc172MiHErr\endcsname{-28.9\%}
\expandafter\newcommand\csname sr35unknownc172RMSE\endcsname{4.97}
\expandafter\newcommand\csname sr35unknownc172VRange\endcsname{8.9}
\expandafter\newcommand\csname sr35unknownc172HRange\endcsname{198}
\expandafter\newcommand\csname sr35unknownc172LoverH\endcsname{0.63}
\expandafter\newcommand\csname sr35unknownc172TrackSource\endcsname{flight\_data.csv}
\expandafter\newcommand\csname sr35unknownc172AdsAvailable\endcsname{1}
\expandafter\newcommand\csname sr35unknownc172TrackAvailable\endcsname{1}
\expandafter\newcommand\csname sr35unknownc172InsetXMin\endcsname{-76.9}
\expandafter\newcommand\csname sr35unknownc172InsetXMax\endcsname{45.5}
\expandafter\newcommand\csname sr35unknownc172InsetXTickLo\endcsname{-60}
\expandafter\newcommand\csname sr35unknownc172InsetXTickHi\endcsname{20}
\expandafter\newcommand\csname sr35unknownc172InsetVMin\endcsname{208}
\expandafter\newcommand\csname sr35unknownc172InsetVMax\endcsname{238}
\expandafter\newcommand\csname sr35unknownc172InsetHMin\endcsname{1099}
\expandafter\newcommand\csname sr35unknownc172InsetHMax\endcsname{1356}
\expandafter\newcommand\csname sr35unknownc172TrackXMin\endcsname{-2425}
\expandafter\newcommand\csname sr35unknownc172TrackXMax\endcsname{3576}
\expandafter\newcommand\csname sr35unknownc172TrackYMin\endcsname{-76}
\expandafter\newcommand\csname sr35unknownc172TrackYMax\endcsname{1020}
\expandafter\newcommand\csname sr35unknownc172TrackXTickLo\endcsname{-2000}
\expandafter\newcommand\csname sr35unknownc172TrackXTickHi\endcsname{3576}
\expandafter\newcommand\csname sr35unknownc172TrackYTickLo\endcsname{0}
\expandafter\newcommand\csname sr35unknownc172TrackYTickHi\endcsname{800}
\expandafter\newcommand\csname sr35unknownc172TrackYTick\endcsname{800}
\expandafter\newcommand\csname sr35unknownc172TrackClosestY\endcsname{699}
\expandafter\newcommand\csname sr36n748spRank\endcsname{36}
\expandafter\newcommand\csname sr36n748spLabel\endcsname{N748SP}
\expandafter\newcommand\csname sr36n748spDataset\endcsname{N748SP lower ridge and auto ridge}
\expandafter\newcommand\csname sr36n748spGeometry\endcsname{far away/non-overhead}
\expandafter\newcommand\csname sr36n748spVCat\endcsname{V varying}
\expandafter\newcommand\csname sr36n748spHCat\endcsname{H constant}
\expandafter\newcommand\csname sr36n748spAudio\endcsname{plateau-like + curve but scattered/ambiguous}
\expandafter\newcommand\csname sr36n748spClosest\endcsname{2026-05-17 20:07:15.83 UTC}
\expandafter\newcommand\csname sr36n748spXMin\endcsname{-29.9}
\expandafter\newcommand\csname sr36n748spXMax\endcsname{34.9}
\expandafter\newcommand\csname sr36n748spYMin\endcsname{250.7}
\expandafter\newcommand\csname sr36n748spYMax\endcsname{365.0}
\expandafter\newcommand\csname sr36n748spFzero\endcsname{293.6}
\expandafter\newcommand\csname sr36n748spXStartTick\endcsname{-30}
\expandafter\newcommand\csname sr36n748spXEndTick\endcsname{30}
\expandafter\newcommand\csname sr36n748spYLowTick\endcsname{259}
\expandafter\newcommand\csname sr36n748spYHighTick\endcsname{349}
\expandafter\newcommand\csname sr36n748spAdsV\endcsname{182.8}
\expandafter\newcommand\csname sr36n748spAdsH\endcsname{1209}
\expandafter\newcommand\csname sr36n748spMiiV\endcsname{227.3}
\expandafter\newcommand\csname sr36n748spMiiH\endcsname{1503}
\expandafter\newcommand\csname sr36n748spMiV\endcsname{159.5}
\expandafter\newcommand\csname sr36n748spMiH\endcsname{516}
\expandafter\newcommand\csname sr36n748spMiiVErr\endcsname{24.3\%}
\expandafter\newcommand\csname sr36n748spMiiHErr\endcsname{24.3\%}
\expandafter\newcommand\csname sr36n748spMiVErr\endcsname{-12.8\%}
\expandafter\newcommand\csname sr36n748spMiHErr\endcsname{-57.3\%}
\expandafter\newcommand\csname sr36n748spRMSE\endcsname{2.34}
\expandafter\newcommand\csname sr36n748spVRange\endcsname{35.8}
\expandafter\newcommand\csname sr36n748spHRange\endcsname{44}
\expandafter\newcommand\csname sr36n748spLoverH\endcsname{0.56}
\expandafter\newcommand\csname sr36n748spTrackSource\endcsname{N748SP-track-EGM96\_avg.kml}
\expandafter\newcommand\csname sr36n748spAdsAvailable\endcsname{1}
\expandafter\newcommand\csname sr36n748spTrackAvailable\endcsname{1}
\expandafter\newcommand\csname sr36n748spInsetXMin\endcsname{-26.7}
\expandafter\newcommand\csname sr36n748spInsetXMax\endcsname{31.7}
\expandafter\newcommand\csname sr36n748spInsetXTickLo\endcsname{-20}
\expandafter\newcommand\csname sr36n748spInsetXTickHi\endcsname{20}
\expandafter\newcommand\csname sr36n748spInsetVMin\endcsname{148}
\expandafter\newcommand\csname sr36n748spInsetVMax\endcsname{199}
\expandafter\newcommand\csname sr36n748spInsetHMin\endcsname{1167}
\expandafter\newcommand\csname sr36n748spInsetHMax\endcsname{1290}
\expandafter\newcommand\csname sr36n748spTrackXMin\endcsname{-3950}
\expandafter\newcommand\csname sr36n748spTrackXMax\endcsname{3351}
\expandafter\newcommand\csname sr36n748spTrackYMin\endcsname{-175}
\expandafter\newcommand\csname sr36n748spTrackYMax\endcsname{774}
\expandafter\newcommand\csname sr36n748spTrackXTickLo\endcsname{-3950}
\expandafter\newcommand\csname sr36n748spTrackXTickHi\endcsname{2000}
\expandafter\newcommand\csname sr36n748spTrackYTickLo\endcsname{0}
\expandafter\newcommand\csname sr36n748spTrackYTickHi\endcsname{600}
\expandafter\newcommand\csname sr36n748spTrackYTick\endcsname{600}
\expandafter\newcommand\csname sr36n748spTrackClosestY\endcsname{666}
\expandafter\newcommand\csname sr37n61ngRank\endcsname{37}
\expandafter\newcommand\csname sr37n61ngLabel\endcsname{N61NG}
\expandafter\newcommand\csname sr37n61ngDataset\endcsname{N61NG}
\expandafter\newcommand\csname sr37n61ngGeometry\endcsname{far away/non-overhead}
\expandafter\newcommand\csname sr37n61ngVCat\endcsname{V constant}
\expandafter\newcommand\csname sr37n61ngHCat\endcsname{H varying}
\expandafter\newcommand\csname sr37n61ngAudio\endcsname{smooth plateau-like + curve}
\expandafter\newcommand\csname sr37n61ngClosest\endcsname{2026-05-17 20:58:10.98 UTC}
\expandafter\newcommand\csname sr37n61ngXMin\endcsname{-34.2}
\expandafter\newcommand\csname sr37n61ngXMax\endcsname{41.2}
\expandafter\newcommand\csname sr37n61ngYMin\endcsname{211.2}
\expandafter\newcommand\csname sr37n61ngYMax\endcsname{451.2}
\expandafter\newcommand\csname sr37n61ngFzero\endcsname{296.4}
\expandafter\newcommand\csname sr37n61ngXStartTick\endcsname{-30}
\expandafter\newcommand\csname sr37n61ngXEndTick\endcsname{40}
\expandafter\newcommand\csname sr37n61ngYLowTick\endcsname{226}
\expandafter\newcommand\csname sr37n61ngYHighTick\endcsname{415}
\expandafter\newcommand\csname sr37n61ngAdsV\endcsname{371.4}
\expandafter\newcommand\csname sr37n61ngAdsH\endcsname{2382}
\expandafter\newcommand\csname sr37n61ngMiiV\endcsname{498.8}
\expandafter\newcommand\csname sr37n61ngMiiH\endcsname{4851}
\expandafter\newcommand\csname sr37n61ngMiV\endcsname{323.2}
\expandafter\newcommand\csname sr37n61ngMiH\endcsname{1478}
\expandafter\newcommand\csname sr37n61ngMiiVErr\endcsname{34.3\%}
\expandafter\newcommand\csname sr37n61ngMiiHErr\endcsname{103.7\%}
\expandafter\newcommand\csname sr37n61ngMiVErr\endcsname{-13.0\%}
\expandafter\newcommand\csname sr37n61ngMiHErr\endcsname{-37.9\%}
\expandafter\newcommand\csname sr37n61ngRMSE\endcsname{4.26}
\expandafter\newcommand\csname sr37n61ngVRange\endcsname{9.4}
\expandafter\newcommand\csname sr37n61ngHRange\endcsname{472}
\expandafter\newcommand\csname sr37n61ngLoverH\endcsname{0.60}
\expandafter\newcommand\csname sr37n61ngTrackSource\endcsname{N61NG-track-EGM96\_avg.kml}
\expandafter\newcommand\csname sr37n61ngAdsAvailable\endcsname{1}
\expandafter\newcommand\csname sr37n61ngTrackAvailable\endcsname{1}
\expandafter\newcommand\csname sr37n61ngInsetXMin\endcsname{-30.5}
\expandafter\newcommand\csname sr37n61ngInsetXMax\endcsname{37.4}
\expandafter\newcommand\csname sr37n61ngInsetXTickLo\endcsname{-20}
\expandafter\newcommand\csname sr37n61ngInsetXTickHi\endcsname{37}
\expandafter\newcommand\csname sr37n61ngInsetVMin\endcsname{352}
\expandafter\newcommand\csname sr37n61ngInsetVMax\endcsname{424}
\expandafter\newcommand\csname sr37n61ngInsetHMin\endcsname{2013}
\expandafter\newcommand\csname sr37n61ngInsetHMax\endcsname{2712}
\expandafter\newcommand\csname sr37n61ngTrackXMin\endcsname{-8189}
\expandafter\newcommand\csname sr37n61ngTrackXMax\endcsname{7962}
\expandafter\newcommand\csname sr37n61ngTrackYMin\endcsname{-2098}
\expandafter\newcommand\csname sr37n61ngTrackYMax\endcsname{1960}
\expandafter\newcommand\csname sr37n61ngTrackXTickLo\endcsname{-5000}
\expandafter\newcommand\csname sr37n61ngTrackXTickHi\endcsname{5000}
\expandafter\newcommand\csname sr37n61ngTrackYTickLo\endcsname{-2000}
\expandafter\newcommand\csname sr37n61ngTrackYTickHi\endcsname{1000}
\expandafter\newcommand\csname sr37n61ngTrackYTick\endcsname{1000}
\expandafter\newcommand\csname sr37n61ngTrackClosestY\endcsname{1425}
\expandafter\newcommand\csname sr38n456csRank\endcsname{38}
\expandafter\newcommand\csname sr38n456csLabel\endcsname{N456CS}
\expandafter\newcommand\csname sr38n456csDataset\endcsname{N456CS}
\expandafter\newcommand\csname sr38n456csGeometry\endcsname{far away/non-overhead}
\expandafter\newcommand\csname sr38n456csVCat\endcsname{V varying}
\expandafter\newcommand\csname sr38n456csHCat\endcsname{H constant}
\expandafter\newcommand\csname sr38n456csAudio\endcsname{smooth plateau-like + curve}
\expandafter\newcommand\csname sr38n456csClosest\endcsname{2026-05-17 21:55:00.18 UTC}
\expandafter\newcommand\csname sr38n456csXMin\endcsname{-52.2}
\expandafter\newcommand\csname sr38n456csXMax\endcsname{69.0}
\expandafter\newcommand\csname sr38n456csYMin\endcsname{200.3}
\expandafter\newcommand\csname sr38n456csYMax\endcsname{339.2}
\expandafter\newcommand\csname sr38n456csFzero\endcsname{252.7}
\expandafter\newcommand\csname sr38n456csXStartTick\endcsname{-50}
\expandafter\newcommand\csname sr38n456csXEndTick\endcsname{60}
\expandafter\newcommand\csname sr38n456csYLowTick\endcsname{209}
\expandafter\newcommand\csname sr38n456csYHighTick\endcsname{319}
\expandafter\newcommand\csname sr38n456csAdsV\endcsname{273.5}
\expandafter\newcommand\csname sr38n456csAdsH\endcsname{1392}
\expandafter\newcommand\csname sr38n456csMiiV\endcsname{250.5}
\expandafter\newcommand\csname sr38n456csMiiH\endcsname{--}
\expandafter\newcommand\csname sr38n456csMiV\endcsname{214.4}
\expandafter\newcommand\csname sr38n456csMiH\endcsname{1356}
\expandafter\newcommand\csname sr38n456csMiiVErr\endcsname{-8.4\%}
\expandafter\newcommand\csname sr38n456csMiiHErr\endcsname{--}
\expandafter\newcommand\csname sr38n456csMiVErr\endcsname{-21.6\%}
\expandafter\newcommand\csname sr38n456csMiHErr\endcsname{-2.6\%}
\expandafter\newcommand\csname sr38n456csRMSE\endcsname{4.19}
\expandafter\newcommand\csname sr38n456csVRange\endcsname{24.0}
\expandafter\newcommand\csname sr38n456csHRange\endcsname{27}
\expandafter\newcommand\csname sr38n456csLoverH\endcsname{1.50}
\expandafter\newcommand\csname sr38n456csTrackSource\endcsname{N456CS-track-EGM96\_avg.kml}
\expandafter\newcommand\csname sr38n456csAdsAvailable\endcsname{1}
\expandafter\newcommand\csname sr38n456csTrackAvailable\endcsname{1}
\expandafter\newcommand\csname sr38n456csInsetXMin\endcsname{-46.2}
\expandafter\newcommand\csname sr38n456csInsetXMax\endcsname{63.0}
\expandafter\newcommand\csname sr38n456csInsetXTickLo\endcsname{-20}
\expandafter\newcommand\csname sr38n456csInsetXTickHi\endcsname{40}
\expandafter\newcommand\csname sr38n456csInsetVMin\endcsname{227}
\expandafter\newcommand\csname sr38n456csInsetVMax\endcsname{312}
\expandafter\newcommand\csname sr38n456csInsetHMin\endcsname{1344}
\expandafter\newcommand\csname sr38n456csInsetHMax\endcsname{1427}
\expandafter\newcommand\csname sr38n456csTrackXMin\endcsname{-7474}
\expandafter\newcommand\csname sr38n456csTrackXMax\endcsname{8104}
\expandafter\newcommand\csname sr38n456csTrackYMin\endcsname{-238}
\expandafter\newcommand\csname sr38n456csTrackYMax\endcsname{3210}
\expandafter\newcommand\csname sr38n456csTrackXTickLo\endcsname{-5000}
\expandafter\newcommand\csname sr38n456csTrackXTickHi\endcsname{5000}
\expandafter\newcommand\csname sr38n456csTrackYTickLo\endcsname{0}
\expandafter\newcommand\csname sr38n456csTrackYTickHi\endcsname{3000}
\expandafter\newcommand\csname sr38n456csTrackYTick\endcsname{3000}
\expandafter\newcommand\csname sr38n456csTrackClosestY\endcsname{2091}
\expandafter\newcommand\csname sr39n292wsRank\endcsname{39}
\expandafter\newcommand\csname sr39n292wsLabel\endcsname{N292WS}
\expandafter\newcommand\csname sr39n292wsDataset\endcsname{N292WS}
\expandafter\newcommand\csname sr39n292wsGeometry\endcsname{far away/non-overhead}
\expandafter\newcommand\csname sr39n292wsVCat\endcsname{V constant}
\expandafter\newcommand\csname sr39n292wsHCat\endcsname{H constant}
\expandafter\newcommand\csname sr39n292wsAudio\endcsname{smooth curve only / incomplete ends}
\expandafter\newcommand\csname sr39n292wsClosest\endcsname{2026-05-16 18:34:22.80 UTC}
\expandafter\newcommand\csname sr39n292wsXMin\endcsname{-78.7}
\expandafter\newcommand\csname sr39n292wsXMax\endcsname{68.3}
\expandafter\newcommand\csname sr39n292wsYMin\endcsname{418.3}
\expandafter\newcommand\csname sr39n292wsYMax\endcsname{702.2}
\expandafter\newcommand\csname sr39n292wsFzero\endcsname{524.8}
\expandafter\newcommand\csname sr39n292wsXStartTick\endcsname{-70}
\expandafter\newcommand\csname sr39n292wsXEndTick\endcsname{60}
\expandafter\newcommand\csname sr39n292wsYLowTick\endcsname{436}
\expandafter\newcommand\csname sr39n292wsYHighTick\endcsname{662}
\expandafter\newcommand\csname sr39n292wsAdsV\endcsname{268.0}
\expandafter\newcommand\csname sr39n292wsAdsH\endcsname{1516}
\expandafter\newcommand\csname sr39n292wsMiiV\endcsname{247.2}
\expandafter\newcommand\csname sr39n292wsMiiH\endcsname{--}
\expandafter\newcommand\csname sr39n292wsMiV\endcsname{217.8}
\expandafter\newcommand\csname sr39n292wsMiH\endcsname{--}
\expandafter\newcommand\csname sr39n292wsMiiVErr\endcsname{-7.7\%}
\expandafter\newcommand\csname sr39n292wsMiiHErr\endcsname{--}
\expandafter\newcommand\csname sr39n292wsMiVErr\endcsname{-18.7\%}
\expandafter\newcommand\csname sr39n292wsMiHErr\endcsname{--}
\expandafter\newcommand\csname sr39n292wsRMSE\endcsname{8.81}
\expandafter\newcommand\csname sr39n292wsVRange\endcsname{4.5}
\expandafter\newcommand\csname sr39n292wsHRange\endcsname{3}
\expandafter\newcommand\csname sr39n292wsLoverH\endcsname{2.03}
\expandafter\newcommand\csname sr39n292wsTrackSource\endcsname{N292WS-track-EGM96\_avg.kml}
\expandafter\newcommand\csname sr39n292wsAdsAvailable\endcsname{1}
\expandafter\newcommand\csname sr39n292wsTrackAvailable\endcsname{1}
\expandafter\newcommand\csname sr39n292wsInsetXMin\endcsname{-71.4}
\expandafter\newcommand\csname sr39n292wsInsetXMax\endcsname{61.0}
\expandafter\newcommand\csname sr39n292wsInsetXTickLo\endcsname{-40}
\expandafter\newcommand\csname sr39n292wsInsetXTickHi\endcsname{40}
\expandafter\newcommand\csname sr39n292wsInsetVMin\endcsname{260}
\expandafter\newcommand\csname sr39n292wsInsetVMax\endcsname{279}
\expandafter\newcommand\csname sr39n292wsInsetHMin\endcsname{1489}
\expandafter\newcommand\csname sr39n292wsInsetHMax\endcsname{1536}
\expandafter\newcommand\csname sr39n292wsTrackXMin\endcsname{-8390}
\expandafter\newcommand\csname sr39n292wsTrackXMax\endcsname{8486}
\expandafter\newcommand\csname sr39n292wsTrackYMin\endcsname{-247}
\expandafter\newcommand\csname sr39n292wsTrackYMax\endcsname{3338}
\expandafter\newcommand\csname sr39n292wsTrackXTickLo\endcsname{-5000}
\expandafter\newcommand\csname sr39n292wsTrackXTickHi\endcsname{5000}
\expandafter\newcommand\csname sr39n292wsTrackYTickLo\endcsname{0}
\expandafter\newcommand\csname sr39n292wsTrackYTickHi\endcsname{3000}
\expandafter\newcommand\csname sr39n292wsTrackYTick\endcsname{3000}
\expandafter\newcommand\csname sr39n292wsTrackClosestY\endcsname{3079}
\expandafter\newcommand\csname sr40n3012nRank\endcsname{40}
\expandafter\newcommand\csname sr40n3012nLabel\endcsname{N3012N}
\expandafter\newcommand\csname sr40n3012nDataset\endcsname{N3012N}
\expandafter\newcommand\csname sr40n3012nGeometry\endcsname{far away/non-overhead}
\expandafter\newcommand\csname sr40n3012nVCat\endcsname{V varying}
\expandafter\newcommand\csname sr40n3012nHCat\endcsname{H varying}
\expandafter\newcommand\csname sr40n3012nAudio\endcsname{smooth curve only / incomplete ends}
\expandafter\newcommand\csname sr40n3012nClosest\endcsname{2026-05-16 16:38:50.30 UTC}
\expandafter\newcommand\csname sr40n3012nXMin\endcsname{-40.1}
\expandafter\newcommand\csname sr40n3012nXMax\endcsname{37.4}
\expandafter\newcommand\csname sr40n3012nYMin\endcsname{291.4}
\expandafter\newcommand\csname sr40n3012nYMax\endcsname{430.2}
\expandafter\newcommand\csname sr40n3012nFzero\endcsname{345.4}
\expandafter\newcommand\csname sr40n3012nXStartTick\endcsname{-40}
\expandafter\newcommand\csname sr40n3012nXEndTick\endcsname{30}
\expandafter\newcommand\csname sr40n3012nYLowTick\endcsname{300}
\expandafter\newcommand\csname sr40n3012nYHighTick\endcsname{410}
\expandafter\newcommand\csname sr40n3012nAdsV\endcsname{215.0}
\expandafter\newcommand\csname sr40n3012nAdsH\endcsname{833}
\expandafter\newcommand\csname sr40n3012nMiiV\endcsname{204.8}
\expandafter\newcommand\csname sr40n3012nMiiH\endcsname{--}
\expandafter\newcommand\csname sr40n3012nMiV\endcsname{178.3}
\expandafter\newcommand\csname sr40n3012nMiH\endcsname{--}
\expandafter\newcommand\csname sr40n3012nMiiVErr\endcsname{-4.8\%}
\expandafter\newcommand\csname sr40n3012nMiiHErr\endcsname{--}
\expandafter\newcommand\csname sr40n3012nMiVErr\endcsname{-17.1\%}
\expandafter\newcommand\csname sr40n3012nMiHErr\endcsname{--}
\expandafter\newcommand\csname sr40n3012nRMSE\endcsname{2.21}
\expandafter\newcommand\csname sr40n3012nVRange\endcsname{16.1}
\expandafter\newcommand\csname sr40n3012nHRange\endcsname{233}
\expandafter\newcommand\csname sr40n3012nLoverH\endcsname{2.38}
\expandafter\newcommand\csname sr40n3012nTrackSource\endcsname{N3012N-track-EGM96\_avg.kml}
\expandafter\newcommand\csname sr40n3012nAdsAvailable\endcsname{1}
\expandafter\newcommand\csname sr40n3012nTrackAvailable\endcsname{1}
\expandafter\newcommand\csname sr40n3012nInsetXMin\endcsname{-36.2}
\expandafter\newcommand\csname sr40n3012nInsetXMax\endcsname{33.6}
\expandafter\newcommand\csname sr40n3012nInsetXTickLo\endcsname{-20}
\expandafter\newcommand\csname sr40n3012nInsetXTickHi\endcsname{20}
\expandafter\newcommand\csname sr40n3012nInsetVMin\endcsname{195}
\expandafter\newcommand\csname sr40n3012nInsetVMax\endcsname{261}
\expandafter\newcommand\csname sr40n3012nInsetHMin\endcsname{563}
\expandafter\newcommand\csname sr40n3012nInsetHMax\endcsname{997}
\expandafter\newcommand\csname sr40n3012nTrackXMin\endcsname{-4212}
\expandafter\newcommand\csname sr40n3012nTrackXMax\endcsname{3831}
\expandafter\newcommand\csname sr40n3012nTrackYMin\endcsname{-271}
\expandafter\newcommand\csname sr40n3012nTrackYMax\endcsname{3658}
\expandafter\newcommand\csname sr40n3012nTrackXTickLo\endcsname{-4000}
\expandafter\newcommand\csname sr40n3012nTrackXTickHi\endcsname{2000}
\expandafter\newcommand\csname sr40n3012nTrackYTickLo\endcsname{0}
\expandafter\newcommand\csname sr40n3012nTrackYTickHi\endcsname{3000}
\expandafter\newcommand\csname sr40n3012nTrackYTick\endcsname{3000}
\expandafter\newcommand\csname sr40n3012nTrackClosestY\endcsname{1975}
\expandafter\newcommand\csname sr41n757ahRank\endcsname{41}
\expandafter\newcommand\csname sr41n757ahLabel\endcsname{N757AH}
\expandafter\newcommand\csname sr41n757ahDataset\endcsname{N757AH 12 May}
\expandafter\newcommand\csname sr41n757ahGeometry\endcsname{near overhead}
\expandafter\newcommand\csname sr41n757ahVCat\endcsname{V constant}
\expandafter\newcommand\csname sr41n757ahHCat\endcsname{H constant}
\expandafter\newcommand\csname sr41n757ahAudio\endcsname{smooth curve only / incomplete ends}
\expandafter\newcommand\csname sr41n757ahClosest\endcsname{2026-05-12 20:02:17.80 UTC}
\expandafter\newcommand\csname sr41n757ahXMin\endcsname{-18.5}
\expandafter\newcommand\csname sr41n757ahXMax\endcsname{12.8}
\expandafter\newcommand\csname sr41n757ahYMin\endcsname{809.2}
\expandafter\newcommand\csname sr41n757ahYMax\endcsname{1030.5}
\expandafter\newcommand\csname sr41n757ahFzero\endcsname{895.9}
\expandafter\newcommand\csname sr41n757ahXStartTick\endcsname{-20}
\expandafter\newcommand\csname sr41n757ahXEndTick\endcsname{10}
\expandafter\newcommand\csname sr41n757ahYLowTick\endcsname{823}
\expandafter\newcommand\csname sr41n757ahYHighTick\endcsname{999}
\expandafter\newcommand\csname sr41n757ahAdsV\endcsname{186.5}
\expandafter\newcommand\csname sr41n757ahAdsH\endcsname{579}
\expandafter\newcommand\csname sr41n757ahMiiV\endcsname{133.7}
\expandafter\newcommand\csname sr41n757ahMiiH\endcsname{268}
\expandafter\newcommand\csname sr41n757ahMiV\endcsname{111.3}
\expandafter\newcommand\csname sr41n757ahMiH\endcsname{201}
\expandafter\newcommand\csname sr41n757ahMiiVErr\endcsname{-28.3\%}
\expandafter\newcommand\csname sr41n757ahMiiHErr\endcsname{-53.7\%}
\expandafter\newcommand\csname sr41n757ahMiVErr\endcsname{-40.3\%}
\expandafter\newcommand\csname sr41n757ahMiHErr\endcsname{-65.2\%}
\expandafter\newcommand\csname sr41n757ahRMSE\endcsname{3.49}
\expandafter\newcommand\csname sr41n757ahVRange\endcsname{6.0}
\expandafter\newcommand\csname sr41n757ahHRange\endcsname{24}
\expandafter\newcommand\csname sr41n757ahLoverH\endcsname{0.24}
\expandafter\newcommand\csname sr41n757ahTrackSource\endcsname{N757AH-track-EGM96\_avg.kml}
\expandafter\newcommand\csname sr41n757ahAdsAvailable\endcsname{1}
\expandafter\newcommand\csname sr41n757ahTrackAvailable\endcsname{1}
\expandafter\newcommand\csname sr41n757ahInsetXMin\endcsname{-16.5}
\expandafter\newcommand\csname sr41n757ahInsetXMax\endcsname{10.8}
\expandafter\newcommand\csname sr41n757ahInsetXTickLo\endcsname{-10}
\expandafter\newcommand\csname sr41n757ahInsetXTickHi\endcsname{5}
\expandafter\newcommand\csname sr41n757ahInsetVMin\endcsname{169}
\expandafter\newcommand\csname sr41n757ahInsetVMax\endcsname{193}
\expandafter\newcommand\csname sr41n757ahInsetHMin\endcsname{537}
\expandafter\newcommand\csname sr41n757ahInsetHMax\endcsname{611}
\expandafter\newcommand\csname sr41n757ahTrackXMin\endcsname{-3180}
\expandafter\newcommand\csname sr41n757ahTrackXMax\endcsname{2905}
\expandafter\newcommand\csname sr41n757ahTrackYMin\endcsname{-115}
\expandafter\newcommand\csname sr41n757ahTrackYMax\endcsname{180}
\expandafter\newcommand\csname sr41n757ahTrackXTickLo\endcsname{-2000}
\expandafter\newcommand\csname sr41n757ahTrackXTickHi\endcsname{2000}
\expandafter\newcommand\csname sr41n757ahTrackYTickLo\endcsname{-50}
\expandafter\newcommand\csname sr41n757ahTrackYTickHi\endcsname{150}
\expandafter\newcommand\csname sr41n757ahTrackYTick\endcsname{150}
\expandafter\newcommand\csname sr41n757ahTrackClosestY\endcsname{140}
\expandafter\newcommand\csname sr42n757ahRank\endcsname{42}
\expandafter\newcommand\csname sr42n757ahLabel\endcsname{N757AH}
\expandafter\newcommand\csname sr42n757ahDataset\endcsname{N757AH 16 May}
\expandafter\newcommand\csname sr42n757ahGeometry\endcsname{far away/non-overhead}
\expandafter\newcommand\csname sr42n757ahVCat\endcsname{V constant}
\expandafter\newcommand\csname sr42n757ahHCat\endcsname{H varying}
\expandafter\newcommand\csname sr42n757ahAudio\endcsname{curve only + scattered/ambiguous}
\expandafter\newcommand\csname sr42n757ahClosest\endcsname{2026-05-16 21:33:46.50 UTC}
\expandafter\newcommand\csname sr42n757ahXMin\endcsname{-40.9}
\expandafter\newcommand\csname sr42n757ahXMax\endcsname{31.5}
\expandafter\newcommand\csname sr42n757ahYMin\endcsname{374.9}
\expandafter\newcommand\csname sr42n757ahYMax\endcsname{524.5}
\expandafter\newcommand\csname sr42n757ahFzero\endcsname{435.3}
\expandafter\newcommand\csname sr42n757ahXStartTick\endcsname{-40}
\expandafter\newcommand\csname sr42n757ahXEndTick\endcsname{30}
\expandafter\newcommand\csname sr42n757ahYLowTick\endcsname{384}
\expandafter\newcommand\csname sr42n757ahYHighTick\endcsname{503}
\expandafter\newcommand\csname sr42n757ahAdsV\endcsname{174.0}
\expandafter\newcommand\csname sr42n757ahAdsH\endcsname{832}
\expandafter\newcommand\csname sr42n757ahMiiV\endcsname{157.1}
\expandafter\newcommand\csname sr42n757ahMiiH\endcsname{--}
\expandafter\newcommand\csname sr42n757ahMiV\endcsname{135.7}
\expandafter\newcommand\csname sr42n757ahMiH\endcsname{--}
\expandafter\newcommand\csname sr42n757ahMiiVErr\endcsname{-9.7\%}
\expandafter\newcommand\csname sr42n757ahMiiHErr\endcsname{--}
\expandafter\newcommand\csname sr42n757ahMiVErr\endcsname{-22.0\%}
\expandafter\newcommand\csname sr42n757ahMiHErr\endcsname{--}
\expandafter\newcommand\csname sr42n757ahRMSE\endcsname{2.45}
\expandafter\newcommand\csname sr42n757ahVRange\endcsname{13.7}
\expandafter\newcommand\csname sr42n757ahHRange\endcsname{52}
\expandafter\newcommand\csname sr42n757ahLoverH\endcsname{1.40}
\expandafter\newcommand\csname sr42n757ahTrackSource\endcsname{N757AH-track-EGM96\_avg.kml}
\expandafter\newcommand\csname sr42n757ahAdsAvailable\endcsname{1}
\expandafter\newcommand\csname sr42n757ahTrackAvailable\endcsname{1}
\expandafter\newcommand\csname sr42n757ahInsetXMin\endcsname{-37.3}
\expandafter\newcommand\csname sr42n757ahInsetXMax\endcsname{27.9}
\expandafter\newcommand\csname sr42n757ahInsetXTickLo\endcsname{-37}
\expandafter\newcommand\csname sr42n757ahInsetXTickHi\endcsname{20}
\expandafter\newcommand\csname sr42n757ahInsetVMin\endcsname{58}
\expandafter\newcommand\csname sr42n757ahInsetVMax\endcsname{292}
\expandafter\newcommand\csname sr42n757ahInsetHMin\endcsname{-29}
\expandafter\newcommand\csname sr42n757ahInsetHMax\endcsname{1244}
\expandafter\newcommand\csname sr42n757ahTrackXMin\endcsname{-16093}
\expandafter\newcommand\csname sr42n757ahTrackXMax\endcsname{135188}
\expandafter\newcommand\csname sr42n757ahTrackYMin\endcsname{-77847}
\expandafter\newcommand\csname sr42n757ahTrackYMax\endcsname{39083}
\expandafter\newcommand\csname sr42n757ahTrackXTickLo\endcsname{0}
\expandafter\newcommand\csname sr42n757ahTrackXTickHi\endcsname{100000}
\expandafter\newcommand\csname sr42n757ahTrackYTickLo\endcsname{-60000}
\expandafter\newcommand\csname sr42n757ahTrackYTickHi\endcsname{20000}
\expandafter\newcommand\csname sr42n757ahTrackYTick\endcsname{20000}
\expandafter\newcommand\csname sr42n757ahTrackClosestY\endcsname{140}
\expandafter\newcommand\csname sr43n906mdRank\endcsname{43}
\expandafter\newcommand\csname sr43n906mdLabel\endcsname{N906MD}
\expandafter\newcommand\csname sr43n906mdDataset\endcsname{N906MD}
\expandafter\newcommand\csname sr43n906mdGeometry\endcsname{far away/non-overhead}
\expandafter\newcommand\csname sr43n906mdVCat\endcsname{V varying}
\expandafter\newcommand\csname sr43n906mdHCat\endcsname{H varying}
\expandafter\newcommand\csname sr43n906mdAudio\endcsname{smooth plateau-like + curve}
\expandafter\newcommand\csname sr43n906mdClosest\endcsname{2026-05-09 22:39:00 UTC}
\expandafter\newcommand\csname sr43n906mdXMin\endcsname{-17.3}
\expandafter\newcommand\csname sr43n906mdXMax\endcsname{15.6}
\expandafter\newcommand\csname sr43n906mdYMin\endcsname{755.6}
\expandafter\newcommand\csname sr43n906mdYMax\endcsname{1184.3}
\expandafter\newcommand\csname sr43n906mdFzero\endcsname{916.9}
\expandafter\newcommand\csname sr43n906mdXStartTick\endcsname{-20}
\expandafter\newcommand\csname sr43n906mdXEndTick\endcsname{10}
\expandafter\newcommand\csname sr43n906mdYLowTick\endcsname{783}
\expandafter\newcommand\csname sr43n906mdYHighTick\endcsname{1123}
\expandafter\newcommand\csname sr43n906mdAdsV\endcsname{152.0}
\expandafter\newcommand\csname sr43n906mdAdsH\endcsname{252}
\expandafter\newcommand\csname sr43n906mdMiiV\endcsname{199.7}
\expandafter\newcommand\csname sr43n906mdMiiH\endcsname{--}
\expandafter\newcommand\csname sr43n906mdMiV\endcsname{186.3}
\expandafter\newcommand\csname sr43n906mdMiH\endcsname{--}
\expandafter\newcommand\csname sr43n906mdMiiVErr\endcsname{31.4\%}
\expandafter\newcommand\csname sr43n906mdMiiHErr\endcsname{--}
\expandafter\newcommand\csname sr43n906mdMiVErr\endcsname{22.5\%}
\expandafter\newcommand\csname sr43n906mdMiHErr\endcsname{--}
\expandafter\newcommand\csname sr43n906mdRMSE\endcsname{5.65}
\expandafter\newcommand\csname sr43n906mdVRange\endcsname{46.6}
\expandafter\newcommand\csname sr43n906mdHRange\endcsname{264}
\expandafter\newcommand\csname sr43n906mdLoverH\endcsname{2.27}
\expandafter\newcommand\csname sr43n906mdTrackSource\endcsname{N906MD-track-EGM96\_avg.kml}
\expandafter\newcommand\csname sr43n906mdAdsAvailable\endcsname{1}
\expandafter\newcommand\csname sr43n906mdTrackAvailable\endcsname{1}
\expandafter\newcommand\csname sr43n906mdInsetXMin\endcsname{-15.3}
\expandafter\newcommand\csname sr43n906mdInsetXMax\endcsname{13.6}
\expandafter\newcommand\csname sr43n906mdInsetXTickLo\endcsname{-10}
\expandafter\newcommand\csname sr43n906mdInsetXTickHi\endcsname{10}
\expandafter\newcommand\csname sr43n906mdInsetVMin\endcsname{97}
\expandafter\newcommand\csname sr43n906mdInsetVMax\endcsname{121}
\expandafter\newcommand\csname sr43n906mdInsetHMin\endcsname{178}
\expandafter\newcommand\csname sr43n906mdInsetHMax\endcsname{288}
\expandafter\newcommand\csname sr43n906mdTrackXMin\endcsname{-649}
\expandafter\newcommand\csname sr43n906mdTrackXMax\endcsname{1181}
\expandafter\newcommand\csname sr43n906mdTrackYMin\endcsname{-109}
\expandafter\newcommand\csname sr43n906mdTrackYMax\endcsname{1477}
\expandafter\newcommand\csname sr43n906mdTrackXTickLo\endcsname{-500}
\expandafter\newcommand\csname sr43n906mdTrackXTickHi\endcsname{1000}
\expandafter\newcommand\csname sr43n906mdTrackYTickLo\endcsname{0}
\expandafter\newcommand\csname sr43n906mdTrackYTickHi\endcsname{1477}
\expandafter\newcommand\csname sr43n906mdTrackYTick\endcsname{1477}
\expandafter\newcommand\csname sr43n906mdTrackClosestY\endcsname{577}
\expandafter\newcommand\csname sr44n1387uRank\endcsname{44}
\expandafter\newcommand\csname sr44n1387uLabel\endcsname{N1387U}
\expandafter\newcommand\csname sr44n1387uDataset\endcsname{N1387U}
\expandafter\newcommand\csname sr44n1387uGeometry\endcsname{near overhead}
\expandafter\newcommand\csname sr44n1387uVCat\endcsname{V varying}
\expandafter\newcommand\csname sr44n1387uHCat\endcsname{H varying}
\expandafter\newcommand\csname sr44n1387uAudio\endcsname{smooth curve only / incomplete ends}
\expandafter\newcommand\csname sr44n1387uClosest\endcsname{}
\expandafter\newcommand\csname sr44n1387uXMin\endcsname{-21.8}
\expandafter\newcommand\csname sr44n1387uXMax\endcsname{18.5}
\expandafter\newcommand\csname sr44n1387uYMin\endcsname{508.0}
\expandafter\newcommand\csname sr44n1387uYMax\endcsname{654.1}
\expandafter\newcommand\csname sr44n1387uFzero\endcsname{565.9}
\expandafter\newcommand\csname sr44n1387uXStartTick\endcsname{-20}
\expandafter\newcommand\csname sr44n1387uXEndTick\endcsname{20}
\expandafter\newcommand\csname sr44n1387uYLowTick\endcsname{517}
\expandafter\newcommand\csname sr44n1387uYHighTick\endcsname{633}
\expandafter\newcommand\csname sr44n1387uAdsV\endcsname{137.9}
\expandafter\newcommand\csname sr44n1387uAdsH\endcsname{252}
\expandafter\newcommand\csname sr44n1387uMiiV\endcsname{111.5}
\expandafter\newcommand\csname sr44n1387uMiiH\endcsname{--}
\expandafter\newcommand\csname sr44n1387uMiV\endcsname{91.5}
\expandafter\newcommand\csname sr44n1387uMiH\endcsname{--}
\expandafter\newcommand\csname sr44n1387uMiiVErr\endcsname{-19.2\%}
\expandafter\newcommand\csname sr44n1387uMiiHErr\endcsname{--}
\expandafter\newcommand\csname sr44n1387uMiVErr\endcsname{-33.7\%}
\expandafter\newcommand\csname sr44n1387uMiHErr\endcsname{--}
\expandafter\newcommand\csname sr44n1387uRMSE\endcsname{5.78}
\expandafter\newcommand\csname sr44n1387uVRange\endcsname{82.5}
\expandafter\newcommand\csname sr44n1387uHRange\endcsname{366}
\expandafter\newcommand\csname sr44n1387uLoverH\endcsname{1.13}
\expandafter\newcommand\csname sr44n1387uTrackSource\endcsname{N1387U-track-EGM96\_avg.kml}
\expandafter\newcommand\csname sr44n1387uAdsAvailable\endcsname{1}
\expandafter\newcommand\csname sr44n1387uTrackAvailable\endcsname{1}
\expandafter\newcommand\csname sr44n1387uInsetXMin\endcsname{-19.8}
\expandafter\newcommand\csname sr44n1387uInsetXMax\endcsname{16.5}
\expandafter\newcommand\csname sr44n1387uInsetXTickLo\endcsname{-20}
\expandafter\newcommand\csname sr44n1387uInsetXTickHi\endcsname{10}
\expandafter\newcommand\csname sr44n1387uInsetVMin\endcsname{132}
\expandafter\newcommand\csname sr44n1387uInsetVMax\endcsname{143}
\expandafter\newcommand\csname sr44n1387uInsetHMin\endcsname{157}
\expandafter\newcommand\csname sr44n1387uInsetHMax\endcsname{316}
\expandafter\newcommand\csname sr44n1387uTrackXMin\endcsname{-2343}
\expandafter\newcommand\csname sr44n1387uTrackXMax\endcsname{1289}
\expandafter\newcommand\csname sr44n1387uTrackYMin\endcsname{-117}
\expandafter\newcommand\csname sr44n1387uTrackYMax\endcsname{1582}
\expandafter\newcommand\csname sr44n1387uTrackXTickLo\endcsname{-2000}
\expandafter\newcommand\csname sr44n1387uTrackXTickHi\endcsname{1000}
\expandafter\newcommand\csname sr44n1387uTrackYTickLo\endcsname{0}
\expandafter\newcommand\csname sr44n1387uTrackYTickHi\endcsname{1500}
\expandafter\newcommand\csname sr44n1387uTrackYTick\endcsname{1500}
\expandafter\newcommand\csname sr44n1387uTrackClosestY\endcsname{299}
\expandafter\newcommand\csname sr45n708pdRank\endcsname{45}
\expandafter\newcommand\csname sr45n708pdLabel\endcsname{N708PD}
\expandafter\newcommand\csname sr45n708pdDataset\endcsname{N708PD}
\expandafter\newcommand\csname sr45n708pdGeometry\endcsname{near overhead}
\expandafter\newcommand\csname sr45n708pdVCat\endcsname{V varying}
\expandafter\newcommand\csname sr45n708pdHCat\endcsname{H varying}
\expandafter\newcommand\csname sr45n708pdAudio\endcsname{plateau-like + curve but scattered/ambiguous}
\expandafter\newcommand\csname sr45n708pdClosest\endcsname{}
\expandafter\newcommand\csname sr45n708pdXMin\endcsname{-16.7}
\expandafter\newcommand\csname sr45n708pdXMax\endcsname{19.2}
\expandafter\newcommand\csname sr45n708pdYMin\endcsname{640.3}
\expandafter\newcommand\csname sr45n708pdYMax\endcsname{1039.9}
\expandafter\newcommand\csname sr45n708pdFzero\endcsname{801.8}
\expandafter\newcommand\csname sr45n708pdXStartTick\endcsname{-10}
\expandafter\newcommand\csname sr45n708pdXEndTick\endcsname{20}
\expandafter\newcommand\csname sr45n708pdYLowTick\endcsname{666}
\expandafter\newcommand\csname sr45n708pdYHighTick\endcsname{983}
\expandafter\newcommand\csname sr45n708pdAdsV\endcsname{141.0}
\expandafter\newcommand\csname sr45n708pdAdsH\endcsname{172}
\expandafter\newcommand\csname sr45n708pdMiiV\endcsname{254.0}
\expandafter\newcommand\csname sr45n708pdMiiH\endcsname{648}
\expandafter\newcommand\csname sr45n708pdMiV\endcsname{197.9}
\expandafter\newcommand\csname sr45n708pdMiH\endcsname{512}
\expandafter\newcommand\csname sr45n708pdMiiVErr\endcsname{80.1\%}
\expandafter\newcommand\csname sr45n708pdMiiHErr\endcsname{275.8\%}
\expandafter\newcommand\csname sr45n708pdMiVErr\endcsname{40.3\%}
\expandafter\newcommand\csname sr45n708pdMiHErr\endcsname{196.8\%}
\expandafter\newcommand\csname sr45n708pdRMSE\endcsname{8.70}
\expandafter\newcommand\csname sr45n708pdVRange\endcsname{28.7}
\expandafter\newcommand\csname sr45n708pdHRange\endcsname{56}
\expandafter\newcommand\csname sr45n708pdLoverH\endcsname{2.47}
\expandafter\newcommand\csname sr45n708pdTrackSource\endcsname{N708PD-track-EGM96\_avg.kml}
\expandafter\newcommand\csname sr45n708pdAdsAvailable\endcsname{1}
\expandafter\newcommand\csname sr45n708pdTrackAvailable\endcsname{1}
\expandafter\newcommand\csname sr45n708pdInsetXMin\endcsname{-14.7}
\expandafter\newcommand\csname sr45n708pdInsetXMax\endcsname{17.2}
\expandafter\newcommand\csname sr45n708pdInsetXTickLo\endcsname{-10}
\expandafter\newcommand\csname sr45n708pdInsetXTickHi\endcsname{10}
\expandafter\newcommand\csname sr45n708pdInsetVMin\endcsname{136}
\expandafter\newcommand\csname sr45n708pdInsetVMax\endcsname{167}
\expandafter\newcommand\csname sr45n708pdInsetHMin\endcsname{149}
\expandafter\newcommand\csname sr45n708pdInsetHMax\endcsname{194}
\expandafter\newcommand\csname sr45n708pdTrackXMin\endcsname{-2318}
\expandafter\newcommand\csname sr45n708pdTrackXMax\endcsname{2424}
\expandafter\newcommand\csname sr45n708pdTrackYMin\endcsname{-51}
\expandafter\newcommand\csname sr45n708pdTrackYMax\endcsname{695}
\expandafter\newcommand\csname sr45n708pdTrackXTickLo\endcsname{-2000}
\expandafter\newcommand\csname sr45n708pdTrackXTickHi\endcsname{2000}
\expandafter\newcommand\csname sr45n708pdTrackYTickLo\endcsname{0}
\expandafter\newcommand\csname sr45n708pdTrackYTickHi\endcsname{600}
\expandafter\newcommand\csname sr45n708pdTrackYTick\endcsname{600}
\expandafter\newcommand\csname sr45n708pdTrackClosestY\endcsname{619}
\expandafter\newcommand\csname sr46n4902fRank\endcsname{46}
\expandafter\newcommand\csname sr46n4902fLabel\endcsname{N4902F}
\expandafter\newcommand\csname sr46n4902fDataset\endcsname{C172 N4902F pass 1}
\expandafter\newcommand\csname sr46n4902fGeometry\endcsname{far away/non-overhead}
\expandafter\newcommand\csname sr46n4902fVCat\endcsname{V varying}
\expandafter\newcommand\csname sr46n4902fHCat\endcsname{H constant}
\expandafter\newcommand\csname sr46n4902fAudio\endcsname{smooth plateau-like + curve}
\expandafter\newcommand\csname sr46n4902fClosest\endcsname{2026-05-14 17:26:11 UTC}
\expandafter\newcommand\csname sr46n4902fXMin\endcsname{-52.7}
\expandafter\newcommand\csname sr46n4902fXMax\endcsname{41.9}
\expandafter\newcommand\csname sr46n4902fYMin\endcsname{199.5}
\expandafter\newcommand\csname sr46n4902fYMax\endcsname{306.5}
\expandafter\newcommand\csname sr46n4902fFzero\endcsname{240.0}
\expandafter\newcommand\csname sr46n4902fXStartTick\endcsname{-50}
\expandafter\newcommand\csname sr46n4902fXEndTick\endcsname{40}
\expandafter\newcommand\csname sr46n4902fYLowTick\endcsname{206}
\expandafter\newcommand\csname sr46n4902fYHighTick\endcsname{290}
\expandafter\newcommand\csname sr46n4902fAdsV\endcsname{160.0}
\expandafter\newcommand\csname sr46n4902fAdsH\endcsname{1048}
\expandafter\newcommand\csname sr46n4902fMiiV\endcsname{223.1}
\expandafter\newcommand\csname sr46n4902fMiiH\endcsname{--}
\expandafter\newcommand\csname sr46n4902fMiV\endcsname{196.8}
\expandafter\newcommand\csname sr46n4902fMiH\endcsname{--}
\expandafter\newcommand\csname sr46n4902fMiiVErr\endcsname{39.4\%}
\expandafter\newcommand\csname sr46n4902fMiiHErr\endcsname{--}
\expandafter\newcommand\csname sr46n4902fMiVErr\endcsname{23.0\%}
\expandafter\newcommand\csname sr46n4902fMiHErr\endcsname{--}
\expandafter\newcommand\csname sr46n4902fRMSE\endcsname{1.61}
\expandafter\newcommand\csname sr46n4902fVRange\endcsname{93.9}
\expandafter\newcommand\csname sr46n4902fHRange\endcsname{33}
\expandafter\newcommand\csname sr46n4902fLoverH\endcsname{3.48}
\expandafter\newcommand\csname sr46n4902fTrackSource\endcsname{N4902F-track-EGM96\_avg.kml}
\expandafter\newcommand\csname sr46n4902fAdsAvailable\endcsname{1}
\expandafter\newcommand\csname sr46n4902fTrackAvailable\endcsname{1}
\expandafter\newcommand\csname sr46n4902fInsetXMin\endcsname{-48.0}
\expandafter\newcommand\csname sr46n4902fInsetXMax\endcsname{37.2}
\expandafter\newcommand\csname sr46n4902fInsetXTickLo\endcsname{-40}
\expandafter\newcommand\csname sr46n4902fInsetXTickHi\endcsname{20}
\expandafter\newcommand\csname sr46n4902fInsetVMin\endcsname{137}
\expandafter\newcommand\csname sr46n4902fInsetVMax\endcsname{173}
\expandafter\newcommand\csname sr46n4902fInsetHMin\endcsname{1013}
\expandafter\newcommand\csname sr46n4902fInsetHMax\endcsname{1104}
\expandafter\newcommand\csname sr46n4902fTrackXMin\endcsname{-2676}
\expandafter\newcommand\csname sr46n4902fTrackXMax\endcsname{5245}
\expandafter\newcommand\csname sr46n4902fTrackYMin\endcsname{-295}
\expandafter\newcommand\csname sr46n4902fTrackYMax\endcsname{3985}
\expandafter\newcommand\csname sr46n4902fTrackXTickLo\endcsname{-2000}
\expandafter\newcommand\csname sr46n4902fTrackXTickHi\endcsname{4000}
\expandafter\newcommand\csname sr46n4902fTrackYTickLo\endcsname{0}
\expandafter\newcommand\csname sr46n4902fTrackYTickHi\endcsname{3000}
\expandafter\newcommand\csname sr46n4902fTrackYTick\endcsname{3000}
\expandafter\newcommand\csname sr46n4902fTrackClosestY\endcsname{3644}
\expandafter\newcommand\csname sr47n4902fRank\endcsname{47}
\expandafter\newcommand\csname sr47n4902fLabel\endcsname{N4902F}
\expandafter\newcommand\csname sr47n4902fDataset\endcsname{C172 N4902F pass 2}
\expandafter\newcommand\csname sr47n4902fGeometry\endcsname{near overhead}
\expandafter\newcommand\csname sr47n4902fVCat\endcsname{V varying}
\expandafter\newcommand\csname sr47n4902fHCat\endcsname{H constant}
\expandafter\newcommand\csname sr47n4902fAudio\endcsname{plateau-like + curve but scattered/ambiguous}
\expandafter\newcommand\csname sr47n4902fClosest\endcsname{2026-05-14 17:33:55 UTC}
\expandafter\newcommand\csname sr47n4902fXMin\endcsname{-61.7}
\expandafter\newcommand\csname sr47n4902fXMax\endcsname{74.0}
\expandafter\newcommand\csname sr47n4902fYMin\endcsname{185.5}
\expandafter\newcommand\csname sr47n4902fYMax\endcsname{353.0}
\expandafter\newcommand\csname sr47n4902fFzero\endcsname{246.0}
\expandafter\newcommand\csname sr47n4902fXStartTick\endcsname{-50}
\expandafter\newcommand\csname sr47n4902fXEndTick\endcsname{70}
\expandafter\newcommand\csname sr47n4902fYLowTick\endcsname{196}
\expandafter\newcommand\csname sr47n4902fYHighTick\endcsname{329}
\expandafter\newcommand\csname sr47n4902fAdsV\endcsname{117.5}
\expandafter\newcommand\csname sr47n4902fAdsH\endcsname{897}
\expandafter\newcommand\csname sr47n4902fMiiV\endcsname{324.0}
\expandafter\newcommand\csname sr47n4902fMiiH\endcsname{2146}
\expandafter\newcommand\csname sr47n4902fMiV\endcsname{298.1}
\expandafter\newcommand\csname sr47n4902fMiH\endcsname{1836}
\expandafter\newcommand\csname sr47n4902fMiiVErr\endcsname{175.8\%}
\expandafter\newcommand\csname sr47n4902fMiiHErr\endcsname{139.2\%}
\expandafter\newcommand\csname sr47n4902fMiVErr\endcsname{153.7\%}
\expandafter\newcommand\csname sr47n4902fMiHErr\endcsname{104.7\%}
\expandafter\newcommand\csname sr47n4902fRMSE\endcsname{0.65}
\expandafter\newcommand\csname sr47n4902fVRange\endcsname{94.1}
\expandafter\newcommand\csname sr47n4902fHRange\endcsname{37}
\expandafter\newcommand\csname sr47n4902fLoverH\endcsname{0.39}
\expandafter\newcommand\csname sr47n4902fTrackSource\endcsname{N4902F-track-EGM96\_avg.kml}
\expandafter\newcommand\csname sr47n4902fAdsAvailable\endcsname{1}
\expandafter\newcommand\csname sr47n4902fTrackAvailable\endcsname{1}
\expandafter\newcommand\csname sr47n4902fInsetXMin\endcsname{-54.9}
\expandafter\newcommand\csname sr47n4902fInsetXMax\endcsname{67.3}
\expandafter\newcommand\csname sr47n4902fInsetXTickLo\endcsname{-40}
\expandafter\newcommand\csname sr47n4902fInsetXTickHi\endcsname{40}
\expandafter\newcommand\csname sr47n4902fInsetVMin\endcsname{97}
\expandafter\newcommand\csname sr47n4902fInsetVMax\endcsname{159}
\expandafter\newcommand\csname sr47n4902fInsetHMin\endcsname{846}
\expandafter\newcommand\csname sr47n4902fInsetHMax\endcsname{943}
\expandafter\newcommand\csname sr47n4902fTrackXMin\endcsname{-3149}
\expandafter\newcommand\csname sr47n4902fTrackXMax\endcsname{3810}
\expandafter\newcommand\csname sr47n4902fTrackYMin\endcsname{-1087}
\expandafter\newcommand\csname sr47n4902fTrackYMax\endcsname{447}
\expandafter\newcommand\csname sr47n4902fTrackXTickLo\endcsname{-2000}
\expandafter\newcommand\csname sr47n4902fTrackXTickHi\endcsname{3810}
\expandafter\newcommand\csname sr47n4902fTrackYTickLo\endcsname{-1000}
\expandafter\newcommand\csname sr47n4902fTrackYTickHi\endcsname{447}
\expandafter\newcommand\csname sr47n4902fTrackYTick\endcsname{447}
\expandafter\newcommand\csname sr47n4902fTrackClosestY\endcsname{341}
\expandafter\newcommand\csname sr48n662ezRank\endcsname{48}
\expandafter\newcommand\csname sr48n662ezLabel\endcsname{N662EZ}
\expandafter\newcommand\csname sr48n662ezDataset\endcsname{N662EZ}
\expandafter\newcommand\csname sr48n662ezGeometry\endcsname{far away/non-overhead}
\expandafter\newcommand\csname sr48n662ezVCat\endcsname{V varying}
\expandafter\newcommand\csname sr48n662ezHCat\endcsname{H varying}
\expandafter\newcommand\csname sr48n662ezAudio\endcsname{smooth plateau-like + curve}
\expandafter\newcommand\csname sr48n662ezClosest\endcsname{2026-05-13 23:31:55.80 UTC}
\expandafter\newcommand\csname sr48n662ezXMin\endcsname{-58.5}
\expandafter\newcommand\csname sr48n662ezXMax\endcsname{47.5}
\expandafter\newcommand\csname sr48n662ezYMin\endcsname{193.1}
\expandafter\newcommand\csname sr48n662ezYMax\endcsname{328.6}
\expandafter\newcommand\csname sr48n662ezFzero\endcsname{242.7}
\expandafter\newcommand\csname sr48n662ezXStartTick\endcsname{-50}
\expandafter\newcommand\csname sr48n662ezXEndTick\endcsname{40}
\expandafter\newcommand\csname sr48n662ezYLowTick\endcsname{202}
\expandafter\newcommand\csname sr48n662ezYHighTick\endcsname{309}
\expandafter\newcommand\csname sr48n662ezAdsV\endcsname{312.0}
\expandafter\newcommand\csname sr48n662ezAdsH\endcsname{949}
\expandafter\newcommand\csname sr48n662ezMiiV\endcsname{271.3}
\expandafter\newcommand\csname sr48n662ezMiiH\endcsname{--}
\expandafter\newcommand\csname sr48n662ezMiV\endcsname{244.2}
\expandafter\newcommand\csname sr48n662ezMiH\endcsname{--}
\expandafter\newcommand\csname sr48n662ezMiiVErr\endcsname{-13.1\%}
\expandafter\newcommand\csname sr48n662ezMiiHErr\endcsname{--}
\expandafter\newcommand\csname sr48n662ezMiVErr\endcsname{-21.7\%}
\expandafter\newcommand\csname sr48n662ezMiHErr\endcsname{--}
\expandafter\newcommand\csname sr48n662ezRMSE\endcsname{1.88}
\expandafter\newcommand\csname sr48n662ezVRange\endcsname{18.6}
\expandafter\newcommand\csname sr48n662ezHRange\endcsname{128}
\expandafter\newcommand\csname sr48n662ezLoverH\endcsname{2.29}
\expandafter\newcommand\csname sr48n662ezTrackSource\endcsname{N662EZ-track-EGM96\_avg.kml}
\expandafter\newcommand\csname sr48n662ezAdsAvailable\endcsname{1}
\expandafter\newcommand\csname sr48n662ezTrackAvailable\endcsname{1}
\expandafter\newcommand\csname sr48n662ezInsetXMin\endcsname{-53.3}
\expandafter\newcommand\csname sr48n662ezInsetXMax\endcsname{42.2}
\expandafter\newcommand\csname sr48n662ezInsetXTickLo\endcsname{-40}
\expandafter\newcommand\csname sr48n662ezInsetXTickHi\endcsname{20}
\expandafter\newcommand\csname sr48n662ezInsetVMin\endcsname{276}
\expandafter\newcommand\csname sr48n662ezInsetVMax\endcsname{346}
\expandafter\newcommand\csname sr48n662ezInsetHMin\endcsname{880}
\expandafter\newcommand\csname sr48n662ezInsetHMax\endcsname{1068}
\expandafter\newcommand\csname sr48n662ezTrackXMin\endcsname{-8737}
\expandafter\newcommand\csname sr48n662ezTrackXMax\endcsname{7288}
\expandafter\newcommand\csname sr48n662ezTrackYMin\endcsname{-176}
\expandafter\newcommand\csname sr48n662ezTrackYMax\endcsname{2375}
\expandafter\newcommand\csname sr48n662ezTrackXTickLo\endcsname{-5000}
\expandafter\newcommand\csname sr48n662ezTrackXTickHi\endcsname{5000}
\expandafter\newcommand\csname sr48n662ezTrackYTickLo\endcsname{0}
\expandafter\newcommand\csname sr48n662ezTrackYTickHi\endcsname{2000}
\expandafter\newcommand\csname sr48n662ezTrackYTick\endcsname{2000}
\expandafter\newcommand\csname sr48n662ezTrackClosestY\endcsname{2175}
\expandafter\newcommand\csname sr49n183ejRank\endcsname{49}
\expandafter\newcommand\csname sr49n183ejLabel\endcsname{N183EJ}
\expandafter\newcommand\csname sr49n183ejDataset\endcsname{N183EJ}
\expandafter\newcommand\csname sr49n183ejGeometry\endcsname{far away/non-overhead}
\expandafter\newcommand\csname sr49n183ejVCat\endcsname{V constant}
\expandafter\newcommand\csname sr49n183ejHCat\endcsname{H constant}
\expandafter\newcommand\csname sr49n183ejAudio\endcsname{smooth curve only / incomplete ends}
\expandafter\newcommand\csname sr49n183ejClosest\endcsname{2026-05-16 21:40:25 UTC}
\expandafter\newcommand\csname sr49n183ejXMin\endcsname{-57.9}
\expandafter\newcommand\csname sr49n183ejXMax\endcsname{34.5}
\expandafter\newcommand\csname sr49n183ejYMin\endcsname{254.7}
\expandafter\newcommand\csname sr49n183ejYMax\endcsname{394.5}
\expandafter\newcommand\csname sr49n183ejFzero\endcsname{299.8}
\expandafter\newcommand\csname sr49n183ejXStartTick\endcsname{-50}
\expandafter\newcommand\csname sr49n183ejXEndTick\endcsname{30}
\expandafter\newcommand\csname sr49n183ejYLowTick\endcsname{264}
\expandafter\newcommand\csname sr49n183ejYHighTick\endcsname{375}
\expandafter\newcommand\csname sr49n183ejAdsV\endcsname{271.0}
\expandafter\newcommand\csname sr49n183ejAdsH\endcsname{1963}
\expandafter\newcommand\csname sr49n183ejMiiV\endcsname{276.1}
\expandafter\newcommand\csname sr49n183ejMiiH\endcsname{--}
\expandafter\newcommand\csname sr49n183ejMiV\endcsname{198.6}
\expandafter\newcommand\csname sr49n183ejMiH\endcsname{--}
\expandafter\newcommand\csname sr49n183ejMiiVErr\endcsname{1.9\%}
\expandafter\newcommand\csname sr49n183ejMiiHErr\endcsname{--}
\expandafter\newcommand\csname sr49n183ejMiVErr\endcsname{-26.7\%}
\expandafter\newcommand\csname sr49n183ejMiHErr\endcsname{--}
\expandafter\newcommand\csname sr49n183ejRMSE\endcsname{2.62}
\expandafter\newcommand\csname sr49n183ejVRange\endcsname{2.0}
\expandafter\newcommand\csname sr49n183ejHRange\endcsname{7}
\expandafter\newcommand\csname sr49n183ejLoverH\endcsname{3.01}
\expandafter\newcommand\csname sr49n183ejTrackSource\endcsname{N183EJ-track-EGM96\_avg.kml}
\expandafter\newcommand\csname sr49n183ejAdsAvailable\endcsname{1}
\expandafter\newcommand\csname sr49n183ejTrackAvailable\endcsname{1}
\expandafter\newcommand\csname sr49n183ejInsetXMin\endcsname{-53.3}
\expandafter\newcommand\csname sr49n183ejInsetXMax\endcsname{29.9}
\expandafter\newcommand\csname sr49n183ejInsetXTickLo\endcsname{-40}
\expandafter\newcommand\csname sr49n183ejInsetXTickHi\endcsname{20}
\expandafter\newcommand\csname sr49n183ejInsetVMin\endcsname{247}
\expandafter\newcommand\csname sr49n183ejInsetVMax\endcsname{296}
\expandafter\newcommand\csname sr49n183ejInsetHMin\endcsname{1817}
\expandafter\newcommand\csname sr49n183ejInsetHMax\endcsname{2019}
\expandafter\newcommand\csname sr49n183ejTrackXMin\endcsname{-6084}
\expandafter\newcommand\csname sr49n183ejTrackXMax\endcsname{6744}
\expandafter\newcommand\csname sr49n183ejTrackYMin\endcsname{-264}
\expandafter\newcommand\csname sr49n183ejTrackYMax\endcsname{3569}
\expandafter\newcommand\csname sr49n183ejTrackXTickLo\endcsname{-4000}
\expandafter\newcommand\csname sr49n183ejTrackXTickHi\endcsname{4000}
\expandafter\newcommand\csname sr49n183ejTrackYTickLo\endcsname{0}
\expandafter\newcommand\csname sr49n183ejTrackYTickHi\endcsname{3000}
\expandafter\newcommand\csname sr49n183ejTrackYTick\endcsname{3000}
\expandafter\newcommand\csname sr49n183ejTrackClosestY\endcsname{3305}
\expandafter\newcommand\csname sr50n958jxRank\endcsname{50}
\expandafter\newcommand\csname sr50n958jxLabel\endcsname{N958JX}
\expandafter\newcommand\csname sr50n958jxDataset\endcsname{N958JX}
\expandafter\newcommand\csname sr50n958jxGeometry\endcsname{overhead}
\expandafter\newcommand\csname sr50n958jxVCat\endcsname{V constant}
\expandafter\newcommand\csname sr50n958jxHCat\endcsname{H varying}
\expandafter\newcommand\csname sr50n958jxAudio\endcsname{smooth curve only / incomplete ends}
\expandafter\newcommand\csname sr50n958jxClosest\endcsname{2026-05-16 18:31:37 UTC}
\expandafter\newcommand\csname sr50n958jxXMin\endcsname{-23.6}
\expandafter\newcommand\csname sr50n958jxXMax\endcsname{17.8}
\expandafter\newcommand\csname sr50n958jxYMin\endcsname{91.5}
\expandafter\newcommand\csname sr50n958jxYMax\endcsname{271.0}
\expandafter\newcommand\csname sr50n958jxFzero\endcsname{142.1}
\expandafter\newcommand\csname sr50n958jxXStartTick\endcsname{-20}
\expandafter\newcommand\csname sr50n958jxXEndTick\endcsname{20}
\expandafter\newcommand\csname sr50n958jxYLowTick\endcsname{103}
\expandafter\newcommand\csname sr50n958jxYHighTick\endcsname{245}
\expandafter\newcommand\csname sr50n958jxAdsV\endcsname{776.0}
\expandafter\newcommand\csname sr50n958jxAdsH\endcsname{4985}
\expandafter\newcommand\csname sr50n958jxMiiV\endcsname{538.6}
\expandafter\newcommand\csname sr50n958jxMiiH\endcsname{1750}
\expandafter\newcommand\csname sr50n958jxMiV\endcsname{464.6}
\expandafter\newcommand\csname sr50n958jxMiH\endcsname{1583}
\expandafter\newcommand\csname sr50n958jxMiiVErr\endcsname{-30.6\%}
\expandafter\newcommand\csname sr50n958jxMiiHErr\endcsname{-64.9\%}
\expandafter\newcommand\csname sr50n958jxMiVErr\endcsname{-40.1\%}
\expandafter\newcommand\csname sr50n958jxMiHErr\endcsname{-68.2\%}
\expandafter\newcommand\csname sr50n958jxRMSE\endcsname{4.38}
\expandafter\newcommand\csname sr50n958jxVRange\endcsname{0.5}
\expandafter\newcommand\csname sr50n958jxHRange\endcsname{183}
\expandafter\newcommand\csname sr50n958jxLoverH\endcsname{0.01}
\expandafter\newcommand\csname sr50n958jxTrackSource\endcsname{N958JX-track-EGM96\_avg.kml}
\expandafter\newcommand\csname sr50n958jxAdsAvailable\endcsname{1}
\expandafter\newcommand\csname sr50n958jxTrackAvailable\endcsname{1}
\expandafter\newcommand\csname sr50n958jxInsetXMin\endcsname{-21.5}
\expandafter\newcommand\csname sr50n958jxInsetXMax\endcsname{15.8}
\expandafter\newcommand\csname sr50n958jxInsetXTickLo\endcsname{-20}
\expandafter\newcommand\csname sr50n958jxInsetXTickHi\endcsname{10}
\expandafter\newcommand\csname sr50n958jxInsetVMin\endcsname{754}
\expandafter\newcommand\csname sr50n958jxInsetVMax\endcsname{776}
\expandafter\newcommand\csname sr50n958jxInsetHMin\endcsname{4850}
\expandafter\newcommand\csname sr50n958jxInsetHMax\endcsname{5134}
\expandafter\newcommand\csname sr50n958jxTrackXMin\endcsname{-13407}
\expandafter\newcommand\csname sr50n958jxTrackXMax\endcsname{12262}
\expandafter\newcommand\csname sr50n958jxTrackYMin\endcsname{-40}
\expandafter\newcommand\csname sr50n958jxTrackYMax\endcsname{96}
\expandafter\newcommand\csname sr50n958jxTrackXTickLo\endcsname{-10000}
\expandafter\newcommand\csname sr50n958jxTrackXTickHi\endcsname{10000}
\expandafter\newcommand\csname sr50n958jxTrackYTickLo\endcsname{-20}
\expandafter\newcommand\csname sr50n958jxTrackYTickHi\endcsname{80}
\expandafter\newcommand\csname sr50n958jxTrackYTick\endcsname{80}
\expandafter\newcommand\csname sr50n958jxTrackClosestY\endcsname{56}
\expandafter\newcommand\csname sr51n451eeRank\endcsname{51}
\expandafter\newcommand\csname sr51n451eeLabel\endcsname{N451EE}
\expandafter\newcommand\csname sr51n451eeDataset\endcsname{N451EE high ridge}
\expandafter\newcommand\csname sr51n451eeGeometry\endcsname{near overhead}
\expandafter\newcommand\csname sr51n451eeVCat\endcsname{V varying}
\expandafter\newcommand\csname sr51n451eeHCat\endcsname{H varying}
\expandafter\newcommand\csname sr51n451eeAudio\endcsname{smooth plateau-like + curve}
\expandafter\newcommand\csname sr51n451eeClosest\endcsname{2026-05-13 20:43:39 UTC}
\expandafter\newcommand\csname sr51n451eeXMin\endcsname{-15.9}
\expandafter\newcommand\csname sr51n451eeXMax\endcsname{24.3}
\expandafter\newcommand\csname sr51n451eeYMin\endcsname{356.2}
\expandafter\newcommand\csname sr51n451eeYMax\endcsname{766.5}
\expandafter\newcommand\csname sr51n451eeFzero\endcsname{504.7}
\expandafter\newcommand\csname sr51n451eeXStartTick\endcsname{-10}
\expandafter\newcommand\csname sr51n451eeXEndTick\endcsname{20}
\expandafter\newcommand\csname sr51n451eeYLowTick\endcsname{382}
\expandafter\newcommand\csname sr51n451eeYHighTick\endcsname{708}
\expandafter\newcommand\csname sr51n451eeAdsV\endcsname{294.2}
\expandafter\newcommand\csname sr51n451eeAdsH\endcsname{588}
\expandafter\newcommand\csname sr51n451eeMiiV\endcsname{360.8}
\expandafter\newcommand\csname sr51n451eeMiiH\endcsname{512}
\expandafter\newcommand\csname sr51n451eeMiV\endcsname{333.8}
\expandafter\newcommand\csname sr51n451eeMiH\endcsname{500}
\expandafter\newcommand\csname sr51n451eeMiiVErr\endcsname{22.6\%}
\expandafter\newcommand\csname sr51n451eeMiiHErr\endcsname{-12.9\%}
\expandafter\newcommand\csname sr51n451eeMiVErr\endcsname{13.5\%}
\expandafter\newcommand\csname sr51n451eeMiHErr\endcsname{-14.9\%}
\expandafter\newcommand\csname sr51n451eeRMSE\endcsname{7.36}
\expandafter\newcommand\csname sr51n451eeVRange\endcsname{17.1}
\expandafter\newcommand\csname sr51n451eeHRange\endcsname{129}
\expandafter\newcommand\csname sr51n451eeLoverH\endcsname{0.65}
\expandafter\newcommand\csname sr51n451eeTrackSource\endcsname{N451EE-track-EGM96\_avg.kml}
\expandafter\newcommand\csname sr51n451eeAdsAvailable\endcsname{1}
\expandafter\newcommand\csname sr51n451eeTrackAvailable\endcsname{1}
\expandafter\newcommand\csname sr51n451eeInsetXMin\endcsname{-13.9}
\expandafter\newcommand\csname sr51n451eeInsetXMax\endcsname{22.3}
\expandafter\newcommand\csname sr51n451eeInsetXTickLo\endcsname{-10}
\expandafter\newcommand\csname sr51n451eeInsetXTickHi\endcsname{20}
\expandafter\newcommand\csname sr51n451eeInsetVMin\endcsname{295}
\expandafter\newcommand\csname sr51n451eeInsetVMax\endcsname{307}
\expandafter\newcommand\csname sr51n451eeInsetHMin\endcsname{568}
\expandafter\newcommand\csname sr51n451eeInsetHMax\endcsname{608}
\expandafter\newcommand\csname sr51n451eeTrackXMin\endcsname{-5027}
\expandafter\newcommand\csname sr51n451eeTrackXMax\endcsname{5102}
\expandafter\newcommand\csname sr51n451eeTrackYMin\endcsname{-68}
\expandafter\newcommand\csname sr51n451eeTrackYMax\endcsname{920}
\expandafter\newcommand\csname sr51n451eeTrackXTickLo\endcsname{-4000}
\expandafter\newcommand\csname sr51n451eeTrackXTickHi\endcsname{4000}
\expandafter\newcommand\csname sr51n451eeTrackYTickLo\endcsname{0}
\expandafter\newcommand\csname sr51n451eeTrackYTickHi\endcsname{800}
\expandafter\newcommand\csname sr51n451eeTrackYTick\endcsname{800}
\expandafter\newcommand\csname sr51n451eeTrackClosestY\endcsname{387}
\expandafter\newcommand\csname sr53n120rdRank\endcsname{53}
\expandafter\newcommand\csname sr53n120rdLabel\endcsname{N120RD}
\expandafter\newcommand\csname sr53n120rdDataset\endcsname{N120RD earlier track}
\expandafter\newcommand\csname sr53n120rdGeometry\endcsname{far away/non-overhead}
\expandafter\newcommand\csname sr53n120rdVCat\endcsname{V varying}
\expandafter\newcommand\csname sr53n120rdHCat\endcsname{H constant}
\expandafter\newcommand\csname sr53n120rdAudio\endcsname{smooth plateau-like + curve}
\expandafter\newcommand\csname sr53n120rdClosest\endcsname{2026-05-16 19:01:11 UTC}
\expandafter\newcommand\csname sr53n120rdXMin\endcsname{-29.6}
\expandafter\newcommand\csname sr53n120rdXMax\endcsname{45.7}
\expandafter\newcommand\csname sr53n120rdYMin\endcsname{429.1}
\expandafter\newcommand\csname sr53n120rdYMax\endcsname{687.7}
\expandafter\newcommand\csname sr53n120rdFzero\endcsname{527.1}
\expandafter\newcommand\csname sr53n120rdXStartTick\endcsname{-30}
\expandafter\newcommand\csname sr53n120rdXEndTick\endcsname{40}
\expandafter\newcommand\csname sr53n120rdYLowTick\endcsname{446}
\expandafter\newcommand\csname sr53n120rdYHighTick\endcsname{651}
\expandafter\newcommand\csname sr53n120rdAdsV\endcsname{219.7}
\expandafter\newcommand\csname sr53n120rdAdsH\endcsname{369}
\expandafter\newcommand\csname sr53n120rdMiiV\endcsname{230.0}
\expandafter\newcommand\csname sr53n120rdMiiH\endcsname{--}
\expandafter\newcommand\csname sr53n120rdMiV\endcsname{199.0}
\expandafter\newcommand\csname sr53n120rdMiH\endcsname{--}
\expandafter\newcommand\csname sr53n120rdMiiVErr\endcsname{4.7\%}
\expandafter\newcommand\csname sr53n120rdMiiHErr\endcsname{--}
\expandafter\newcommand\csname sr53n120rdMiVErr\endcsname{-9.4\%}
\expandafter\newcommand\csname sr53n120rdMiHErr\endcsname{--}
\expandafter\newcommand\csname sr53n120rdRMSE\endcsname{6.02}
\expandafter\newcommand\csname sr53n120rdVRange\endcsname{64.3}
\expandafter\newcommand\csname sr53n120rdHRange\endcsname{42}
\expandafter\newcommand\csname sr53n120rdLoverH\endcsname{2.68}
\expandafter\newcommand\csname sr53n120rdTrackSource\endcsname{N120RD-track-press\_alt\_uncorrected.kml}
\expandafter\newcommand\csname sr53n120rdAdsAvailable\endcsname{1}
\expandafter\newcommand\csname sr53n120rdTrackAvailable\endcsname{1}
\expandafter\newcommand\csname sr53n120rdInsetXMin\endcsname{-25.9}
\expandafter\newcommand\csname sr53n120rdInsetXMax\endcsname{41.9}
\expandafter\newcommand\csname sr53n120rdInsetXTickLo\endcsname{-20}
\expandafter\newcommand\csname sr53n120rdInsetXTickHi\endcsname{40}
\expandafter\newcommand\csname sr53n120rdInsetVMin\endcsname{195}
\expandafter\newcommand\csname sr53n120rdInsetVMax\endcsname{284}
\expandafter\newcommand\csname sr53n120rdInsetHMin\endcsname{288}
\expandafter\newcommand\csname sr53n120rdInsetHMax\endcsname{450}
\expandafter\newcommand\csname sr53n120rdTrackXMin\endcsname{-4441}
\expandafter\newcommand\csname sr53n120rdTrackXMax\endcsname{3811}
\expandafter\newcommand\csname sr53n120rdTrackYMin\endcsname{-1196}
\expandafter\newcommand\csname sr53n120rdTrackYMax\endcsname{1201}
\expandafter\newcommand\csname sr53n120rdTrackXTickLo\endcsname{-4000}
\expandafter\newcommand\csname sr53n120rdTrackXTickHi\endcsname{2000}
\expandafter\newcommand\csname sr53n120rdTrackYTickLo\endcsname{-1000}
\expandafter\newcommand\csname sr53n120rdTrackYTickHi\endcsname{1000}
\expandafter\newcommand\csname sr53n120rdTrackYTick\endcsname{1000}
\expandafter\newcommand\csname sr53n120rdTrackClosestY\endcsname{1036}
\expandafter\newcommand\csname sr56n263jxRank\endcsname{56}
\expandafter\newcommand\csname sr56n263jxLabel\endcsname{N263JX}
\expandafter\newcommand\csname sr56n263jxDataset\endcsname{N263JX}
\expandafter\newcommand\csname sr56n263jxGeometry\endcsname{far away/non-overhead}
\expandafter\newcommand\csname sr56n263jxVCat\endcsname{V constant}
\expandafter\newcommand\csname sr56n263jxHCat\endcsname{H varying}
\expandafter\newcommand\csname sr56n263jxAudio\endcsname{plateau-like + curve but scattered/ambiguous}
\expandafter\newcommand\csname sr56n263jxClosest\endcsname{2026-05-17 18:28:28 UTC}
\expandafter\newcommand\csname sr56n263jxXMin\endcsname{-66.4}
\expandafter\newcommand\csname sr56n263jxXMax\endcsname{31.8}
\expandafter\newcommand\csname sr56n263jxYMin\endcsname{232.3}
\expandafter\newcommand\csname sr56n263jxYMax\endcsname{412.0}
\expandafter\newcommand\csname sr56n263jxFzero\endcsname{286.0}
\expandafter\newcommand\csname sr56n263jxXStartTick\endcsname{-60}
\expandafter\newcommand\csname sr56n263jxXEndTick\endcsname{30}
\expandafter\newcommand\csname sr56n263jxYLowTick\endcsname{246}
\expandafter\newcommand\csname sr56n263jxYHighTick\endcsname{386}
\expandafter\newcommand\csname sr56n263jxAdsV\endcsname{365.0}
\expandafter\newcommand\csname sr56n263jxAdsH\endcsname{1856}
\expandafter\newcommand\csname sr56n263jxMiiV\endcsname{317.9}
\expandafter\newcommand\csname sr56n263jxMiiH\endcsname{1690}
\expandafter\newcommand\csname sr56n263jxMiV\endcsname{245.6}
\expandafter\newcommand\csname sr56n263jxMiH\endcsname{--}
\expandafter\newcommand\csname sr56n263jxMiiVErr\endcsname{-12.9\%}
\expandafter\newcommand\csname sr56n263jxMiiHErr\endcsname{-8.9\%}
\expandafter\newcommand\csname sr56n263jxMiVErr\endcsname{-32.7\%}
\expandafter\newcommand\csname sr56n263jxMiHErr\endcsname{--}
\expandafter\newcommand\csname sr56n263jxRMSE\endcsname{3.67}
\expandafter\newcommand\csname sr56n263jxVRange\endcsname{9.9}
\expandafter\newcommand\csname sr56n263jxHRange\endcsname{502}
\expandafter\newcommand\csname sr56n263jxLoverH\endcsname{0.93}
\expandafter\newcommand\csname sr56n263jxTrackSource\endcsname{N263JX-track-EGM96\_avg.kml}
\expandafter\newcommand\csname sr56n263jxAdsAvailable\endcsname{1}
\expandafter\newcommand\csname sr56n263jxTrackAvailable\endcsname{1}
\expandafter\newcommand\csname sr56n263jxInsetXMin\endcsname{-61.5}
\expandafter\newcommand\csname sr56n263jxInsetXMax\endcsname{26.9}
\expandafter\newcommand\csname sr56n263jxInsetXTickLo\endcsname{-40}
\expandafter\newcommand\csname sr56n263jxInsetXTickHi\endcsname{20}
\expandafter\newcommand\csname sr56n263jxInsetVMin\endcsname{366}
\expandafter\newcommand\csname sr56n263jxInsetVMax\endcsname{393}
\expandafter\newcommand\csname sr56n263jxInsetHMin\endcsname{1615}
\expandafter\newcommand\csname sr56n263jxInsetHMax\endcsname{2135}
\expandafter\newcommand\csname sr56n263jxTrackXMin\endcsname{-11820}
\expandafter\newcommand\csname sr56n263jxTrackXMax\endcsname{8220}
\expandafter\newcommand\csname sr56n263jxTrackYMin\endcsname{-144}
\expandafter\newcommand\csname sr56n263jxTrackYMax\endcsname{1944}
\expandafter\newcommand\csname sr56n263jxTrackXTickLo\endcsname{-10000}
\expandafter\newcommand\csname sr56n263jxTrackXTickHi\endcsname{5000}
\expandafter\newcommand\csname sr56n263jxTrackYTickLo\endcsname{0}
\expandafter\newcommand\csname sr56n263jxTrackYTickHi\endcsname{1500}
\expandafter\newcommand\csname sr56n263jxTrackYTick\endcsname{1500}
\expandafter\newcommand\csname sr56n263jxTrackClosestY\endcsname{1713}
\expandafter\newcommand\csname sr57n231dtRank\endcsname{57}
\expandafter\newcommand\csname sr57n231dtLabel\endcsname{N231DT}
\expandafter\newcommand\csname sr57n231dtDataset\endcsname{N231DT}
\expandafter\newcommand\csname sr57n231dtGeometry\endcsname{far away/non-overhead}
\expandafter\newcommand\csname sr57n231dtVCat\endcsname{V varying}
\expandafter\newcommand\csname sr57n231dtHCat\endcsname{H varying}
\expandafter\newcommand\csname sr57n231dtAudio\endcsname{smooth curve only / incomplete ends}
\expandafter\newcommand\csname sr57n231dtClosest\endcsname{2026-05-17 17:40:41.36 UTC}
\expandafter\newcommand\csname sr57n231dtXMin\endcsname{-45.0}
\expandafter\newcommand\csname sr57n231dtXMax\endcsname{32.7}
\expandafter\newcommand\csname sr57n231dtYMin\endcsname{329.1}
\expandafter\newcommand\csname sr57n231dtYMax\endcsname{428.5}
\expandafter\newcommand\csname sr57n231dtFzero\endcsname{367.6}
\expandafter\newcommand\csname sr57n231dtXStartTick\endcsname{-40}
\expandafter\newcommand\csname sr57n231dtXEndTick\endcsname{30}
\expandafter\newcommand\csname sr57n231dtYLowTick\endcsname{335}
\expandafter\newcommand\csname sr57n231dtYHighTick\endcsname{412}
\expandafter\newcommand\csname sr57n231dtAdsV\endcsname{186.0}
\expandafter\newcommand\csname sr57n231dtAdsH\endcsname{875}
\expandafter\newcommand\csname sr57n231dtMiiV\endcsname{139.2}
\expandafter\newcommand\csname sr57n231dtMiiH\endcsname{--}
\expandafter\newcommand\csname sr57n231dtMiV\endcsname{111.0}
\expandafter\newcommand\csname sr57n231dtMiH\endcsname{--}
\expandafter\newcommand\csname sr57n231dtMiiVErr\endcsname{-25.2\%}
\expandafter\newcommand\csname sr57n231dtMiiHErr\endcsname{--}
\expandafter\newcommand\csname sr57n231dtMiVErr\endcsname{-40.3\%}
\expandafter\newcommand\csname sr57n231dtMiHErr\endcsname{--}
\expandafter\newcommand\csname sr57n231dtRMSE\endcsname{2.44}
\expandafter\newcommand\csname sr57n231dtVRange\endcsname{46.4}
\expandafter\newcommand\csname sr57n231dtHRange\endcsname{69}
\expandafter\newcommand\csname sr57n231dtLoverH\endcsname{2.65}
\expandafter\newcommand\csname sr57n231dtTrackSource\endcsname{N231DT-track-EGM96\_avg.kml}
\expandafter\newcommand\csname sr57n231dtAdsAvailable\endcsname{1}
\expandafter\newcommand\csname sr57n231dtTrackAvailable\endcsname{1}
\expandafter\newcommand\csname sr57n231dtInsetXMin\endcsname{-41.1}
\expandafter\newcommand\csname sr57n231dtInsetXMax\endcsname{28.9}
\expandafter\newcommand\csname sr57n231dtInsetXTickLo\endcsname{-40}
\expandafter\newcommand\csname sr57n231dtInsetXTickHi\endcsname{20}
\expandafter\newcommand\csname sr57n231dtInsetVMin\endcsname{164}
\expandafter\newcommand\csname sr57n231dtInsetVMax\endcsname{227}
\expandafter\newcommand\csname sr57n231dtInsetHMin\endcsname{805}
\expandafter\newcommand\csname sr57n231dtInsetHMax\endcsname{944}
\expandafter\newcommand\csname sr57n231dtTrackXMin\endcsname{-4854}
\expandafter\newcommand\csname sr57n231dtTrackXMax\endcsname{4056}
\expandafter\newcommand\csname sr57n231dtTrackYMin\endcsname{-218}
\expandafter\newcommand\csname sr57n231dtTrackYMax\endcsname{2948}
\expandafter\newcommand\csname sr57n231dtTrackXTickLo\endcsname{-4000}
\expandafter\newcommand\csname sr57n231dtTrackXTickHi\endcsname{2000}
\expandafter\newcommand\csname sr57n231dtTrackYTickLo\endcsname{0}
\expandafter\newcommand\csname sr57n231dtTrackYTickHi\endcsname{2948}
\expandafter\newcommand\csname sr57n231dtTrackYTick\endcsname{2948}
\expandafter\newcommand\csname sr57n231dtTrackClosestY\endcsname{2321}
\expandafter\newcommand\csname sr58n917hrRank\endcsname{58}
\expandafter\newcommand\csname sr58n917hrLabel\endcsname{N917HR}
\expandafter\newcommand\csname sr58n917hrDataset\endcsname{N917HR}
\expandafter\newcommand\csname sr58n917hrGeometry\endcsname{far away/non-overhead}
\expandafter\newcommand\csname sr58n917hrVCat\endcsname{V constant}
\expandafter\newcommand\csname sr58n917hrHCat\endcsname{H varying}
\expandafter\newcommand\csname sr58n917hrAudio\endcsname{smooth plateau-like + curve}
\expandafter\newcommand\csname sr58n917hrClosest\endcsname{2026-05-17 23:40:54.25 UTC}
\expandafter\newcommand\csname sr58n917hrXMin\endcsname{-30.3}
\expandafter\newcommand\csname sr58n917hrXMax\endcsname{41.5}
\expandafter\newcommand\csname sr58n917hrYMin\endcsname{217.8}
\expandafter\newcommand\csname sr58n917hrYMax\endcsname{325.9}
\expandafter\newcommand\csname sr58n917hrFzero\endcsname{262.4}
\expandafter\newcommand\csname sr58n917hrXStartTick\endcsname{-30}
\expandafter\newcommand\csname sr58n917hrXEndTick\endcsname{40}
\expandafter\newcommand\csname sr58n917hrYLowTick\endcsname{225}
\expandafter\newcommand\csname sr58n917hrYHighTick\endcsname{310}
\expandafter\newcommand\csname sr58n917hrAdsV\endcsname{401.3}
\expandafter\newcommand\csname sr58n917hrAdsH\endcsname{1898}
\expandafter\newcommand\csname sr58n917hrMiiV\endcsname{181.4}
\expandafter\newcommand\csname sr58n917hrMiiH\endcsname{--}
\expandafter\newcommand\csname sr58n917hrMiV\endcsname{170.2}
\expandafter\newcommand\csname sr58n917hrMiH\endcsname{--}
\expandafter\newcommand\csname sr58n917hrMiiVErr\endcsname{-54.8\%}
\expandafter\newcommand\csname sr58n917hrMiiHErr\endcsname{--}
\expandafter\newcommand\csname sr58n917hrMiVErr\endcsname{-57.6\%}
\expandafter\newcommand\csname sr58n917hrMiHErr\endcsname{--}
\expandafter\newcommand\csname sr58n917hrRMSE\endcsname{2.21}
\expandafter\newcommand\csname sr58n917hrVRange\endcsname{7.6}
\expandafter\newcommand\csname sr58n917hrHRange\endcsname{246}
\expandafter\newcommand\csname sr58n917hrLoverH\endcsname{0.93}
\expandafter\newcommand\csname sr58n917hrTrackSource\endcsname{N917HR-track-EGM96\_avg.kml}
\expandafter\newcommand\csname sr58n917hrAdsAvailable\endcsname{1}
\expandafter\newcommand\csname sr58n917hrTrackAvailable\endcsname{1}
\expandafter\newcommand\csname sr58n917hrInsetXMin\endcsname{-26.7}
\expandafter\newcommand\csname sr58n917hrInsetXMax\endcsname{38.0}
\expandafter\newcommand\csname sr58n917hrInsetXTickLo\endcsname{-20}
\expandafter\newcommand\csname sr58n917hrInsetXTickHi\endcsname{38}
\expandafter\newcommand\csname sr58n917hrInsetVMin\endcsname{358}
\expandafter\newcommand\csname sr58n917hrInsetVMax\endcsname{409}
\expandafter\newcommand\csname sr58n917hrInsetHMin\endcsname{1702}
\expandafter\newcommand\csname sr58n917hrInsetHMax\endcsname{2080}
\expandafter\newcommand\csname sr58n917hrTrackXMin\endcsname{-8048}
\expandafter\newcommand\csname sr58n917hrTrackXMax\endcsname{9069}
\expandafter\newcommand\csname sr58n917hrTrackYMin\endcsname{-143}
\expandafter\newcommand\csname sr58n917hrTrackYMax\endcsname{1937}
\expandafter\newcommand\csname sr58n917hrTrackXTickLo\endcsname{-5000}
\expandafter\newcommand\csname sr58n917hrTrackXTickHi\endcsname{5000}
\expandafter\newcommand\csname sr58n917hrTrackYTickLo\endcsname{0}
\expandafter\newcommand\csname sr58n917hrTrackYTickHi\endcsname{1500}
\expandafter\newcommand\csname sr58n917hrTrackYTick\endcsname{1500}
\expandafter\newcommand\csname sr58n917hrTrackClosestY\endcsname{1769}
\expandafter\newcommand\csname sr59n67wlRank\endcsname{59}
\expandafter\newcommand\csname sr59n67wlLabel\endcsname{N67WL}
\expandafter\newcommand\csname sr59n67wlDataset\endcsname{N67WL}
\expandafter\newcommand\csname sr59n67wlGeometry\endcsname{near overhead}
\expandafter\newcommand\csname sr59n67wlVCat\endcsname{V constant}
\expandafter\newcommand\csname sr59n67wlHCat\endcsname{H varying}
\expandafter\newcommand\csname sr59n67wlAudio\endcsname{smooth curve only / incomplete ends}
\expandafter\newcommand\csname sr59n67wlClosest\endcsname{2026-05-17 17:02:45 UTC}
\expandafter\newcommand\csname sr59n67wlXMin\endcsname{-65.5}
\expandafter\newcommand\csname sr59n67wlXMax\endcsname{63.6}
\expandafter\newcommand\csname sr59n67wlYMin\endcsname{217.7}
\expandafter\newcommand\csname sr59n67wlYMax\endcsname{405.9}
\expandafter\newcommand\csname sr59n67wlFzero\endcsname{283.3}
\expandafter\newcommand\csname sr59n67wlXStartTick\endcsname{-60}
\expandafter\newcommand\csname sr59n67wlXEndTick\endcsname{60}
\expandafter\newcommand\csname sr59n67wlYLowTick\endcsname{230}
\expandafter\newcommand\csname sr59n67wlYHighTick\endcsname{379}
\expandafter\newcommand\csname sr59n67wlAdsV\endcsname{414.0}
\expandafter\newcommand\csname sr59n67wlAdsH\endcsname{3887}
\expandafter\newcommand\csname sr59n67wlMiiV\endcsname{335.7}
\expandafter\newcommand\csname sr59n67wlMiiH\endcsname{2834}
\expandafter\newcommand\csname sr59n67wlMiV\endcsname{281.1}
\expandafter\newcommand\csname sr59n67wlMiH\endcsname{2194}
\expandafter\newcommand\csname sr59n67wlMiiVErr\endcsname{-18.9\%}
\expandafter\newcommand\csname sr59n67wlMiiHErr\endcsname{-27.1\%}
\expandafter\newcommand\csname sr59n67wlMiVErr\endcsname{-32.1\%}
\expandafter\newcommand\csname sr59n67wlMiHErr\endcsname{-43.6\%}
\expandafter\newcommand\csname sr59n67wlRMSE\endcsname{3.16}
\expandafter\newcommand\csname sr59n67wlVRange\endcsname{13.5}
\expandafter\newcommand\csname sr59n67wlHRange\endcsname{946}
\expandafter\newcommand\csname sr59n67wlLoverH\endcsname{0.08}
\expandafter\newcommand\csname sr59n67wlTrackSource\endcsname{N67WL-track-EGM96\_avg.kml}
\expandafter\newcommand\csname sr59n67wlAdsAvailable\endcsname{1}
\expandafter\newcommand\csname sr59n67wlTrackAvailable\endcsname{1}
\expandafter\newcommand\csname sr59n67wlInsetXMin\endcsname{-59.1}
\expandafter\newcommand\csname sr59n67wlInsetXMax\endcsname{57.2}
\expandafter\newcommand\csname sr59n67wlInsetXTickLo\endcsname{-40}
\expandafter\newcommand\csname sr59n67wlInsetXTickHi\endcsname{40}
\expandafter\newcommand\csname sr59n67wlInsetVMin\endcsname{356}
\expandafter\newcommand\csname sr59n67wlInsetVMax\endcsname{423}
\expandafter\newcommand\csname sr59n67wlInsetHMin\endcsname{3275}
\expandafter\newcommand\csname sr59n67wlInsetHMax\endcsname{4370}
\expandafter\newcommand\csname sr59n67wlTrackXMin\endcsname{-10764}
\expandafter\newcommand\csname sr59n67wlTrackXMax\endcsname{10766}
\expandafter\newcommand\csname sr59n67wlTrackYMin\endcsname{-3521}
\expandafter\newcommand\csname sr59n67wlTrackYMax\endcsname{521}
\expandafter\newcommand\csname sr59n67wlTrackXTickLo\endcsname{-10000}
\expandafter\newcommand\csname sr59n67wlTrackXTickHi\endcsname{10000}
\expandafter\newcommand\csname sr59n67wlTrackYTickLo\endcsname{-3000}
\expandafter\newcommand\csname sr59n67wlTrackYTickHi\endcsname{0}
\expandafter\newcommand\csname sr59n67wlTrackYTick\endcsname{0}
\expandafter\newcommand\csname sr59n67wlTrackClosestY\endcsname{242}
\expandafter\newcommand\csname sr60n173jeRank\endcsname{60}
\expandafter\newcommand\csname sr60n173jeLabel\endcsname{N173JE}
\expandafter\newcommand\csname sr60n173jeDataset\endcsname{N173JE}
\expandafter\newcommand\csname sr60n173jeGeometry\endcsname{far away/non-overhead}
\expandafter\newcommand\csname sr60n173jeVCat\endcsname{V varying}
\expandafter\newcommand\csname sr60n173jeHCat\endcsname{H varying}
\expandafter\newcommand\csname sr60n173jeAudio\endcsname{smooth curve only / incomplete ends}
\expandafter\newcommand\csname sr60n173jeClosest\endcsname{2026-05-17 17:28:06.70 UTC}
\expandafter\newcommand\csname sr60n173jeXMin\endcsname{-56.6}
\expandafter\newcommand\csname sr60n173jeXMax\endcsname{57.3}
\expandafter\newcommand\csname sr60n173jeYMin\endcsname{496.7}
\expandafter\newcommand\csname sr60n173jeYMax\endcsname{1000.5}
\expandafter\newcommand\csname sr60n173jeFzero\endcsname{672.1}
\expandafter\newcommand\csname sr60n173jeXStartTick\endcsname{-50}
\expandafter\newcommand\csname sr60n173jeXEndTick\endcsname{50}
\expandafter\newcommand\csname sr60n173jeYLowTick\endcsname{529}
\expandafter\newcommand\csname sr60n173jeYHighTick\endcsname{929}
\expandafter\newcommand\csname sr60n173jeAdsV\endcsname{143.0}
\expandafter\newcommand\csname sr60n173jeAdsH\endcsname{247}
\expandafter\newcommand\csname sr60n173jeMiiV\endcsname{400.8}
\expandafter\newcommand\csname sr60n173jeMiiH\endcsname{4728}
\expandafter\newcommand\csname sr60n173jeMiV\endcsname{289.4}
\expandafter\newcommand\csname sr60n173jeMiH\endcsname{1832}
\expandafter\newcommand\csname sr60n173jeMiiVErr\endcsname{180.3\%}
\expandafter\newcommand\csname sr60n173jeMiiHErr\endcsname{1812.2\%}
\expandafter\newcommand\csname sr60n173jeMiVErr\endcsname{102.4\%}
\expandafter\newcommand\csname sr60n173jeMiHErr\endcsname{640.8\%}
\expandafter\newcommand\csname sr60n173jeRMSE\endcsname{6.55}
\expandafter\newcommand\csname sr60n173jeVRange\endcsname{44.9}
\expandafter\newcommand\csname sr60n173jeHRange\endcsname{180}
\expandafter\newcommand\csname sr60n173jeLoverH\endcsname{8.23}
\expandafter\newcommand\csname sr60n173jeTrackSource\endcsname{N173JE-track-EGM96\_avg.kml}
\expandafter\newcommand\csname sr60n173jeAdsAvailable\endcsname{1}
\expandafter\newcommand\csname sr60n173jeTrackAvailable\endcsname{1}
\expandafter\newcommand\csname sr60n173jeInsetXMin\endcsname{-50.9}
\expandafter\newcommand\csname sr60n173jeInsetXMax\endcsname{51.6}
\expandafter\newcommand\csname sr60n173jeInsetXTickLo\endcsname{-40}
\expandafter\newcommand\csname sr60n173jeInsetXTickHi\endcsname{40}
\expandafter\newcommand\csname sr60n173jeInsetVMin\endcsname{95}
\expandafter\newcommand\csname sr60n173jeInsetVMax\endcsname{177}
\expandafter\newcommand\csname sr60n173jeInsetHMin\endcsname{114}
\expandafter\newcommand\csname sr60n173jeInsetHMax\endcsname{310}
\expandafter\newcommand\csname sr60n173jeTrackXMin\endcsname{-3483}
\expandafter\newcommand\csname sr60n173jeTrackXMax\endcsname{1130}
\expandafter\newcommand\csname sr60n173jeTrackYMin\endcsname{-338}
\expandafter\newcommand\csname sr60n173jeTrackYMax\endcsname{4562}
\expandafter\newcommand\csname sr60n173jeTrackXTickLo\endcsname{-3000}
\expandafter\newcommand\csname sr60n173jeTrackXTickHi\endcsname{0}
\expandafter\newcommand\csname sr60n173jeTrackYTickLo\endcsname{0}
\expandafter\newcommand\csname sr60n173jeTrackYTickHi\endcsname{4000}
\expandafter\newcommand\csname sr60n173jeTrackYTick\endcsname{4000}
\expandafter\newcommand\csname sr60n173jeTrackClosestY\endcsname{2032}

\newcommand{\cs}{c_s}
\newcommand{\kms}{\mathrm{km\,h^{-1}}}
\newcommand{\ms}{\mathrm{m\,s^{-1}}}
\newcommand{\cv}[2]{\csname s#1#2\endcsname}
\newcommand{\paperinsetfont}{\fontsize{5.0}{5.5}\selectfont}
\newcommand{\papertrackfont}{\fontsize{5.0}{5.5}\selectfont}
\newcolumntype{L}[1]{>{\raggedright\arraybackslash}p{#1}}
\newcommand{\PaperFrequencyTick}[1]{\pgfmathprintnumber[fixed,precision=1]{#1}}

\newcommand{\PaperSetCase}[3]{\expandafter\def\csname paper#1#2\endcsname{#3}}
\newcommand{\pcv}[2]{%
  \ifcsname paper#1#2\endcsname
    \csname paper#1#2\endcsname
  \else
    \cv{#1}{#2}%
  \fi}
\newif\ifPaperUseContrastAudio
\newcommand{\PaperChooseSmallAudioStyle}[1]{%
  \def\PaperCurrentCase{#1}%
  \def\PaperContrastAudioCase{r06n373df}%
  \ifx\PaperCurrentCase\PaperContrastAudioCase
    \PaperUseContrastAudiotrue
  \else
    \PaperUseContrastAudiofalse
  \fi}
\PaperSetCase{r04n784mb}{TrackXMin}{-4410}
\PaperSetCase{r04n784mb}{TrackXMax}{4350}
\PaperSetCase{r04n784mb}{TrackYMin}{-60}
\PaperSetCase{r04n784mb}{TrackYMax}{1060}
\PaperSetCase{r04n784mb}{TrackXTickLo}{-3000}
\PaperSetCase{r04n784mb}{TrackXTickHi}{3000}
\PaperSetCase{r04n784mb}{TrackYTickLo}{0}
\PaperSetCase{r04n784mb}{TrackYTickHi}{1000}
\PaperSetCase{r04n784mb}{TrackYTick}{1000}
\PaperSetCase{r04n784mb}{FMax}{320.3}
\PaperSetCase{r04n784mb}{FMin}{203.5}
\PaperSetCase{r04n784mb}{Fzero}{248.9}
\PaperSetCase{r04n784mb}{FMaxX}{-37.4}
\PaperSetCase{r04n784mb}{FMinX}{37.9}
\PaperSetCase{r06n373df}{FMax}{296.2}
\PaperSetCase{r06n373df}{FMin}{184.4}
\PaperSetCase{r06n373df}{Fzero}{227.3}
\PaperSetCase{r06n373df}{FMaxX}{-16.0}
\PaperSetCase{r06n373df}{FMinX}{23.6}
\PaperSetCase{r06n373df}{TrackXMin}{-3080}
\PaperSetCase{r06n373df}{TrackXMax}{2020}
\PaperSetCase{r06n373df}{TrackYMin}{-20}
\PaperSetCase{r06n373df}{TrackYMax}{240}
\PaperSetCase{r06n373df}{TrackXTickLo}{-2000}
\PaperSetCase{r06n373df}{TrackXTickHi}{1000}
\PaperSetCase{r06n373df}{TrackYTickLo}{0}
\PaperSetCase{r06n373df}{TrackYTickHi}{200}
\PaperSetCase{r06n373df}{TrackYTick}{200}
\PaperSetCase{r05n413ak}{FMax}{313.5}
\PaperSetCase{r05n413ak}{FMin}{218.8}
\PaperSetCase{r05n413ak}{Fzero}{257.7}
\PaperSetCase{r05n413ak}{FMaxX}{-43.3}
\PaperSetCase{r05n413ak}{FMinX}{90.1}
\PaperSetCase{r05n413ak}{TrackXMin}{-3690}
\PaperSetCase{r05n413ak}{TrackXMax}{6630}
\PaperSetCase{r05n413ak}{TrackYMin}{-30}
\PaperSetCase{r05n413ak}{TrackYMax}{520}
\PaperSetCase{r05n413ak}{TrackXTickLo}{-3000}
\PaperSetCase{r05n413ak}{TrackXTickHi}{6000}
\PaperSetCase{r05n413ak}{TrackYTickLo}{0}
\PaperSetCase{r05n413ak}{TrackYTickHi}{400}
\PaperSetCase{r05n413ak}{TrackYTick}{400}
\PaperSetCase{r04n784mb}{TrackTitleX}{0.32}
\PaperSetCase{r04n784mb}{TrackTitleY}{0.22}
\PaperSetCase{r06n373df}{TrackTitleX}{0.38}
\PaperSetCase{r06n373df}{TrackTitleY}{0.52}
\PaperSetCase{r05n413ak}{TrackTitleX}{0.70}
\PaperSetCase{r05n413ak}{TrackTitleY}{0.25}
\PaperSetCase{r20n394fx}{TrackTitleX}{0.38}
\PaperSetCase{r20n394fx}{TrackTitleY}{0.52}
\PaperSetCase{r20n394fx}{FMax}{492.6}
\PaperSetCase{r20n394fx}{FMin}{264.4}
\PaperSetCase{r20n394fx}{Fzero}{344.1}
\PaperSetCase{r20n394fx}{FMaxX}{-11.7}
\PaperSetCase{r20n394fx}{FMinX}{35.5}
\PaperSetCase{r11n60rs}{TrackTitleX}{0.38}
\PaperSetCase{r11n60rs}{TrackTitleY}{0.36}
\PaperSetCase{r11n60rs}{FMax}{465.8}
\PaperSetCase{r11n60rs}{FMin}{330.5}
\PaperSetCase{r11n60rs}{Fzero}{386.6}
\PaperSetCase{r11n60rs}{FMaxX}{-44.0}
\PaperSetCase{r11n60rs}{FMinX}{33.4}
\PaperSetCase{r24n9658k}{TrackTitleX}{0.70}
\PaperSetCase{r24n9658k}{TrackTitleY}{0.83}
\PaperSetCase{r24n9658k}{FMax}{689.4}
\PaperSetCase{r24n9658k}{FMin}{523.8}
\PaperSetCase{r24n9658k}{Fzero}{595.3}
\PaperSetCase{r24n9658k}{FMaxX}{-37.7}
\PaperSetCase{r24n9658k}{FMinX}{44.7}

\newcommand{\PaperSmallPlotWidth}{1.1\linewidth}
\newcommand{\PaperSmallPlotHeight}{5.55cm}

\newcommand{\PaperSmallAdsInsetWidth}{2.35cm}
\newcommand{\PaperSmallAdsInsetHeight}{1.22cm}
\newcommand{\PaperSmallTrackInsetWidth}{2.35cm}
\newcommand{\PaperSmallTrackInsetHeight}{1.16cm}

\newcommand{\PaperPanelAAdsInsetX}{0.985}
\newcommand{\PaperPanelAAdsInsetY}{0.972}
\newcommand{\PaperPanelAGpsInsetX}{0.985}
\newcommand{\PaperPanelAGpsInsetY}{0.51}
\newcommand{\PaperPanelBAdsInsetX}{0.985}
\newcommand{\PaperPanelBAdsInsetY}{0.972}
\newcommand{\PaperPanelBGpsInsetX}{0.985}
\newcommand{\PaperPanelBGpsInsetY}{0.51}
\newcommand{\PaperSmallAdsInsetX}{0.985}
\newcommand{\PaperSmallAdsInsetY}{0.972}
\newcommand{\PaperSmallTrackInsetX}{0.985}
\newcommand{\PaperSmallTrackInsetY}{0.51}
\PaperSetCase{r06n373df}{SmallAdsInsetX}{\PaperPanelAAdsInsetX}
\PaperSetCase{r06n373df}{SmallAdsInsetY}{\PaperPanelAAdsInsetY}
\PaperSetCase{r06n373df}{SmallTrackInsetX}{\PaperPanelAGpsInsetX}
\PaperSetCase{r06n373df}{SmallTrackInsetY}{\PaperPanelAGpsInsetY}
\PaperSetCase{r05n413ak}{SmallAdsInsetX}{\PaperPanelBAdsInsetX}
\PaperSetCase{r05n413ak}{SmallAdsInsetY}{\PaperPanelBAdsInsetY}
\PaperSetCase{r05n413ak}{SmallTrackInsetX}{\PaperPanelBGpsInsetX}
\PaperSetCase{r05n413ak}{SmallTrackInsetY}{\PaperPanelBGpsInsetY}
\PaperSetCase{r04n784mb}{SmallTrackInsetX}{0.50}
\PaperSetCase{r04n784mb}{SmallTrackInsetY}{0.39}
\PaperSetCase{r11n60rs}{SmallTrackInsetX}{0.50}
\PaperSetCase{r11n60rs}{SmallTrackInsetY}{0.39}
\PaperSetCase{r24n9658k}{SmallTrackInsetX}{0.50}
\PaperSetCase{r24n9658k}{SmallTrackInsetY}{0.39}

\newcommand{\PaperCaseSmallAdsInsetX}[1]{%
  \ifcsname paper#1SmallAdsInsetX\endcsname
    \csname paper#1SmallAdsInsetX\endcsname
  \else
    \PaperSmallAdsInsetX
  \fi}
\newcommand{\PaperCaseSmallAdsInsetY}[1]{%
  \ifcsname paper#1SmallAdsInsetY\endcsname
    \csname paper#1SmallAdsInsetY\endcsname
  \else
    \PaperSmallAdsInsetY
  \fi}
\newcommand{\PaperCaseSmallTrackInsetX}[1]{%
  \ifcsname paper#1SmallTrackInsetX\endcsname
    \csname paper#1SmallTrackInsetX\endcsname
  \else
    \PaperSmallTrackInsetX
  \fi}
\newcommand{\PaperCaseSmallTrackInsetY}[1]{%
  \ifcsname paper#1SmallTrackInsetY\endcsname
    \csname paper#1SmallTrackInsetY\endcsname
  \else
    \PaperSmallTrackInsetY
  \fi}

\newcommand{\PaperSmallXLabelX}{0.52}
\newcommand{\PaperSmallXLabelY}{-0.13}
\newcommand{\PaperSmallYLabelX}{-0.050}
\newcommand{\PaperSmallYLabelY}{0.52}

\newcommand{\PaperAdsTitleX}{0.50}
\newcommand{\PaperAdsTitleY}{0.98}

\newcommand{\PaperTrackTitleText}{ground track}
\newcommand{\PaperInsetBackgroundOpacity}{0.90}
\tikzset{
  paper inset label background/.style={fill=none,inner sep=0.1pt},
  paper inset title background/.style={fill=white,draw=none,inner sep=0.1pt}
}

\newcommand{\PaperSmallFzeroLabelRelX}{0.018}
\newcommand{\PaperFzeroLabelYOffset}{1.0pt}
\newcommand{\PaperFzeroLabelFont}{\scriptsize}
\newcommand{\PaperFzeroLabelInnerSep}{0.6pt}
\newcommand{\PaperFzeroLabelText}[1]{%
  \ifcsname paper#1FzeroLabel\endcsname
    \csname paper#1FzeroLabel\endcsname
  \else
    \pcv{#1}{Fzero} ($f_{0,\mathrm I}$)%
  \fi}
\newcommand{\PaperFzeroMarker}[2]{%
  \coordinate (paperfzerox) at (rel axis cs:#2,0);
  \coordinate (paperfzeroy) at (axis cs:0,\pcv{#1}{Fzero});
  \node[anchor=south west,font=\PaperFzeroLabelFont,fill=white,
        inner sep=\PaperFzeroLabelInnerSep,yshift=\PaperFzeroLabelYOffset]
    at (paperfzerox |- paperfzeroy) {\PaperFzeroLabelText{#1}};
}

\PaperSetCase{r11n60rs}{FMinLabelLeft}{1}

\PaperSetCase{r04n784mb}{TrackXMin}{-4662}
\PaperSetCase{r04n784mb}{TrackXMax}{3401}
\PaperSetCase{r04n784mb}{TrackYMin}{-120}
\PaperSetCase{r04n784mb}{TrackYMax}{1119}
\PaperSetCase{r04n784mb}{TrackXTickLo}{-3000}
\PaperSetCase{r04n784mb}{TrackXTickHi}{2500}
\PaperSetCase{r04n784mb}{TrackYTickLo}{0}
\PaperSetCase{r04n784mb}{TrackYTickHi}{1000}
\PaperSetCase{r04n784mb}{TrackYTick}{1000}
\PaperSetCase{r04n784mb}{TrackClosestY}{994}
\PaperSetCase{r04n784mb}{InsetVMin}{323}
\PaperSetCase{r04n784mb}{InsetVMax}{340}
\PaperSetCase{r04n784mb}{InsetHMin}{2014}
\PaperSetCase{r04n784mb}{InsetHMax}{2056}
\PaperSetCase{r04n784mb}{InsetHRefT}{0.36}
\PaperSetCase{r04n784mb}{InsetHRef}{2034.8}
\PaperSetCase{r06n373df}{TrackXMin}{-1970}
\PaperSetCase{r06n373df}{TrackXMax}{2068}
\PaperSetCase{r06n373df}{TrackYMin}{-26}
\PaperSetCase{r06n373df}{TrackYMax}{240}
\PaperSetCase{r06n373df}{TrackXTickLo}{-1500}
\PaperSetCase{r06n373df}{TrackXTickHi}{1500}
\PaperSetCase{r06n373df}{TrackYTickLo}{0}
\PaperSetCase{r06n373df}{TrackYTickHi}{200}
\PaperSetCase{r06n373df}{TrackYTick}{200}
\PaperSetCase{r06n373df}{TrackClosestY}{203}
\PaperSetCase{r06n373df}{InsetVMin}{324}
\PaperSetCase{r06n373df}{InsetVMax}{335}
\PaperSetCase{r06n373df}{InsetHMin}{905}
\PaperSetCase{r06n373df}{InsetHMax}{932}
\PaperSetCase{r06n373df}{InsetHRefT}{-0.33}
\PaperSetCase{r06n373df}{InsetHRef}{918.8}
\PaperSetCase{r20n394fx}{TrackXMin}{-2433}
\PaperSetCase{r20n394fx}{TrackXMax}{4314}
\PaperSetCase{r20n394fx}{TrackYMin}{-47}
\PaperSetCase{r20n394fx}{TrackYMax}{436}
\PaperSetCase{r20n394fx}{TrackXTickLo}{-1500}
\PaperSetCase{r20n394fx}{TrackXTickHi}{3000}
\PaperSetCase{r20n394fx}{TrackYTickLo}{0}
\PaperSetCase{r20n394fx}{TrackYTickHi}{400}
\PaperSetCase{r20n394fx}{TrackYTick}{400}
\PaperSetCase{r20n394fx}{TrackClosestY}{389}
\PaperSetCase{r20n394fx}{InsetVMin}{447}
\PaperSetCase{r20n394fx}{InsetVMax}{492}
\PaperSetCase{r20n394fx}{InsetHMin}{1690}
\PaperSetCase{r20n394fx}{InsetHMax}{1732}
\PaperSetCase{r20n394fx}{InsetHRefT}{-0.05}
\PaperSetCase{r20n394fx}{InsetHRef}{1714.8}
\PaperSetCase{r05n413ak}{TrackXMin}{-3934}
\PaperSetCase{r05n413ak}{TrackXMax}{5331}
\PaperSetCase{r05n413ak}{TrackYMin}{-64}
\PaperSetCase{r05n413ak}{TrackYMax}{598}
\PaperSetCase{r05n413ak}{TrackXTickLo}{-3000}
\PaperSetCase{r05n413ak}{TrackXTickHi}{4000}
\PaperSetCase{r05n413ak}{TrackYTickLo}{0}
\PaperSetCase{r05n413ak}{TrackYTickHi}{300}
\PaperSetCase{r05n413ak}{TrackYTick}{300}
\PaperSetCase{r05n413ak}{TrackClosestY}{304}
\PaperSetCase{r05n413ak}{InsetVMin}{217}
\PaperSetCase{r05n413ak}{InsetVMax}{256}
\PaperSetCase{r05n413ak}{InsetHMin}{1007}
\PaperSetCase{r05n413ak}{InsetHMax}{1593}
\PaperSetCase{r05n413ak}{InsetHRefT}{1.71}
\PaperSetCase{r05n413ak}{InsetHRef}{1297.2}
\PaperSetCase{r11n60rs}{TrackXMin}{-4364}
\PaperSetCase{r11n60rs}{TrackXMax}{3074}
\PaperSetCase{r11n60rs}{TrackYMin}{-303}
\PaperSetCase{r11n60rs}{TrackYMax}{2827}
\PaperSetCase{r11n60rs}{TrackXTickLo}{-3000}
\PaperSetCase{r11n60rs}{TrackXTickHi}{2000}
\PaperSetCase{r11n60rs}{TrackYTickLo}{0}
\PaperSetCase{r11n60rs}{TrackYTickHi}{2000}
\PaperSetCase{r11n60rs}{TrackYTick}{2000}
\PaperSetCase{r11n60rs}{TrackClosestY}{2469}
\PaperSetCase{r11n60rs}{InsetVMin}{203}
\PaperSetCase{r11n60rs}{InsetVMax}{343}
\PaperSetCase{r11n60rs}{InsetHMin}{1342}
\PaperSetCase{r11n60rs}{InsetHMax}{1404}
\PaperSetCase{r11n60rs}{InsetHRefT}{-1.24}
\PaperSetCase{r11n60rs}{InsetHRef}{1356.8}
\PaperSetCase{r24n9658k}{TrackXMin}{-3166}
\PaperSetCase{r24n9658k}{TrackXMax}{2289}
\PaperSetCase{r24n9658k}{TrackYMin}{-107}
\PaperSetCase{r24n9658k}{TrackYMax}{995}
\PaperSetCase{r24n9658k}{TrackXTickLo}{-2500}
\PaperSetCase{r24n9658k}{TrackXTickHi}{1500}
\PaperSetCase{r24n9658k}{TrackYTickLo}{0}
\PaperSetCase{r24n9658k}{TrackYTickHi}{500}
\PaperSetCase{r24n9658k}{TrackYTick}{500}
\PaperSetCase{r24n9658k}{TrackClosestY}{501}
\PaperSetCase{r24n9658k}{InsetVMin}{154}
\PaperSetCase{r24n9658k}{InsetVMax}{208}
\PaperSetCase{r24n9658k}{InsetHMin}{434}
\PaperSetCase{r24n9658k}{InsetHMax}{774}
\PaperSetCase{r24n9658k}{InsetHRefT}{2.29}
\PaperSetCase{r24n9658k}{InsetHRef}{632.5}
\newcommand{\PaperSmallEndpointMarkers}[1]{%
  \ifcsname paper#1FMax\endcsname
    \addplot[black!58,densely dashed,line width=0.55pt,forget plot]
      coordinates {(\pcv{#1}{XMin},\pcv{#1}{FMax})
                   (\pcv{#1}{FMaxX},\pcv{#1}{FMax})};
    \addplot[black!58,densely dashed,line width=0.55pt,forget plot]
      coordinates {(\pcv{#1}{FMinX},\pcv{#1}{FMin})
                   (\pcv{#1}{XMax},\pcv{#1}{FMin})};
    \node[anchor=south west,font=\fontsize{5.5}{5.8}\selectfont,
          fill=white,inner sep=0.35pt,yshift=4.5pt]
      at (axis cs:\pcv{#1}{XMin},\pcv{#1}{FMax}) {$f_{\max}$};
    \ifcsname paper#1FMinLabelLeft\endcsname
      \node[anchor=south west,font=\fontsize{5.5}{5.8}\selectfont,
            fill=none,inner sep=0.35pt,yshift=1.0pt]
        at (axis cs:\pcv{#1}{XMin},\pcv{#1}{FMin}) {$f_{\min}$};
    \else
      \node[anchor=north east,font=\fontsize{5.5}{5.8}\selectfont,
            fill=white,inner sep=0.35pt]
        at (axis cs:\pcv{#1}{XMax},\pcv{#1}{FMin}) {$f_{\min}$};
    \fi
  \fi
}

\newif\ifPaperShowMainResultBox
\PaperShowMainResultBoxtrue

\newcommand{\PaperLegendDot}[1]{%
  \tikz[baseline=-0.55ex]\fill[#1] (0,0) circle[radius=1.25pt];}
\newcommand{\PaperLegendLine}[2]{%
  \tikz[baseline=-0.55ex]\draw[#1,line width=#2] (0,0) -- (0.55,0);}
\newcommand{\PaperLegendDashedLine}{%
  \tikz[baseline=-0.55ex]\draw[black!55,densely dashed,line width=0.8pt] (0,0) -- (0.55,0);}
\newcommand{\PaperLegend}{%
  {\scriptsize
  \renewcommand{\arraystretch}{1.12}%
  \setlength{\tabcolsep}{0.62em}%
  \begin{tabular}{@{}c@{\hspace{1.15em}}c@{\hspace{1.15em}}c@{}}
    \PaperLegendDot{black}\,observed sound ridge
      & \PaperLegendLine{mainred}{1.2pt}\,Method I
      & \PaperLegendLine{methodiigreen}{1.45pt}\,Method II\\
    \PaperLegendDashedLine\,$\tau=0$: $t_{pc,\mathrm{II}}$
      & \PaperLegendLine{mainorange}{1.0pt}\,ADS-B $V$
      & \PaperLegendLine{adsviolet}{1.0pt}\,ADS-B $H$
  \end{tabular}}%
}

\newcommand{\PaperAdsInset}[3]{%
  \begin{tikzpicture}
    \begin{axis}[
      width=#2,
      height=#3,
      scale only axis,
      axis background/.style={fill=white,draw=none,
        fill opacity=\PaperInsetBackgroundOpacity},
      xmin=\pcv{#1}{InsetXMin},
      xmax=\pcv{#1}{InsetXMax},
      ymin=\pcv{#1}{InsetVMin},
      ymax=\pcv{#1}{InsetVMax},
      axis y line*=left,
      axis x line*=bottom,
      xlabel={$t(\mathrm{s})$},
      ylabel={$V(\mathrm{km/h})$},
      xlabel style={font=\paperinsetfont,fill=white,draw=none,inner sep=0.1pt},
      ylabel style={font=\paperinsetfont,text=mainorange!85!black,fill=white,draw=none,inner sep=0.1pt},
      clip=true,
      grid=both,
      grid style={gray!14},
      tick label style={font=\paperinsetfont,fill=white,draw=none,inner sep=0.1pt},
      xtick={\pcv{#1}{InsetXTickLo},0,\pcv{#1}{InsetXTickHi}},
      ytick={\pcv{#1}{InsetVMin},\pcv{#1}{InsetVMax}},
      scaled ticks=false,
      tick align=outside,
      major tick length=1.1pt,
      axis line style={black!55},
      tick style={black!55},
      xticklabel style={font=\paperinsetfont,/pgf/number format/1000 sep={}},
      yticklabel style={text=mainorange!85!black,font=\paperinsetfont,/pgf/number format/1000 sep={}}]
      \addplot[mainorange,line width=0.85pt]
        table[x=time_rel_s,y=speed_kmh,col sep=comma]
        {\posterroot/paper_data/all_cases/#1_ads.csv};
      \coordinate (paperadstitle) at (rel axis cs:\PaperAdsTitleX,\PaperAdsTitleY);
    \end{axis}
    \begin{axis}[
      width=#2,
      height=#3,
      scale only axis,
      axis background/.style={fill=none,draw=none},
      xmin=\pcv{#1}{InsetXMin},
      xmax=\pcv{#1}{InsetXMax},
      ymin=\pcv{#1}{InsetHMin},
      ymax=\pcv{#1}{InsetHMax},
      axis y line*=right,
      axis x line=none,
      ylabel={$H(\mathrm{m})$},
      ylabel style={font=\paperinsetfont,text=adsviolet,fill=white,draw=none,inner sep=0.1pt},
      clip=true,
      tick label style={font=\paperinsetfont,fill=white,draw=none,inner sep=0.1pt},
      ytick={\pcv{#1}{InsetHMin},\pcv{#1}{InsetHMax}},
      scaled ticks=false,
      tick align=outside,
      major tick length=1.1pt,
      axis line style={black!55},
      tick style={black!55},
      yticklabel style={text=adsviolet,font=\paperinsetfont,/pgf/number format/1000 sep={}}]
      \addplot[adsviolet,line width=0.85pt]
        table[x=time_rel_s,y=height_m,col sep=comma]
        {\posterroot/paper_data/all_cases/#1_ads.csv};
      \ifcsname paper#1InsetHRef\endcsname
      \addplot[only marks,mark=*,mark size=1.3pt,draw=adsviolet,fill=adsviolet]
        coordinates {(\pcv{#1}{InsetHRefT},\pcv{#1}{InsetHRef})};
      \fi
    \end{axis}
    \node[anchor=north,font=\paperinsetfont,paper inset title background]
      at (paperadstitle) {ADS-B};
  \end{tikzpicture}%
}

\newcommand{\PaperTrackInset}[3]{%
  \begin{tikzpicture}
    \begin{axis}[
      width=#2,
      height=#3,
      scale only axis,
      axis background/.style={fill=white,draw=none,
        fill opacity=\PaperInsetBackgroundOpacity},
      xmin=\pcv{#1}{TrackXMin},
      xmax=\pcv{#1}{TrackXMax},
      ymin=\pcv{#1}{TrackYMin},
      ymax=\pcv{#1}{TrackYMax},
      clip=true,
      ylabel={$y(\mathrm{m})$},
      ylabel style={font=\papertrackfont,fill=white,draw=none,inner sep=0.1pt},
      grid=both,
      grid style={gray!14},
      tick label style={font=\papertrackfont,fill=white,draw=none,inner sep=0.1pt},
      xtick={\pcv{#1}{TrackXTickLo},\pcv{#1}{TrackXTickHi}},
      extra x ticks={({\pcv{#1}{TrackXTickLo}+\pcv{#1}{TrackXTickHi}})/2},
      extra x tick labels={$x(\mathrm{m})$},
      extra x tick style={grid=none,tick style={draw=none},
        major tick length=0pt,ticklabel style={font=\papertrackfont,fill=white,draw=none,inner sep=0.1pt}},
      ytick={\pcv{#1}{TrackYTickLo},\pcv{#1}{TrackYTickHi}},
      scaled ticks=false,
      tick align=outside,
      major tick length=1.0pt,
      axis line style={black!55},
      tick style={black!55},
      xticklabel style={font=\papertrackfont,/pgf/number format/1000 sep={}},
      yticklabel style={font=\papertrackfont,/pgf/number format/1000 sep={}}]
      \addplot[black!35,densely dashed,line width=0.45pt]
        coordinates {(0,0) (0,\pcv{#1}{TrackClosestY})};
      \addplot[trackblue,line width=0.70pt]
        table[x=x_m,y=y_m,col sep=comma]
        {\posterroot/paper_data/all_cases/#1_track.csv};
      \IfFileExists{\posterroot/paper_data/all_cases/#1_track_samples.csv}{%
      \addplot[only marks,mark=*,mark size=0.45pt,draw=trackblue,fill=trackblue]
        table[x=x_m,y=y_m,col sep=comma]
        {\posterroot/paper_data/all_cases/#1_track_samples.csv};}{%
      \addplot[only marks,mark=*,mark size=0.45pt,draw=trackblue,fill=trackblue]
        table[x=x_m,y=y_m,col sep=comma]
        {\posterroot/paper_data/all_cases/#1_track.csv};}
      \addplot[only marks,mark=o,mark size=1.25pt,draw=black,line width=0.40pt]
        coordinates {(0,\pcv{#1}{TrackClosestY})};
      \node[font=\bfseries\fontsize{7}{7}\selectfont,anchor=center]
        at (axis cs:0,0) {*};
      \coordinate (papertracktitle) at (rel axis cs:\pcv{#1}{TrackTitleX},\pcv{#1}{TrackTitleY});
    \end{axis}
    \node[anchor=center,font=\papertrackfont,paper inset title background]
      at (papertracktitle) {\PaperTrackTitleText};
  \end{tikzpicture}%
}

\newcommand{\PaperSmallAudioPlotDefault}[1]{%
      \addplot[paper small audio default]
        table[x=time_rel_s,y=frequency_hz,col sep=comma]
        {\posterroot/paper_data/all_cases/#1_ridge.csv};}
\newcommand{\PaperSmallAudioPlotContrast}[1]{%
      \addplot[paper small audio contrast]
        table[x=time_rel_s,y=frequency_hz,col sep=comma]
        {\posterroot/paper_data/all_cases/#1_ridge.csv};}

\newcommand{\PaperSmallCasePlot}[2]{%
  \PaperChooseSmallAudioStyle{#1}%
  \begin{tikzpicture}
    \begin{axis}[
      width=\PaperSmallPlotWidth,
      height=\PaperSmallPlotHeight,
      xmin=\pcv{#1}{XMin},
      xmax=\pcv{#1}{XMax},
      ymin=\pcv{#1}{YMin},
      ymax=\pcv{#1}{YMax},
      xlabel={$\tau$ (s)},
      ylabel={$f$ (Hz)},
      xlabel style={font=\scriptsize,at={(axis description cs:\PaperSmallXLabelX,\PaperSmallXLabelY)},anchor=north},
      ylabel style={font=\scriptsize,at={(axis description cs:\PaperSmallYLabelX,\PaperSmallYLabelY)},anchor=south},
      xtick={\pcv{#1}{XStartTick},0,\pcv{#1}{XEndTick}},
      ytick={\pcv{#1}{FMin},\pcv{#1}{Fzero},\pcv{#1}{FMax}},
      yticklabels={\PaperFrequencyTick{\pcv{#1}{FMin}},,\PaperFrequencyTick{\pcv{#1}{FMax}}},
      grid=both,
      grid style={gray!14},
      tick label style={font=\scriptsize},
      label style={font=\scriptsize},
      tick align=outside,
      major tick length=1.6pt,
      axis line style={black!65},
      tick style={black!65},
      clip=false]
      \ifPaperUseContrastAudio\else
        \PaperSmallAudioPlotDefault{#1}
      \fi
      \addplot[mainred,line width=0.95pt]
        table[x=time_rel_s,y=endpoint_hz,col sep=comma]
        {\posterroot/paper_data/all_cases/#1_fit.csv};
      \ifPaperUseContrastAudio
        \addplot[paper small method ii contrast]
          table[x=time_rel_s,y=full_fit_hz,col sep=comma]
          {\posterroot/paper_data/all_cases/#1_fit.csv};
      \else
        \addplot[paper small method ii default]
          table[x=time_rel_s,y=full_fit_hz,col sep=comma]
          {\posterroot/paper_data/all_cases/#1_fit.csv};
      \fi
      \addplot[black!50,densely dashed,line width=0.60pt,forget plot]
        coordinates {(0,\pcv{#1}{YMin}) (0,\pcv{#1}{YMax})};
      \addplot[black!42,densely dashed,line width=0.60pt,forget plot]
        coordinates {(\pcv{#1}{XMin},\pcv{#1}{Fzero}) (\pcv{#1}{XMax},\pcv{#1}{Fzero})};
      \PaperFzeroMarker{#1}{\PaperSmallFzeroLabelRelX}
      \PaperSmallEndpointMarkers{#1}
      \ifPaperUseContrastAudio
        \PaperSmallAudioPlotContrast{#1}
      \fi
      \coordinate (adsne) at (rel axis cs:\PaperCaseSmallAdsInsetX{#1},\PaperCaseSmallAdsInsetY{#1});
      \coordinate (trackne) at (rel axis cs:\PaperCaseSmallTrackInsetX{#1},\PaperCaseSmallTrackInsetY{#1});
    \end{axis}
    \node[anchor=north east,fill=none,text opacity=1,inner sep=0.8pt]
      at (adsne) {\def\PaperInsetBackgroundOpacity{0.90}%
        \PaperAdsInset{#1}{\PaperSmallAdsInsetWidth}{\PaperSmallAdsInsetHeight}};
    \node[anchor=north east,fill=none,text opacity=1,inner sep=0.6pt]
      at (trackne) {\def\PaperInsetBackgroundOpacity{0.76}%
        \PaperTrackInset{#1}{\PaperSmallTrackInsetWidth}{\PaperSmallTrackInsetHeight}};
  \end{tikzpicture}%
}

\title{Aircraft speed and height from Doppler-shifted sound and flight-track data}
\author{%
Praveen Pathak$^{1,2}$, David Rakestraw$^2$, and Vijay Singh$^3$\\[3pt]
\small $^1$Homi Bhabha Centre for Science Education--TIFR, Mumbai, India\\
\small $^2$Lawrence Livermore National Laboratory, Livermore, CA, USA\\
\small $^3$UM--DAE Centre for Excellence in Basic Sciences, University of Mumbai,\\
\small Kalina Campus, Santacruz (East), Mumbai 400098, India}
\date{}

\begin{document}
\maketitle

The changing pitch of a passing aircraft is a familiar example of the Doppler effect. We use audio recordings of 40 aircraft passages to estimate the speed and height by two methods based on the Doppler effect. The endpoint method is an analytical model that uses the limiting frequencies at the start and end of each recording and the slope near the middle of the frequency–time curve. The second method fits the whole curve. Results from both methods are compared with independent flight-tracking data. The endpoint method consistently underestimates speed and height because the recorded frequencies have not approached their limiting values. The full curve fit agrees with the tracking data within
about 3\% in speed and 12\% in height for steady passages, with poorer agreement when speed
or height changes. Software
developed with AI assistance made it practical to analyze this many passages. The comparison
offers undergraduate students an example of testing model assumptions and identifying
systematic errors in real measurements.

\section{Introduction}
The Doppler effect has long been used in physics teaching because students can hear the change in pitch directly. It appears in school-level derivations \cite{morehouse1997,hoyt2002},
pedagogical models \cite{kaura2017}, classroom demonstrations, and quantitative measurement activities
\cite{gagne1996,schier2011,torres2006}.
Smartphone audio has made such measurements increasingly accessible
\cite{soares2021,pathak2020,staacks2018}.

Aircraft provide a more challenging real-world example. As an aircraft passes, its changing pitch can be followed quantitatively. However, the source is distant, its path rarely passes directly overhead, and its sound may contain several harmonics. During a recording, a harmonic may weaken or become obscured by background noise. We therefore select one harmonic whose frequency can be followed over time in the spectrogram. Costa and Mocellin showed
that aircraft speed can be estimated from the approaching and receding
frequencies visible in an FFT spectrogram of recorded aircraft noise
\cite{costa2007}. More technical passive-acoustic studies have estimated
aircraft speed and altitude and reconstructed aircraft trajectories
\cite{ferguson1992,martin2013}.

\revb{For teaching, the endpoint approach of Costa and Mocellin is attractive. Adding the
slope of the frequency curve near closest approach gives the closest distance, and hence
the height.\cite{usapho2016,inpho2026} We have used this approach as a 30-minute exercise in a workshop for university
and college physics instructors. Such an exercise rests on a single
passage. It offers no independent check of the estimates and cannot show how changes in
aircraft speed or height during the passage affect them. Testing the approach requires
many passages and independent knowledge of the aircraft motion.}

Flight-track data \revb{provide this independent knowledge and} are publicly available from services such as
Flightradar24 and ADS-B Exchange \cite{fr24,adsbexchange}. Automatic Dependent
Surveillance--Broadcast (ADS-B) messages provide aircraft position, ground
speed, and altitude at discrete times. This study combines phone recordings of 40  aircraft passages with
corresponding flight-track data to examine two methods for estimating aircraft motion.
Method~I, the endpoint method, uses endpoint frequencies and the slope near closest approach, while Method~II
fits the selected frequency--time curve and accounts for sound-travel delay. Both
methods estimate speed and minimum slant distance. Flight-track data supply the
horizontal offset needed to calculate height and provide reference values for
comparison. We examine how incomplete frequency curves, flight geometry, and departures
from steady motion affect the estimates. We also discuss uncertainties associated with
frequency extraction, sound propagation, and the reference data. The analysis shows how well the two methods work with real measurements and what limits their accuracy.

\revb{Reading frequencies from a phone display, as in the workshop, becomes impractical for
40 passages. We therefore used AI assistants to write a browser-based tool that extracts
the frequency--time curve from a recording, and code that fits the curve. These tools have
not yet been tested with the teachers and students. However, with current AI assistants, we expect
these steps to be within reach of instructors and students. The prompts we used are given
in the supplementary material.}
\section{Theoretical Model}
\label{sec:theory}

We model the aircraft as a sound source that emits a constant frequency \(f_0\) while moving at constant speed \(V\) along a straight path at approximately constant height \(H\) above the ground at the phone location.

\begin{figure}[h]
  \centering
  \includegraphics[width=0.54\linewidth]{\posterroot/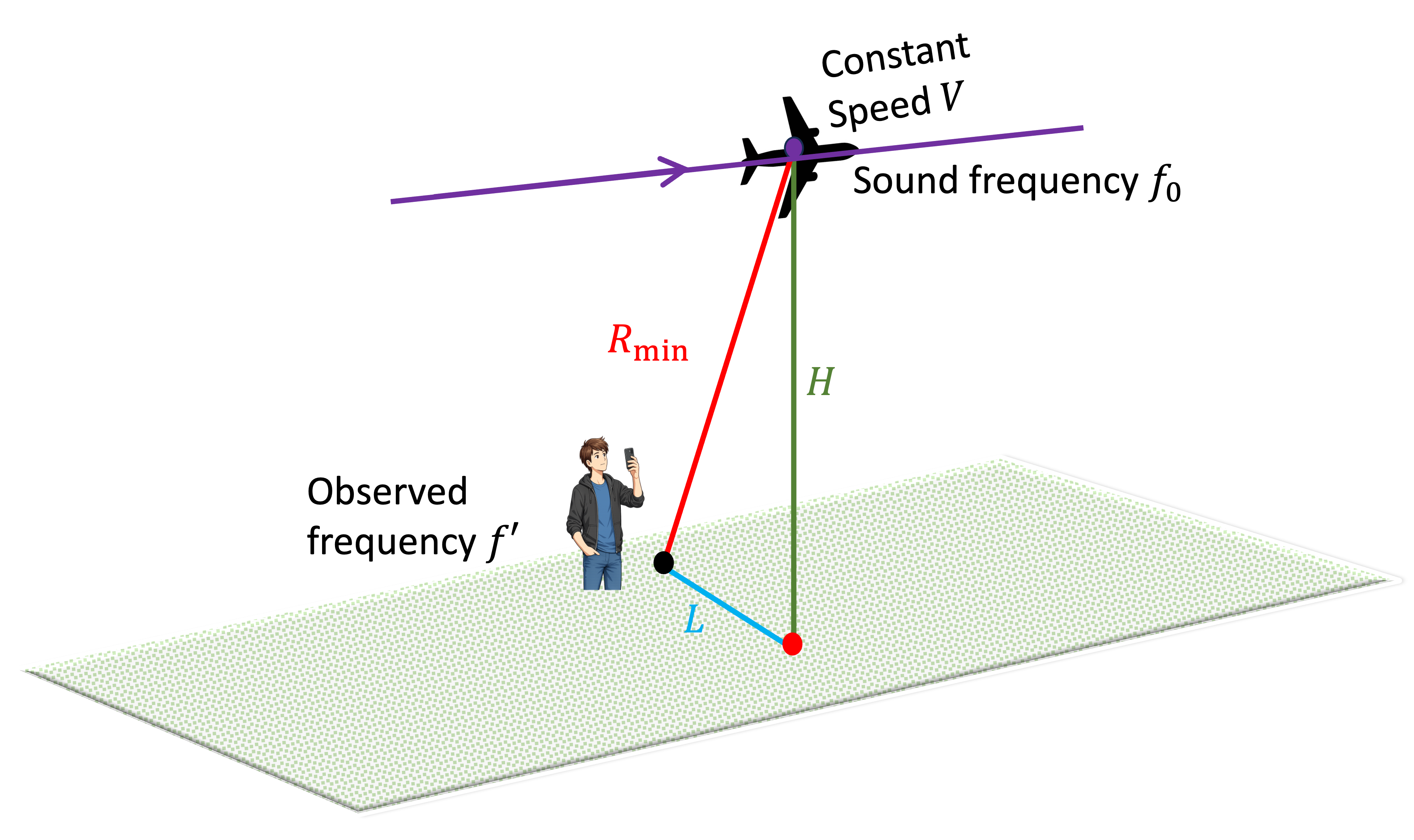}
  \caption{Geometry of the idealized aircraft flyby. The figure
  defines the phone position, the aircraft path, the minimum slant distance
  $R_{\min}$, the ground-track offset $L$, and the aircraft height $H$ at
  closest approach.}
  \label{fig:setup}
\end{figure}

 Here $f_0$ is the unshifted frequency of the selected harmonic. Figure~\ref{fig:setup} shows the idealized
geometry for this straight, constant-height passage.
Let $R_{\min}$ be the minimum slant
distance between the aircraft and the phone, and $L$ the minimum horizontal distance
between the phone and the aircraft's ground track. $V_r$ is the radial velocity
relative to the phone, taken as positive when the aircraft moves away. The
speed of sound is denoted by $\cs$. The time symbols used below are:
\begin{align*}
t_e: & \text{time at which a sound wave is emitted by the aircraft} \\
t_p: & \text{time at which that sound wave is received by the phone} 
\end{align*}

Let $R_e$ denote the aircraft--phone distance when the sound is emitted at
time $t_e$. The subscript $c$ denotes the closest-approach event. At emission time
$t_e=t_{ec}$, $R_e=R_{\min}$ and $V_r=0$. The sound emitted then reaches the
phone later, at $t_p=t_{pc}$.

For a straight, constant-speed passage,
\begin{align}
R_e &= \left[R_{\min}^2+V^2(t_e-t_{ec})^2\right]^{1/2}, \label{eq:Re}\\
V_r &= \frac{dR_e}{dt_e}
     = \frac{V^2(t_e-t_{ec})}{R_e}. \label{eq:Vr}
\end{align}
With this sign convention, \(V_r\) is negative when the aircraft is approaching the observer and positive when it is receding, and is evaluated at the emission time \(t_e\). The frequency received by the phone is
\begin{equation}
f(t_p)=\frac{f_0\cs}{\cs+V_r} .
\label{eq:doppler}
\end{equation}

\begin{figure}[h]
  \centering
  \begin{tikzpicture}
    \begin{axis}[
      width=0.72\linewidth,
      height=0.40\linewidth,
      xmin=-105,xmax=105,
      ymin=235,ymax=470,
      xlabel={$t_p-t_{pc}$},
      xlabel style={at={(axis description cs:1,0)},anchor=north east},
      ylabel={Received frequency, $f$},
      axis line style={line width=0.6pt},
       tick style={draw=none},
       xtick={0},
       xticklabels={$0$},
       ytick=\empty,
      grid=none]
     % \fill[mainblue!8] (axis cs:-105,235) rectangle (axis cs:0,470);
      %\fill[mainorange!9] (axis cs:0,235) rectangle (axis cs:105,470);
      \addplot[gray!70,densely dashed,line width=0.8pt]
        coordinates {(-105,350)(105,350)};
      \addplot[mainblue,line width=1.5pt]
        table[x=time_s,y=frequency_hz,col sep=comma]
        {\posterroot/paper_data/fig2_ideal_trace.csv};
       \addplot[black,dashed,line width=0.8pt]
         coordinates {(0,235)(0,470)};
       \node[anchor=west] at (axis cs:-101,353) {$f_0$};
       \node[anchor=south east] at (axis cs:-78,447) {$f_{\max}$};
       \node[anchor=north west] at (axis cs:78,295) {$f_{\min}$};
       \node[anchor=west] at (axis cs:3,450) {$t_{pc}$};
      \node[anchor=south] at (axis cs:-43,300) {approach};
      \node[anchor=south] at (axis cs:43,305) {recession};
    \end{axis}
  \end{tikzpicture}
  \caption{Ideal received frequency--time curve for a straight,
  constant-speed aircraft passage. The dashed line marks $t_p=t_{pc}$, where
  $V_r=0$ and $f=f_0$.}
  \label{fig:ideal-ft}
\end{figure}
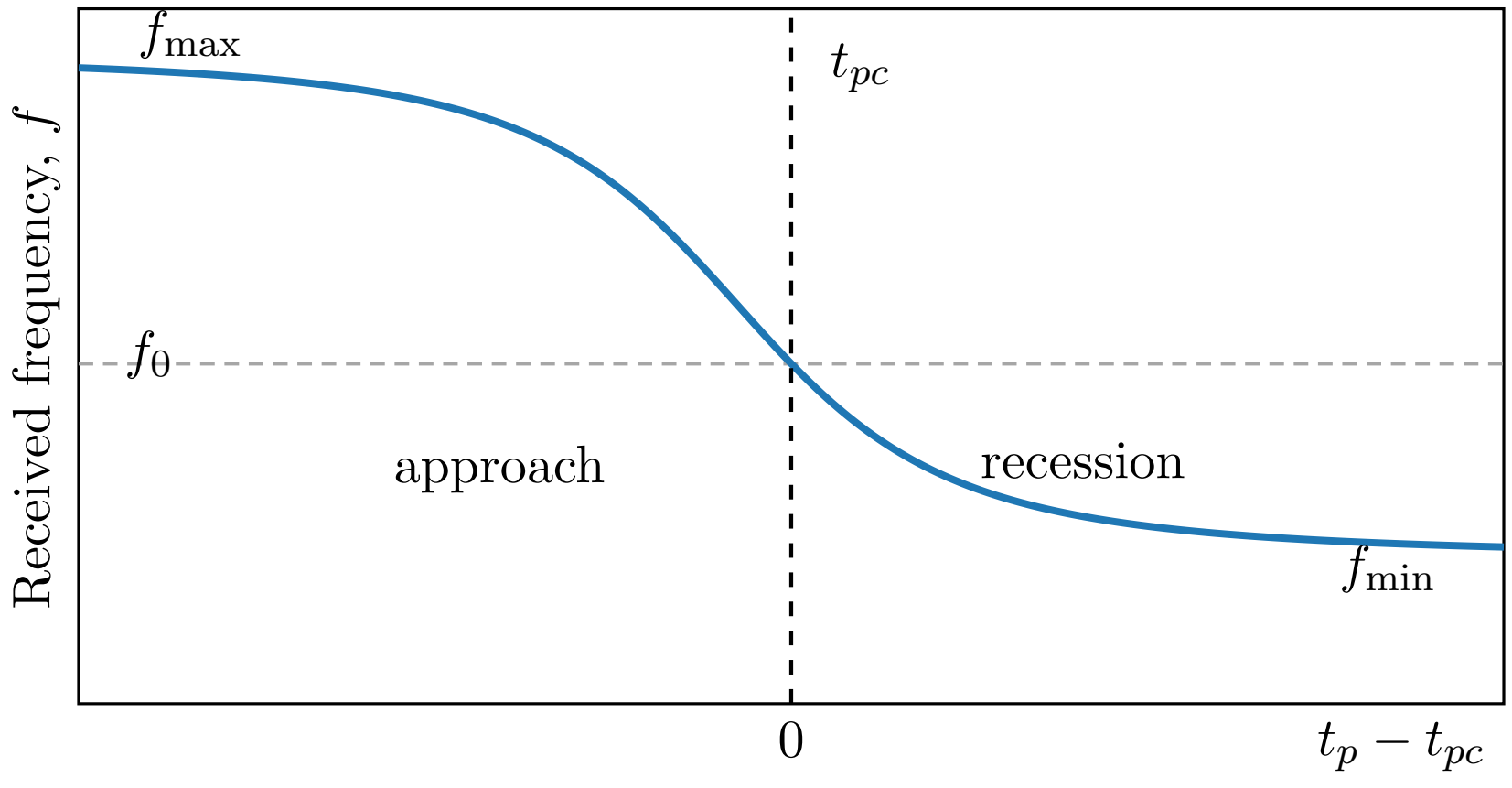

The corresponding phone time includes the propagation delay:
\begin{equation}
t_p=t_e+\frac{R_e}{\cs}.
\label{eq:delay}
\end{equation}
The relation between the two closest-approach times is
\begin{equation}
t_{pc}=t_{ec}+\frac{R_{\min}}{\cs}.
\label{eq:tpc}
\end{equation} 
 Figure~\ref{fig:ideal-ft} shows the ideal frequency--time curve
predicted by this model, plotted against shifted phone time $t_p-t_{pc}$.
At $t_p=t_{pc}$, $f(t_{pc})=f_0$.

There are two ways to obtain the aircraft speed $V$ and minimum slant distance
$R_{\min}$ from the received frequency--time data.

\subsection{Method I: endpoint method}
This method is especially suitable for a classroom activity because it uses only a few
readings from the graph. Far before closest approach, the line of sight is almost
along the flight path and the aircraft is approaching the phone, so

\[
V_r \approx -V.
\]

Far after closest approach, the line of sight is again almost along the flight
path, but the aircraft is receding, so

\[
V_r \approx +V.
\]
Thus, this method assumes that the radial speed at the two
endpoints is approximately the full aircraft speed. Equation~\eqref{eq:doppler}
then gives two frequencies, $f_{\max}$ and $f_{\min}$, which represent the
approach and recession endpoint frequencies:
\begin{equation}
f_{\max}=\frac{f_0\cs}{\cs-V}, \qquad
f_{\min}=\frac{f_0\cs}{\cs+V}.
\label{eq:endpoints}
\end{equation}

The endpoint frequencies are indicated in Fig.~\ref{fig:ideal-ft}. The two
endpoint relations contain the two unknowns $f_0$ and $V$. Let
$q=f_{\max}/f_{\min}$. The speed is then
\begin{equation}
V=\cs\frac{q-1}{q+1}.
\label{eq:endpoint-speed}
\end{equation}

The corresponding emitted frequency is
\begin{equation}
f_0=f_{\max}\frac{\cs-V}{\cs}.
\label{eq:endpoint-f0}
\end{equation}
To obtain the minimum slant distance, differentiate the observed frequency with
respect to $t_p$. Because $V_r$ is a function of $t_e$, the chain rule gives
\begin{equation}
\frac{df}{dt_p}
=-\frac{f_0\cs}{(\cs+V_r)^2}
  \frac{dV_r}{dt_e}\frac{dt_e}{dt_p}.
\label{eq:frequency-slope}
\end{equation}
The derivative of the radial velocity follows from Eq.~\eqref{eq:Vr}:
\begin{equation}
\frac{dV_r}{dt_e}
=\frac{V^2}{R_e}-\frac{V^2(t_e-t_{ec})}{R_e^2}\frac{dR_e}{dt_e}.
\label{eq:radial-acceleration}
\end{equation}
At closest approach,
\begin{equation}
\left.\frac{dV_r}{dt_e}\right|_{t_{ec}}=\frac{V^2}{R_{\min}}.
\label{eq:radial-acceleration-closest}
\end{equation}
Differentiating Eq.~\eqref{eq:delay} gives
\[
\frac{dt_p}{dt_e}=1+\frac{V_r}{\cs}.
\]
At closest approach, $dt_p/dt_e=1$, and therefore $dt_e/dt_p=1$.
Substituting these results into Eq.~\eqref{eq:frequency-slope} gives the slope measured directly from the
phone-time frequency plot:
\begin{equation}
\left.\frac{df}{dt_p}\right|_{t_{pc}}
=-\frac{f_0V^2}{\cs R_{\min}}.
\label{eq:slope}
\end{equation}

Equations~\eqref{eq:endpoint-speed}, \eqref{eq:endpoint-f0}, and
\eqref{eq:slope} give estimates of the speed and minimum slant distance. The
values $f_{\max}$ and $f_{\min}$ are read from the nearly flat regions of the
frequency--time graph. The point where the curve crosses $f_0$ and the local
slope near that point are also read from the graph.

The Method~I curve in Fig.~\ref{fig:evidence-set} is drawn from these estimates
using Eqs.~\eqref{eq:Re}--\eqref{eq:delay}, which include the sound-travel delay.

In practice, the aircraft signal often fades into the noise, or the recording ends, before $|V_r|$ comes close to $V$. The visible endpoint frequencies therefore need not be the limiting frequencies assumed in Eq.~\eqref{eq:endpoints}. At the finite endpoints, the magnitude of the radial velocity is
still smaller than the aircraft speed. Consequently, the measured approach
frequency $f_{\max}$ is lower than its ideal limiting value, while the
measured recession frequency $f_{\min}$ is higher than its ideal limiting
value. Both frequencies therefore lie closer to $f_0$. Thus, the ratio
$q=f_{\max}/f_{\min}$ has a smaller numerator and a larger denominator,
pushing $q$ toward 1. Equation~\eqref{eq:endpoint-speed} then gives a velocity
smaller than the actual aircraft velocity. Since
Eq.~\eqref{eq:slope} makes $R_{\min}$ scale approximately as $V^2$ for a fixed
local slope, a low speed estimate produces a low height estimate as well.

\subsection{Method II: nonlinear fit}
Method II uses a nonlinear fit to the complete received frequency--time data.
The distance $R_e$, radial velocity $V_r$, and received frequency $f$ in
Eqs.~\eqref{eq:Re}--\eqref{eq:doppler} depend on $f_0$, $V$, $R_{\min}$, and
$t_e-t_{ec}$. Subtracting Eq.~\eqref{eq:tpc} from
Eq.~\eqref{eq:delay} gives
\begin{equation}
t_p=t_{pc}+(t_e-t_{ec})+\frac{R_e-R_{\min}}{\cs}.
\label{eq:fit-time}
\end{equation}

Thus, the four fit parameters are $f_0$, $V$, $R_{\min}$, and $t_{pc}$. The
parameter $t_{pc}$ is the phone time at which sound emitted at closest approach is received. For each measured point $(t_{p,i},f_i)$, Eqs.~\eqref{eq:fit-time} and
\eqref{eq:Re} are solved numerically for $t_{e,i}-t_{ec}$. Equations
\eqref{eq:Vr} and \eqref{eq:doppler} then give the predicted received
frequency $\hat f_i$ at phone time $t_{p,i}$. This procedure retains the
sound-travel delay.

The four parameters are determined by an unweighted nonlinear least-squares fit
that minimizes
\begin{equation}
\sum_i\left[f_i-\hat f_i\right]^2.
\label{eq:least-squares}
\end{equation}
The fitted values of $V$ and $R_{\min}$ are reported in the results.

\subsection{Height calculation}
Both methods give the minimum slant distance $R_{\min}$. For a
directly overhead pass, $R_{\min}=H$.  For an aircraft whose ground track
is offset from the phone, the calculation also requires $L$, the minimum horizontal distance
between the phone and the aircraft's ground track. The aircraft height above
the phone location is then, as shown in Fig.~\ref{fig:setup},
\begin{equation}
H=\left(R_{\min}^2-L^2\right)^{1/2}.
\label{eq:height}
\end{equation}
$L$ is determined from the phone coordinates and the ADS-B track. This
calculation is physically meaningful only when $R_{\min}>L$. For a near-overhead
pass, $L\ll H$ and hence $R_{\min}\approx H$. For example, at $L/H=0.3$,
$R_{\min}$ exceeds $H$ by about $4.4\%$.

The recorded passages include overhead, near-overhead, and more strongly offset flight paths. ADS-B
data provide reference speed and height values for comparison with Methods I
and II. Equation~\eqref{eq:height}
works for the straight, constant-height model. During a gradual climb, descent,
or change of direction, the minimum slant distance and minimum horizontal offset
may occur at slightly different positions, making the calculated
height less certain. For these passages, the reported height is an estimate based on the constant-height model.  The ADS-B reference speed is the median speed while the recorded sound was
emitted. The reference height is the height at closest horizontal approach, where $L$ is
measured.

Six cases are presented below to illustrate the effects of incomplete frequency curves, changing aircraft motion, and ground-track geometry. The complete data
set for the 40 analyzed passages is provided with the supplementary materials
\cite{upload}.

\section{Recording and Extracting Frequencies}

\rev{The passages were recorded with a smartphone placed outdoors within about 5~km of
a general-aviation airport. The phone was left recording for a few hours at a time on
six days, about 40 hours in all. The passages were later located in the audio using its
time stamps and a public flight-tracking map. Weekends gave the most passages because
more small aircraft fly then. Most were small propeller aircraft, but turboprops, business jets, a regional airliner, and a
helicopter were also recorded. The aircraft passed at slant distances of 0.3--5.7~km, and
the harmonics used lie between about 100 and 1200~Hz. Instructors near a small airport
can collect similar recordings, and others can use the shared data set.\cite{upload}} Aircraft sound is not a single pure tone. Periodic motion associated with the engine, propeller, or fan can produce repeated pressure variations. When the resulting periodic waveform is not purely sinusoidal, its spectrum contains a fundamental frequency together with harmonics at integer multiples of that frequency. The Doppler shift acts on each component, so the harmonics appear in the spectrogram as several similarly shaped curves. A bright curve showing the frequency of one harmonic over time is called a \emph{ridge}. We select one sufficiently clear ridge, which may correspond to the fundamental or a higher harmonic. Background sound and
wind noise add other features. The analysis therefore uses a spectrogram.
We excluded recordings in which wind, noise, or poor spectral resolution made the ridge too difficult to follow.  Aircraft are identified below by their registration numbers; N784MB is one example. 
\begin{figure}[h]
  \centering
  \begin{subfigure}{0.49\linewidth}
    \centering
    \includegraphics[width=\linewidth]{\posterroot/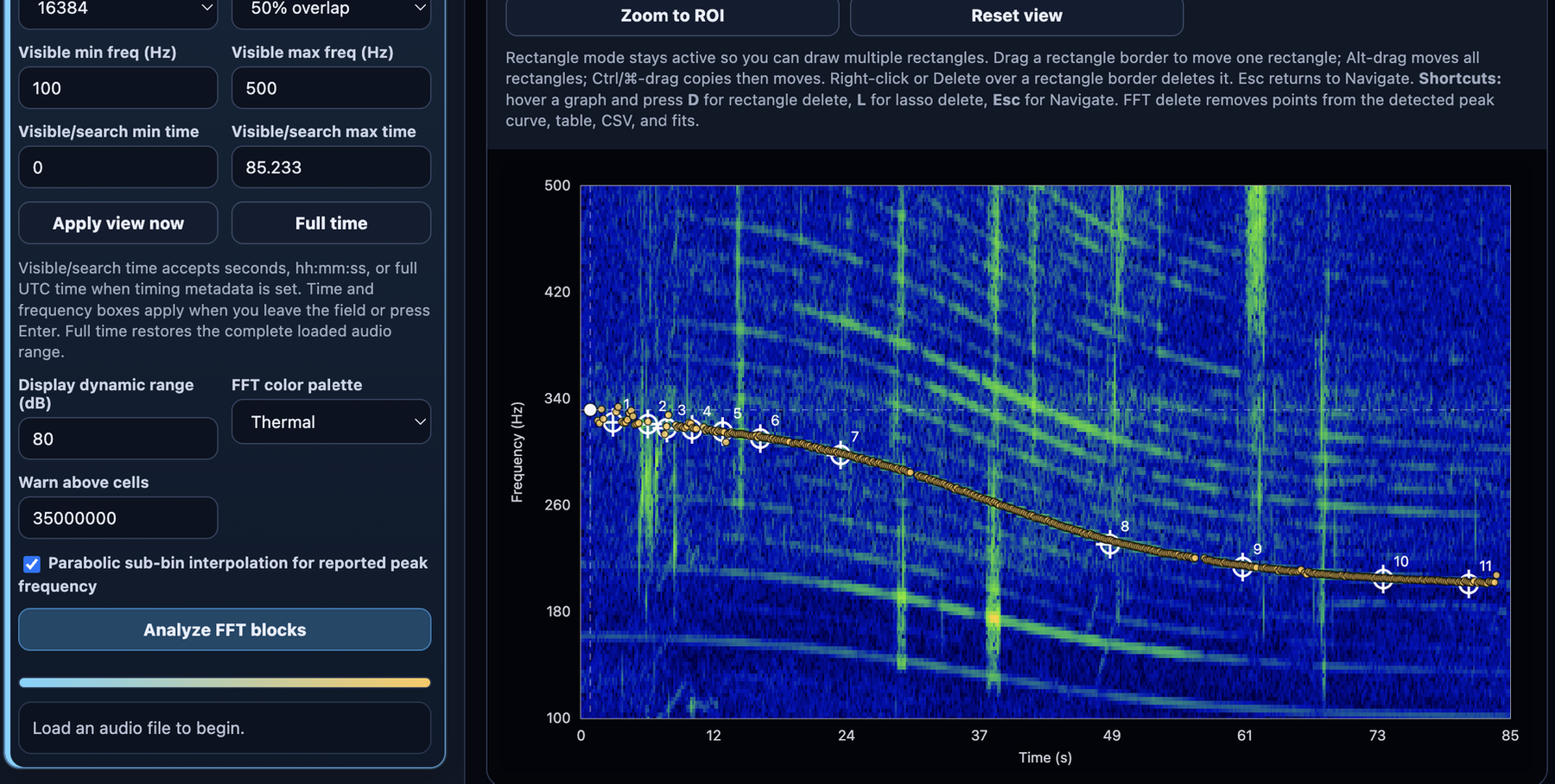}
    \caption*{\scriptsize (a)}
  \end{subfigure}\hfill
  \begin{subfigure}{0.49\linewidth}
    \centering
    \includegraphics[width=\linewidth]{\posterroot/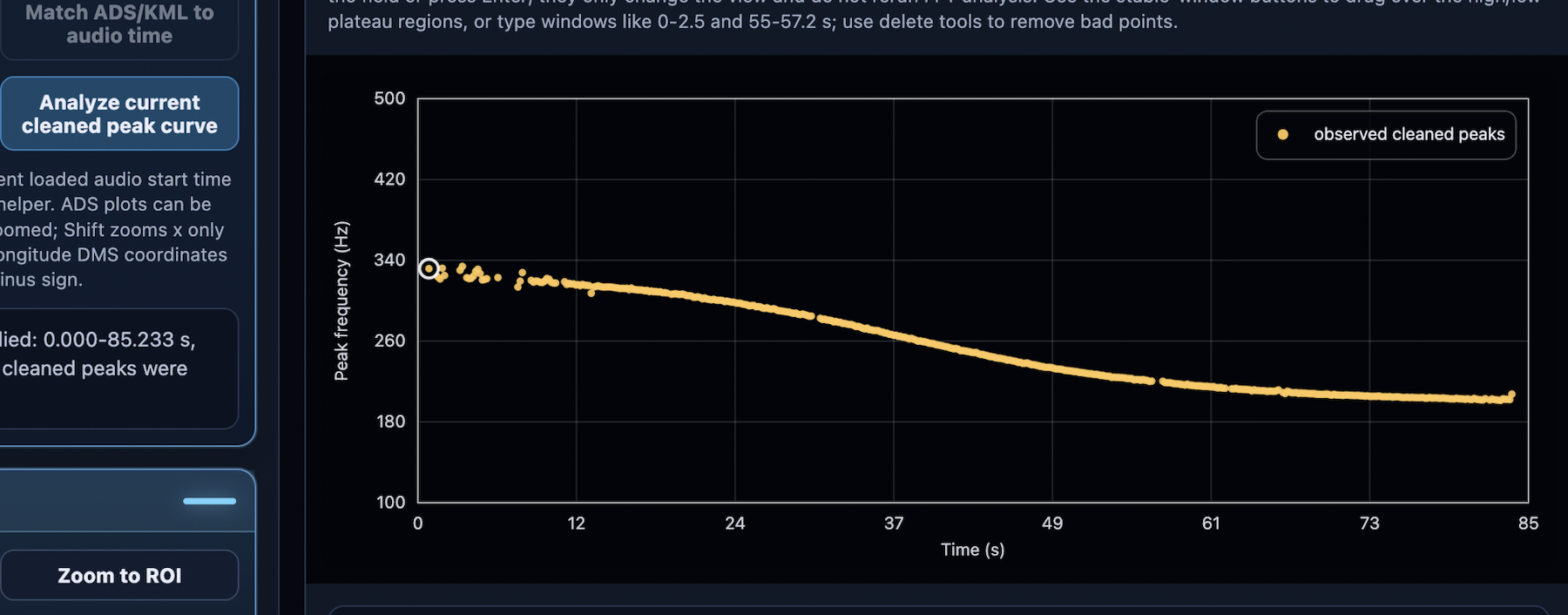}
    \caption*{\scriptsize (b)}
  \end{subfigure}
  \caption{The browser FFT tool converts a phone recording into quantitative data by
  selecting one traceable harmonic ridge from the spectrogram. (a) Spectrogram for N784MB with a visible Doppler-shifted frequency curve. The
    white guide marks show where the user points the extraction tool along the
    ridge. (b) The selected frequency--time curve after cleaning and export.}
  \label{fig:spectrogram}
\end{figure}

Simply selecting the strongest peak at each time can make the extracted curve jump between harmonics.  We developed a browser-based tool to display an uploaded audio file as a spectrogram and extract a selected frequency ridge. ChatGPT helped write the code, and the user guides the extraction by hand.
The left panel of Fig.~\ref{fig:spectrogram} shows an example for N784MB with several
harmonic ridges and background noise in an FFT spectrogram of aircraft sound.
The Doppler analysis needs one relatively clean frequency--time ridge. The user marks a few approximate points along that ridge to define a guide path. The program then searches near the marked path for the strongest
local spectral peak and exports the resulting frequency--time data as a CSV
file. The process from uploading the audio to exporting
the ridge takes only a few minutes. \revb{Two versions of the tool are provided with the
supplementary material.\cite{upload} The extraction tool ends with the CSV file. The full
tool adds the flight-track comparison and both methods of Sec.~\ref{sec:theory}, including
the full fit with sound-travel delay.}

Each recording was associated with clock time so that it could be matched to the
ADS-B track and the relevant aircraft could be identified. \revb{Instructors who make their
own recordings should use an app that saves the start time of each recording in the file
name.\cite{easyvoice} Converted to UTC, this time can be matched with the
flight track. The shared audio files
carry their UTC start time in the file name, which both tools read automatically. The phone
location is given with the data, so the closest approach and the ground-track offset $L$
can be found from the flight track.}
\section{Results and Analysis}

\begin{figure}[!htbp]
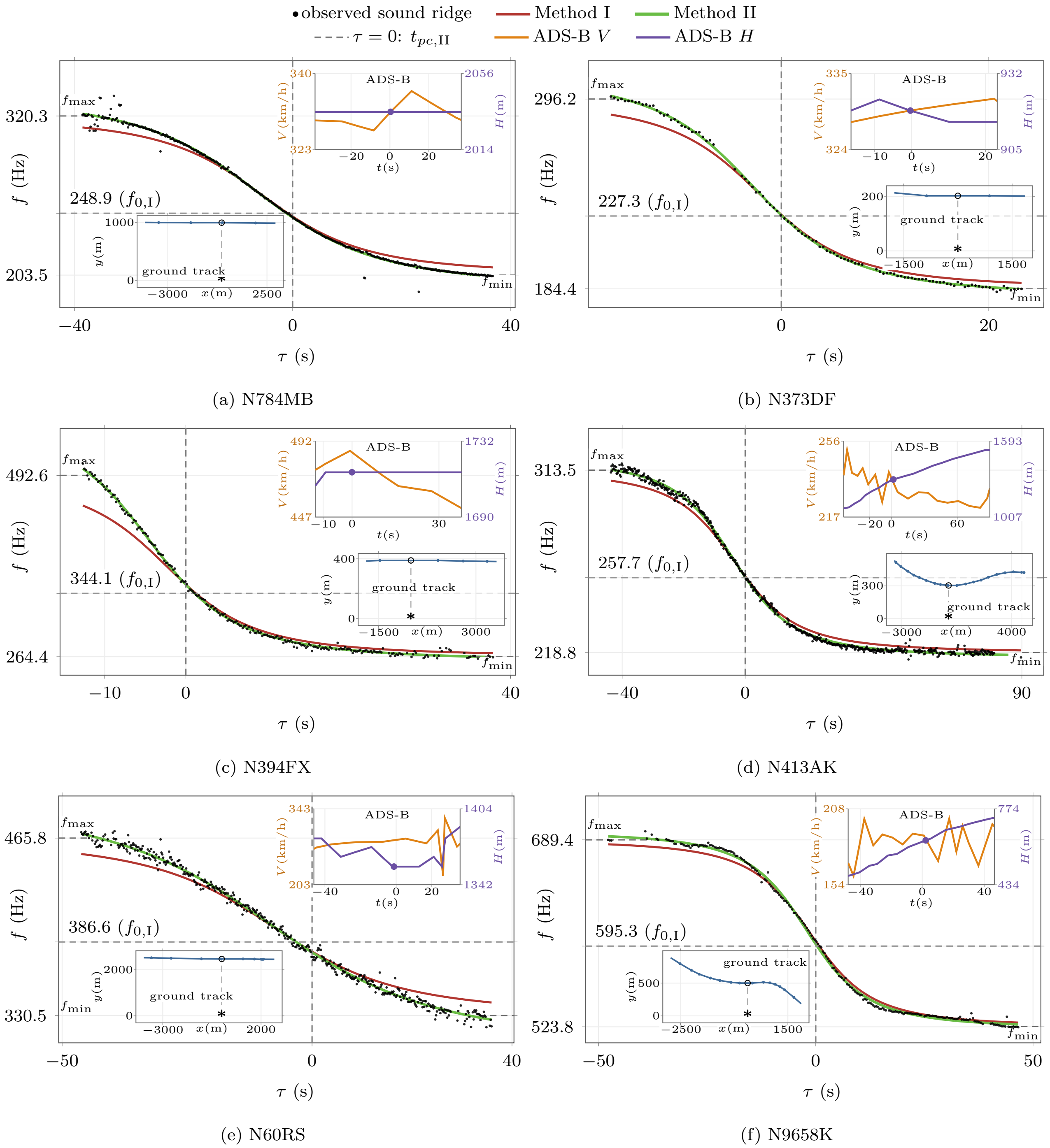

  \centering
  \PaperLegend\par \vspace{0.10cm}
  \begin{subfigure}{0.49\linewidth}
    \centering
    \makebox[\linewidth][c]{\PaperSmallCasePlot{r04n784mb}{(a) N784MB}}
    \caption*{\scriptsize (a) N784MB}
  \end{subfigure}\hfill
  \begin{subfigure}{0.49\linewidth}
    \centering
    \makebox[\linewidth][c]{\PaperSmallCasePlot{r06n373df}{(b) N373DF}}
    \caption*{\scriptsize (b) N373DF}
  \end{subfigure}
  \par\vspace{0.18cm}
  \begin{subfigure}{0.49\linewidth}
    \centering
    \makebox[\linewidth][c]{\PaperSmallCasePlot{r20n394fx}{(c) N394FX}}
    \caption*{\scriptsize (c) N394FX}
  \end{subfigure}\hfill
  \begin{subfigure}{0.49\linewidth}
    \centering
    \makebox[\linewidth][c]{\PaperSmallCasePlot{r05n413ak}{(d) N413AK}}
    \caption*{\scriptsize (d) N413AK}
  \end{subfigure}
  \par\vspace{0.18cm}
  \begin{subfigure}{0.49\linewidth}
    \centering
    \makebox[\linewidth][c]{\PaperSmallCasePlot{r11n60rs}{(e) N60RS}}
    \caption*{\scriptsize (e) N60RS}
  \end{subfigure}\hfill
  \begin{subfigure}{0.49\linewidth}
    \centering
    \makebox[\linewidth][c]{\PaperSmallCasePlot{r24n9658k}{(f) N9658K}}
    \caption*{\scriptsize (f) N9658K}
  \end{subfigure}
  \caption{Six aircraft passages. Each main panel compares the extracted
  frequency--time data with the curves from Methods I and II. The upper inset shows the
  ADS-B speed and height, with a dot marking the reference height. The lower inset shows
  the ADS-B ground track, with $x$ along and $y$ across the flight direction and the phone
  (\textbf{*}) at the origin. The horizontal dashed line marks $f_{0,\mathrm I}$ from
  Method I. Time $\tau$ is measured from the Method II closest-approach time, marked by
  the vertical dashed line.}
   \label{fig:evidence-set}
\end{figure}

Figure~\ref{fig:evidence-set} and Table~\ref{tab:cases} summarize the analysis
of six representative aircraft passages. For Method~I, $f_{\max}$ and $f_{\min}$
are the medians of a sufficient number of points at each end of the curve. The slope comes
from a straight-line fit to the points near $f_0$. Each main panel shows the frequency--time data
extracted from the aircraft sound recording and the curves obtained using Methods I and
II. The upper inset shows the ADS-B speed and height variation during the analyzed time
interval, indicating how closely the passage follows the constant-speed,
constant-height model. The lower inset shows the ADS-B ground track relative to
the phone, whose location is marked by an asterisk (\textbf{*}). This inset
shows whether the path is approximately straight and provides the offset $L$
used in the height calculation. Table~\ref{tab:cases} compares the speed,
height, and minimum slant distance obtained with both methods with the ADS-B
reference values.

 \FloatBarrier
Panels (a) and (b) have clear ridges, nearly steady ADS-B speed and height,
and approximately straight tracks. Method~II recovers both quantities closely
in these favorable passages. The two passages have different
values of \(L\), illustrating how the ground-track offset must be included when
calculating height from the minimum slant distance. For N373DF,
\(L/H\simeq0.22\), so \(R_{\min}\simeq H\) is a useful approximation. For
N784MB, \(L/H\simeq0.49\), so the independently determined \(L\) is needed in
Eq.~\eqref{eq:height}. Method~I gives low estimates in both
panels because the recordings do not extend far enough from closest approach
to reach the limiting frequencies. N784MB illustrates this built-in practical
limitation: ADS-B gives \(330~\mathrm{km\,h^{-1}}\), whereas Method~I gives
\(273~\mathrm{km\,h^{-1}}\). From Eq.~(2), \(\lvert V_r\rvert\) reaches
\(0.95V\) only for sound emitted about \(3R_{\min}/V\) before or after closest
approach. Here, \(R_{\min}/V\approx24~\mathrm{s}\), so about \(73~\mathrm{s}\)
on each side would be required, much more than is available.

Panel (c) shows how missing data on one side of the curve affect Method~I.
N394FX has steady height and a moderate speed variation of about \(7\%\), but
only a short part of the approach side is available. The high-frequency
endpoint is therefore not well determined, and Method~I gives low speed and
height values. Method~II uses all the available frequency--time data and gives
a speed close to the ADS-B value, although its calculated height remains low.
Panel (d), N413AK, has a clearly traceable frequency curve but climbs
substantially during the passage. The changing height affects the fitted
\(R_{\min}\) and the height calculated from it. Its fitted speed remains close
to ADS-B, while the reported height is the approximate value obtained by
applying the constant-height model to these data.

Panels (e) and (f) illustrate additional limitations when speed
or height changes unevenly and the flight path departs from the straight,
constant-motion model. N60RS has varying speed but comparatively steady height,
together with widely
scattered frequency points. Method II gives a speed close to the
ADS-B median for the analyzed time. Because $L$ is large
compared with $H$ ($L/H\simeq1.82$), a small change in the fitted
$R_{\min}$ produces a larger change in the calculated height. Method I does not
give a physical height because its estimate has $R_{\min}<L$. N9658K varies in
both speed and height. Its ground track also curves during the
analyzed time, violating the straight-path assumption. Both methods consequently give values
from a constant-motion model applied to the recorded data. The
speed estimates remain useful, but both height estimates are low.

\begin{table}[!htbp]
  \centering
  \scriptsize
  \setlength{\tabcolsep}{1.5pt}
  \renewcommand{\arraystretch}{1.10}
  \caption{Comparison of the six panels in Fig.~\ref{fig:evidence-set}. Speeds
  are in $\kms$, and $L$, $H$, and $R_{\min}$ are in meters, rounded to the nearest
  integer. Each height entry is $H$ ($R_{\min}$). A dash means that $R_{\min}$ is smaller
  than $L$, so no real height follows from Eq.~\eqref{eq:height}. The uncertainties come
  from the scatter of the frequency data.}
  \label{tab:cases}
  \begin{tabular*}{\linewidth}{@{\extracolsep{\fill}}c r rr rr rr L{0.20\linewidth}@{}}
    \toprule
    & & \multicolumn{2}{c}{ADS-B data}
    & \multicolumn{2}{c}{Method I}
    & \multicolumn{2}{c}{Method II}
    & \\
    \cmidrule(lr){3-4}\cmidrule(lr){5-6}\cmidrule(lr){7-8}
    Panel & $L$ & $V$ & $H\,(R_{\min})$
      & $V$ & $H\,(R_{\min})$
      & $V$ & $H\,(R_{\min})$ & Remarks\\
    \midrule
    (a) & 994 & 330 & 2035 (2264) & 273$\pm3$ & 1200$\pm45$ (1558$\pm35$) & 328$\pm3$ & 1994$\pm72$ (2228$\pm64$)
      & Steady motion; clear ridge; Method I does not reach the limiting frequencies.\\
    (b) & 203 & 330 & 919 (941) & 285$\pm1$ & 739$\pm20$ (766$\pm19$) & 328$\pm2$ & 941$\pm15$ (963$\pm15$)
      & Steady motion; near overhead; $R_{\min}\simeq H$.\\
    (c) & 389 & 475 & 1715 (1758) & 369$\pm3$ & 1175$\pm41$ (1237$\pm39$) & 456$\pm3$ & 1515$\pm32$ (1564$\pm31$)
      & Height steady; speed range about 7\% of median; missing approach data affect Method I.\\
    (d) & 304 & 229 & 1297 (1332) & 218$\pm1$ & 1267$\pm22$ (1303$\pm21$) & 236$\pm1$ & 1357$\pm16$ (1390$\pm16$)
      & Speed nearly steady; height changes during the interval.\\
    (e) & 2469 & 281 & 1357 (2817) & 208$\pm2$ & -- (1571$\pm57$) & 287$\pm4$ & 1576$\pm207$ (2929$\pm111$)
      & Speed varies; frequency points are scattered; large $L/H$ makes height more sensitive.\\
    (f) & 501 & 188 & 632 (807) & 167$\pm1$ & 486$\pm17$ (698$\pm12$) & 175$\pm1$ & 521$\pm17$ (723$\pm12$)
      & Both speed and height vary; the model gives values for the recorded passage.\\
    \bottomrule
  \end{tabular*}
  \par\vspace{0.4em}
  \parbox{0.96\linewidth}{\scriptsize For passages with changing
  altitude, the reported height is the approximate value obtained by applying
  the constant-height model to the recorded passage.}
\end{table}

\section{Discussion and Classroom Use}
\label{sec:disc}

The combined analysis shows that a phone recording and a browser-based FFT tool
can provide physically meaningful estimates of aircraft motion when a clear
harmonic ridge can be traced and the passage is reasonably consistent with the
model. Speed and minimum slant distance can be obtained from the sound recording alone.
For an offset passage, converting $R_{\min}$ to height also
requires the independently determined ground-track offset $L$. Method~I estimates
speed and minimum slant distance from a few readings of the frequency--time graph.
Using endpoint frequencies that have yet to reach their limiting values can give
low speed and height estimates. Method~II fits the selected frequency--time curve
and includes sound-travel delay. The comparison illustrates the capabilities and limitations of both methods.
For the eight passages with nearly steady speed and height, Method~II speeds
typically agree with ADS-B within 3\%, and heights within 12\%. Method~I gives speeds about
19\% lower and heights about 40\% lower, as expected from the finite recording length.
Passages with changing speed or height show larger differences.

The varied cases in the data set are useful for teaching.
Some passages give clear results, while others show the effects of an incomplete
frequency curve, changing motion, or unfavorable geometry. \rev{The supplementary material forms a rich data set. For each passage it contains the audio recording, the extracted frequency curve, the matching flight track, and the results of both methods. Students can use it to repeat the analysis}
\cite{upload}.

Several uncertainties affect the comparison. Ridge extraction
depends on spectral resolution, signal-to-noise ratio, and manual tracing.
The Method~I values also depend somewhat on how many points are used at
each end. The
model assumes a straight path, constant speed and height, and a constant emitted
frequency $f_0$; changes in engine or propeller speed may therefore bias both
$V$ and $R_{\min}$. The model assumes still air, so the fitted speed is compared with the ADS-B ground speed. Wind can affect sound propagation and bias the speed and distance estimates which is not a part of this analysis. The
analysis uses $\cs=340\,\ms$ without a local temperature correction. A 10~K temperature
difference changes the sound speed by about 1.7\%. Since $V$ and $R_{\min}$ both scale
linearly with $\cs$, the same fractional error applies to both quantities. The
ADS-B reference itself is discretely sampled. The phone elevations used for the
reference heights are given with the supplementary data. The sound is timestamped by the phone and the flight track by the ADS-B
receivers. The closest-approach times from the two typically agree within one or two
seconds, so any clock offset is small. We estimate $L$ from the flight-track data. Errors in ADS-B positions and in matching the track to the recording make $L$ uncertain. These effects limit the accuracy of the comparison.

ChatGPT and Claude helped write the code and were used in the analysis, which
made a study of this size practical. The authors chose the passages, traced the ridges, and
interpreted the results. This is one example of how generative AI can help instructors
build teaching material around real-world systems.\cite{eblenzayas2026}

\revb{The} recordings and flight-track data support several levels of
activity. Students can first listen to a passage, look at its spectrogram, and identify the
approaching and receding Doppler shifts. They can then use Method~I to estimate $V$ and
$R_{\min}$ from a few readings of the graph.
\revb{In our workshop, participants did this with the Audio Spectrum tool of the phyphox
app,\cite{staacks2018} reading the frequencies from their phone screens while a recording
of an aircraft passing overhead was played. After a guided derivation, they made the
approximations themselves. Many needed help to see why the two limiting frequencies alone
determine the speed, so this step deserves time in class. With the shared data, students
can then compare their estimates with ADS-B.} More
advanced students can fit the full curve with Method~II and see how the ground-track offset
and changing motion affect the results. \revb{We recommend that students do this fit
in an AI chat window, giving it the exported frequency--time data and short prompts such as
those in the supplementary material. Students then have to state the model, including the
sound-travel delay, in their own words. The full tool serves as a check on their results.
A useful exercise is to repeat the fit without the delay. For most passages the speed and
height change by only a few percent, and the fit looks almost as good. The closest
approach, however, always falls later than the flight track shows, by nearly the time the
sound takes to reach the phone. For N784MB it falls about 5~s late. The delay is part of
the physics, so a correct model must include it, however small its effect on the final
numbers. The comparison also shows students that a good-looking fit does not prove a
model right.}

The data also suit open-ended projects. Students can test how data selection,
fitting choices, and model assumptions change the results, and use the ADS-B values to find
where the simple model breaks down.

\section*{Acknowledgments}

P. Pathak acknowledges support from the Fulbright Fellowship administered by
the Institute of International Education, USA.

This work was partially performed under the auspices of the U.S. Department of Energy by Lawrence Livermore National Laboratory under Contract DE-AC52-07NA27344.

\section*{Author Declarations}
\subsection*{Data Availability}
The sound recordings, frequency--time data, flight-track files, \revb{results, browser tools,
and AI prompt guide} that support this study are openly available on Zenodo.\cite{upload}

\end{document}